\documentclass[twocolumn,floats,floatfix,amssymb,nofootinbib,prd,superscriptaddress]{revtex4-1}
\usepackage{graphicx,amssymb,amsmath,amsthm,amsfonts,epsfig}
\usepackage[linktocpage,colorlinks=true,citecolor=OliveGreen,linkcolor=Maroon,urlcolor=Maroon]{hyperref}
\usepackage[usenames]{color}
\usepackage{epstopdf}
\usepackage{bm}
\usepackage{dcolumn}
\usepackage[utf8]{inputenc}
\usepackage{latexsym}
\usepackage{rotating}
\usepackage{hyperref}
\usepackage{tabularx}
\usepackage{braket}
\usepackage{color}
\usepackage{longtable}
\usepackage{enumerate}
\usepackage{footmisc}
\usepackage{array}
\usepackage{tensor}
\usepackage{mathtools}
\usepackage{url}
\usepackage[dvipsnames]{xcolor}
\renewcommand{\vec}[1]{\boldsymbol{#1}}

\newcommand{\ben}{\begin{enumerate}}
\newcommand{\een}{\end{enumerate}}

\def\be{\begin{equation}}
\def\ee{\end{equation}}
\def\bea{\begin{eqnarray}}
\def\eea{\end{eqnarray}}
\newcommand{\beq}{\begin{eqnarray}}
\newcommand{\eeq}{\end{eqnarray}} 
\newcommand{\ba}{\begin{align}}
\newcommand{\ea}{\end{align}}
\def\ba{\bar{a}}

\begin{document}
\title{Superradiance:~Axionic Couplings and Plasma Effects}
\author{Thomas F.M.~Spieksma}
\affiliation{Niels Bohr International Academy, Niels Bohr Institute, Blegdamsvej 17, 2100 Copenhagen, Denmark}
\author{Enrico Cannizzaro}
\affiliation{Dipartimento di Fisica, ``Sapienza" Universit\`{a} di Roma \& Sezione INFN Roma1, Piazzale Aldo Moro 5, 00185, Roma, Italy}
\author{Taishi Ikeda}
\affiliation{Niels Bohr International Academy, Niels Bohr Institute, Blegdamsvej 17, 2100 Copenhagen, Denmark}
\author{Vitor Cardoso}
\affiliation{Niels Bohr International Academy, Niels Bohr Institute, Blegdamsvej 17, 2100 Copenhagen, Denmark}
\affiliation{CENTRA, Departamento de F\'{\i}sica, Instituto Superior T\'ecnico -- IST, Universidade de Lisboa -- UL,
Avenida Rovisco Pais 1, 1049 Lisboa, Portugal}
\author{Yifan Chen}
\affiliation{Niels Bohr International Academy, Niels Bohr Institute, Blegdamsvej 17, 2100 Copenhagen, Denmark}
\begin{abstract}
Spinning black holes can transfer a significant fraction of their energy to ultralight bosonic fields via superradiance, condensing them in a co-rotating structure or ``cloud.'' This mechanism turns black holes into powerful particle detectors for bosons with extremely feeble interactions. To explore its full potential, the couplings between such particles and the Maxwell field in the presence of plasma need to be understood.

In this work, we study these couplings using numerical relativity. We first focus on the coupled axion-Maxwell system evolving on a black hole background. By taking into account the axionic coupling concurrently with the growth of the cloud, we observe for the first time that a new stage emerges:~that of a stationary state where a constant flux of electromagnetic waves is fed by superradiance, for which we find accurate analytical estimates. Moreover, we show that the existence of electromagnetic instabilities in the presence of plasma is entirely controlled by the axionic coupling; even for dense plasmas, an instability is triggered for high enough couplings.
\end{abstract}
\maketitle
\tableofcontents
%%%%%%%%%%%%%%%%%%%%%%%%%%%%%%%%%
\section{Introduction}
%%%%%%%%%%%%%%%%%%%%%%%%%%%%%%%%%
Black holes (BHs) populate the cosmos with a wide range of masses varying across 8 or more orders of magnitude. Accretion of baryonic matter or mergers with other BHs and stars increase the BH mass and may provide it with a considerable spin, turning BHs into engines that may power other phenomena. One of such mechanisms relies on superradiance (SR)~\cite{ZelDovich1971,ZelDovich1972,Starobinsky:1973aij,Brito:2015oca} and makes BHs ideal tools for particle physics. Indeed, in the presence of new light, bosonic degrees of freedom, spinning BHs are unstable, and spin down while depositing a substantial fraction of their energy into a bosonic cloud~\cite{Arvanitaki:2009fg,Arvanitaki:2010sy,Brito:2014wla,Brito:2015oca}. This mechanism does not require any significant initial abundance of the bosonic field, as it relies on a linear instability:~any small initial field amplitude will lead to a sizeable and potentially measurable effect. Therefore, the SR mechanism is a compelling way to search for new degrees of freedom that may or may not be a significant component of dark matter~\cite{Bergstrom:2009ib,Hui:2016ltb,Marsh:2015xka,Fairbairn:2014zta}.

The precise development of the instability is well understood in vacuum and in the absence of couplings to the Standard Model~\cite{Arvanitaki:2009fg,Arvanitaki:2010sy,Brito:2014wla,Brito:2015oca,East:2018glu}. However, 
BHs are surrounded by interstellar matter or accretion disks and couplings between bosonic fields and the Standard Model may be non-vanishing. It was argued analytically and with numerical simulations, that axionic couplings to the Maxwell sector might trigger {\it parametric instabilities}, whereby the scalar cloud transfers energy to electromagnetic (EM) radiation~\cite{Rosa:2017ury,SenPlasma,Boskovic:2018lkj,Ikeda:2018nhb}. Additionally, the presence of a surrounding plasma may quench the parametric instability due to the high energy (large ``effective mass'') of typical astrophysical environments~\cite{Cardoso:2020nst,Cannizzaro:2021zbp,Blas:2020kaa}. The previous works left important gaps:~(i) the parametric instability was shown to give rise to periodic bursts of light, but its period and amplitude were not studied. In fact, the effect of a SR growing cloud was also not understood properly.\footnote{As we show here, bursts will in general not occur, but give way to a stationary emission of light.}~(ii)~The role of plasmas in the development of EM instabilities is known poorly, but could have a drastic effect (see e.g.~recent works on dark photon SR~\cite{Caputo:2021efm,Siemonsen:2022ivj}), since the plasma frequency is rather large in most astrophysical circumstances.

The outline of this work is as follows. In Section~\ref{sec:setup}, we set up the relevant equations of motion and we discuss the modeling of the cloud as well as the plasmic environment. In Section~\ref{sec:withoutSR}, we study the evolution of the axion-Maxwell system in the absence of SR. In Section~\ref{sec:withSR}, we do a similar exercise yet now while including SR. In Section~\ref{sec:surroundingplasma}, we describe the influence of a plasma on the EM instability. In Section~\ref{sec:observational}, we discuss possible observational signatures. Finally, we summarize and conclude in Section~\ref{sec:conclusions}.

This work contains a number of appendices with additional details. In Appendix~\ref{app:freescalars}, we study the time evolution of massive scalar fields around BHs. In Appendix~\ref{sec:Waveextraction}, we discuss the wave extraction from our simulations. In Appendix~\ref{app:theplasma}, we motivate the assumptions of our plasma model. In Appendix~\ref{sec:appendDecomp}, we formulate our equations of motion as an initial value problem. In Appendix~\ref{app:convergence}, we show the convergence of our code. In Appendix~\ref{appB:higherorder}, we report higher multipoles. In Appendices~\ref{sec:appendMathieu} and~\ref{app:plasmaMathieu}, we study the Mathieu equation in presence of SR and plasma, respectively. Finally, in Appendix~\ref{sec:selectionrules}, we discuss the spherical harmonics decomposition of the Maxwell equations.

We adopt the mostly positive metric signature and use geometrized units in which $G = c = 1$ and rationalized Heaviside-Lorentz units for the Maxwell equations, unless otherwise stated.

%%%%%%%%%%%%%%%%%%%%%%%%%%%%%%%%%%%%%%%
\section{Setup}\label{sec:setup}
%%%%%%%%%%%%%%%%%%%%%%%%%%%%%%%%%%%%%%%
\begin{figure*}
    \includegraphics[width = 0.95\textwidth]{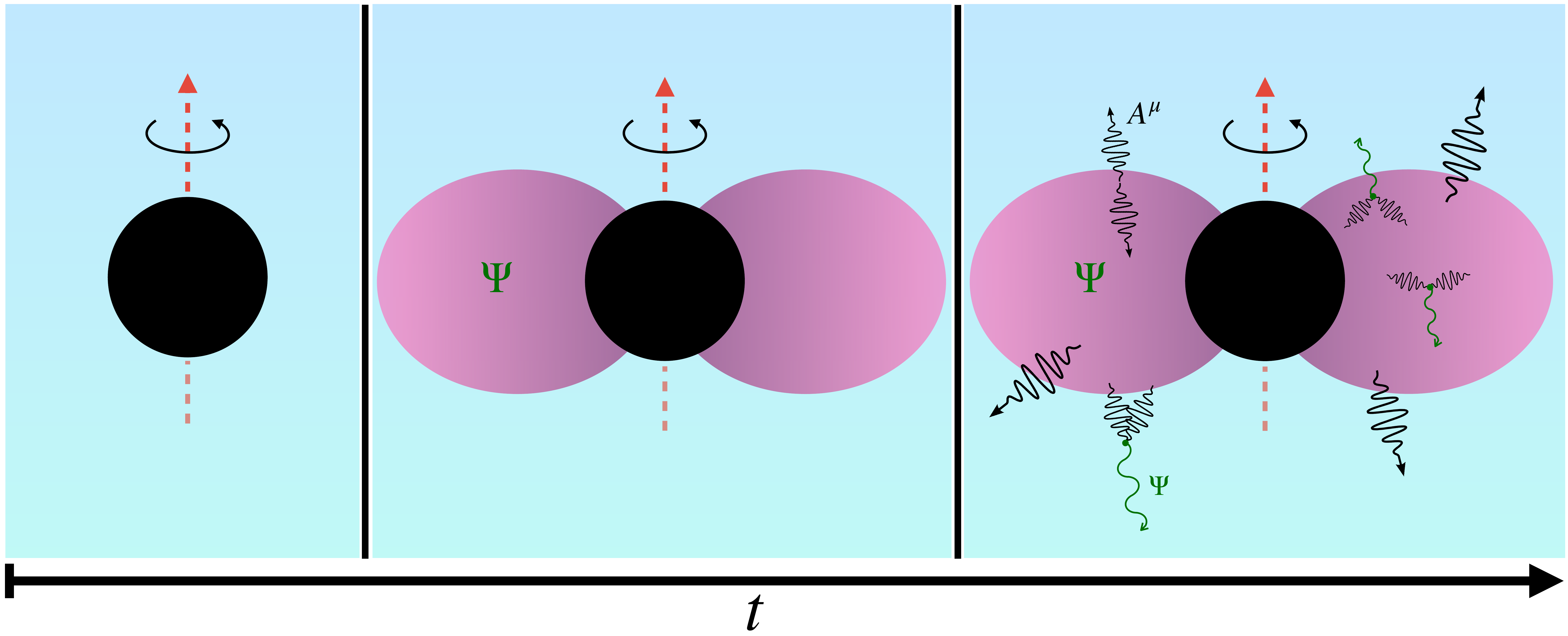}
    \caption{Schematic illustration of our setup. Starting from a spinning BH of mass $M$ ({\it left}), a SR cloud of mass $M_{\rm c}$ is formed in the dominant (dipolar) growing mode ({\it center}). For sufficiently large couplings to the Maxwell sector, the configuration is unstable:~any small EM fluctuation will trigger emission of EM radiation (black lines, {\it right} panel). Some of these waves recombine to create axion waves (green lines). The blue background indicates the presence of a plasma.}
    \label{fig:Evolution}
\end{figure*}
%%%%%%%%%%%%%%%%%%%%%%%%%%%%%%%%%%%%%%
\subsection{The theory}
%%%%%%%%%%%%%%%%%%%%%%%%%%%%%%%%%%%%%%
We consider a real, massive (pseudo)scalar field $\Psi$ with axionic couplings to the EM field. In addition, the EM field is coupled to a cold, collisionless electron-ion plasma. In this setup, the Lagrangian takes on the following form:
\begin{equation}\label{eq:lagrangian}
    \begin{aligned}
        \mathcal{L}&=\frac{R}{16\pi}-\frac{1}{4}F_{\mu\nu}F^{\mu\nu}-\frac{1}{2}\nabla_{\mu}\Psi\nabla^{\mu}\Psi
-\frac{\mu^{2}}{2}\Psi^{2}\\&-\frac{k_{\rm a}}{2}\Psi\,{}^{*}\!F^{\mu\nu}F_{\mu\nu}
+A^\mu j_\mu + \mathcal{L}_{\rm m}\,.
\end{aligned}
\end{equation}
The mass of the scalar field $\Psi$ is given by $m_{\rm a} = \mu \hbar$, $A_{\mu}$ is the vector potential, $F_{\mu\nu} \equiv \nabla_{\mu}A_{\nu}-\nabla_{\nu}A_{\mu}$ is the Maxwell tensor and ${}^{*}\!F^{\mu\nu} \equiv \frac{1}{2}\epsilon^{\mu\nu\rho\sigma}F_{\rho\sigma}$ is its dual. We use the definition $\epsilon^{\mu\nu\rho\sigma}\equiv \frac{1}{\sqrt{-g}}E^{\mu\nu\rho\sigma}$, where $E^{\mu\nu\rho\sigma}$ is the totally anti-symmetric Levi-Civita symbol with $E^{0123}=1$. We define the Lagrangian for the plasma as $\mathcal{L}_{\rm m}$, while $k_{\rm a}$ quantifies the axionic coupling which we take to be constant. There exists a wide variety of theories predicting axions and axion-like particles, and generically $k_{\rm a}$ is independent of the boson mass. Therefore, we take $k_{\rm a}$ to be an additional free parameter of the theory. Notice that we do not consider self-interactions, which could appear as an expansion of the axion's periodic potential. This corresponds to a region $k_{\rm a} f_{\rm a} \geq \mathcal{O}(1)$ predicted in models such as clockwork axions~\cite{Kaplan:2015fuy,Farina:2016tgd} and magnetic monopoles in the anomaly loop~\cite{Sokolov:2021ydn,Sokolov:2022fvs}, where $f_a$ is the decay constant of the axion.\footnote{The strong self-interaction regime was discussed in e.g.~\cite{Yoshino:2012kn, Baryakhtar:2020gao,Omiya:2022gwu,Chia:2022udn}, where transitions to various cloud modes and distortion of bound state wave functions are expected.} In principle, a similar analysis could be performed for scalar couplings ($\mathcal{L} \supset k_{\rm s} \Psi F^{\mu \nu} F_{\mu \nu}$), at least when the coupling strength is weak~\cite{Boskovic:2018lkj}.

Finally, $j_\mu$ is the plasma current, and captures both the contributions of the electrons and the much heavier ions. In this work, we adopt a two-fluid formalism model for the plasma, where electrons and ions are treated as two different fluids, coupled through the Maxwell equations.\footnote{Other plasma models such as kinetic theory or relativistic magneto-hydrodynamics (GRMHD) are not suitable for our purposes. The former is necessary for phenomena where the fluid’s velocity distribution plays an important role. Even though this leads to a correction in the dispersion relation of the photons~\cite{KrallTrivelpiece1973}, they are not relevant for the problem at hand. GRMHD is appropriate for describing the behavior of plasma on long timescales, yet fails to capture the high-frequency oscillations at the plasma frequency scale that provide the photon with an effective mass, i.e.,~it is valid only for $t\gg\omega_{\rm p}^{-1}$, where $\omega_{\rm p}$ is the plasma frequency~\eqref{eq:plasmafreq}. The model we adopt instead, does correctly capture the photon's effective mass, while being computationally lighter.} Hence, the plasma current is given by $j^\mu= \sum_s q_s n_s u^\mu_s$, where the index $s$ represents the sum over the two different species, electrons and ions, and $q_s,n_s,u_s^\mu$ are the charge, number density and four velocity of the fluids, respectively. 

An axion cloud produced from SR can grow to be $\lesssim 10\%$ of the BH mass~\cite{East:2017ovw, Herdeiro:2021znw,Brito:2014wla}. We will consider the cloud's backreaction on the geometry to be small and thus evolve the system on a fixed background. The gravitational coupling $\mu M$ determines the strength of the interaction between the BH and the axion and is a crucial quantity. In order for SR to be efficient on astrophysical timescales, the gravitational coupling must be $\mathcal{O} (1)$. For $\mu M \ll 0.1$, the exponential growth is too slow, while the instability is exponentially suppressed for $\mu M \gg 1$~\cite{Zouros:1979iw,Berti:2009kk}. Consequently, we will perform simulations in the range $\mu M \sim 0.1 -0.3$. 

From the Lagrangian of our theory~\eqref{eq:lagrangian}, we obtain the equations of motion for the scalar and EM field. In order to close the system, we also need to consider the continuity and momentum equation of the fluids, which come from the conservation of the energy-momentum tensor for the Maxwell-plasma sector. Ignoring the backreaction of the fields in the spacetime, we obtain: 
\begin{equation}\label{eq:evoleqns}
\begin{aligned}
\left(\nabla^\mu \nabla_\mu-\mu^2\right) \Psi &= \frac{k_{\mathrm{a}}}{2}\,{ }^*\!F^{\mu \nu} F_{\mu \nu}\,,\\
\nabla_\nu F^{\mu \nu} &=j^{\mu} -2 k_{\mathrm{a}}{ }^*\!F^{\mu \nu} \nabla_\nu \Psi\,,\\
u_s^{\nu}\nabla_{\nu}u_s^{\mu} &= \frac{q_s}{m_s}F^{\mu}_{\ \nu}u^{\nu}_{s}\,,\\ 
\nabla_\mu (n_s u_s^\mu)&=0\,,
\end{aligned}
\end{equation}
where the index $s$ denotes again the particular fluid species. Finally, we impose the Lorenz condition on the vector field
\begin{equation}
    \nabla_{\mu}A^{\mu} = 0\,,
\end{equation}
thereby fixing our gauge freedom.
%%%%%%%%%%%%%%%%%%%%%%%%%%%%%%%%%%%%%%
\subsection{Modeling superradiance}
%%%%%%%%%%%%%%%%%%%%%%%%%%%%%%%%%%%%%%
Even though we are interested in an axion cloud that grows through SR, and thus requires a spinning BH described by the Kerr metric, we will instead mimic SR growth without the need of a spinning BH. The reason is of a practical nature:~timescales to superradiantly grow an axion cloud are larger than $\sim 10^6M$~\cite{Cardoso:2005vk,Dolan:2007mj,Berti:2009kk}, a prohibitively large timescale for our purposes. Therefore, we mimic SR growth following Zel'dovich~\cite{ZelDovich1971,ZelDovich1972,Cardoso:2015zqa}, by adding a simple Lorentz-invariance-violating term to the Klein-Gordon equation,
\begin{equation}\label{eq:ASR}
    \left(\nabla^\mu \nabla_\mu-\mu^2\right) \Psi = C \frac{\partial \Psi}{\partial t}+\frac{k_{\mathrm{a}}}{2}\,{ }^*\!F^{\mu \nu} F_{\mu \nu}\,.
\end{equation}
Here, $C$ is a constant, which in the absence of the axionic coupling gives rise to a linear instability on a timescale of the order $1/C$, where we can tune $C$ to be within our numerical limits. For further details, we refer to Appendix~\ref{sec:app_artificial_super}.
%%%%%%%%%%%%%%%%%%%%%%%%%%%%%%%%%%%%%%
\subsection{Modeling plasma}\label{sec:Plasma}
%%%%%%%%%%%%%%%%%%%%%%%%%%%%%%%%%%%%%%
One of the most important characteristics of plasmas is their peculiar response to external perturbations. When plasma is perturbed by an EM wave, electrons are displaced and start oscillating around their equilibrium position with the so-called {\it plasma frequency}: 
\begin{equation}
    \label{eq:plasmafreq}
     \omega_{\rm p}= \sqrt{\frac{n_{\rm e} q_{\rm e}^2}{m_{\rm e}}}\approx \frac{10^{-12}}{\hbar}\sqrt{\frac{n_{\rm e}}{10^{-3}\text{cm}^{-3}}}\,\text{eV}\,.
\end{equation}
Remarkably, the dispersion relation of the transverse modes of a photon propagating in a plasma are modified by a gap which corresponds to the plasma frequency, i.e.,~$\omega^2=k^2 + \omega_{\rm p}^2$. For this reason, the plasma frequency acts as an effective mass for the transverse polarizations of the photons. This effect is crucial to take into account when studying parametric instabilities as the axion decay into photons could be suppressed in a dense plasma, i.e.,~when $\omega_{\rm p}\gg\mu$. Throughout this work, we work under the following assumptions regarding the plasma. Details and motivations are provided in Appendix~\ref{app:theplasma}.
\begin{enumerate}[(i)]
    \item We drop non-linear terms in the axion-photon-plasma system.
    \item We ignore the oscillations of the ions due to the EM field. 
    \item We consider a locally quasi-neutral plasma as initial data.
    \item We assume a cold and collisionless plasma.
    \item We neglect the gravitational influence on the evolution of the fluid's four velocity as measured by an {\it Eulerian observer}.\footnote{An Eulerian observer is defined as the observer that has its worldline orthogonal to the spacelike hypersurface.\label{ft:Eulerian}} 
\end{enumerate}
%%%%%%%%%%%%%%%%%%%%%%%%%%%%%%%%%%%%%%%%%%%%%%%%%%%%%%
\subsection{Numerical procedure}
%%%%%%%%%%%%%%%%%%%%%%%%%%%%%%%%%%%%%%%%%%%%%%%%%%%%%
\begin{table}
\centering
\renewcommand*{\arraystretch}{1.3}
\resizebox{0.95\linewidth}{!}{%
\begin{tabular}{ | c || c| c | c | c | } 
  \hline
  Run & $k_{\rm a}\Psi_{0}$ & $\mu M$ &  $10^{3}CM$ & $ 10^{4} E_{0}M/\Psi_{0}$\\ 
  \hline \hline
  $\mathcal{I}_{1}$ & $0.0$ & $0.3$ & $0.0$ &  $8.1$ \\ 
  \hline
  $\mathcal{I}_{2}$ & $0.0295$ & $0.3$ & $0.0$ &  $8.1$ \\ 
  \hline
  $\mathcal{I}_{3}$ & $0.147$ & $0.3$ & $0.0$ & $8.1$ \\ 
  \hline \hline
  $\mathcal{J}_{1}$ & $0.0737$ & $0.2$ & $4.0$ &  $100.0$ \\ 
  \hline
  $\mathcal{J}_{2}$ & $0.0737$ & $0.2$ & $4.0$ &  $1.0$ \\ 
  \hline
  $\mathcal{J}_{3}$ & $0.0737$ & $0.2$ & $4.0$ &  $0.01$ \\ 
  \hline
  $\mathcal{J}_{4}$ & $0.0737$ & $0.2$ & $4.0$ &  $0.0001$ \\ 
  \hline \hline
  $\mathcal{J}_{5}$ & $0.0737$ & $0.2$ & $4.0$ &  $8.1$ \\ 
  \hline
  $\mathcal{J}_{6}$ & $0.0563$ & $0.2$ & $4.0$ &  $8.1$ \\ 
  \hline
  $\mathcal{J}_{7}$ & $0.0328$ & $0.2$ & $4.0$ &  $8.1$ \\ 
  \hline
  $\mathcal{J}_{8}$ & $0.00737$ & $0.2$ & $4.0$ & $8.1$ \\ 
  \hline \hline
  $\mathcal{J}_{9}$ & $0.0737$ & $0.2$ & $0.08$ &  $8.1$ \\ 
  \hline
  $\mathcal{J}_{10}$ & $0.0737$ & $0.2$ & $0.2$ &  $8.1$ \\ 
  \hline
  $\mathcal{J}_{11}$ & $0.0737$ & $0.2$ & $0.8$ &  $8.1$ \\ 
  \hline
  $\mathcal{J}_{12}$ & $0.0737$ & $0.2$ & $1.0$ &  $8.1$ \\ 
  \hline
  $\mathcal{J}_{13}$ & $0.0737$ & $0.2$ & $2.0$ &  $8.1$ \\ 
  \hline
  $\mathcal{J}_{14}$ & $0.0737$ & $0.2$ & $8.0$ &  $8.1$ \\ 
  \hline \hline
   &  &  &  &  $\omega_{\rm p} M$ \\ 
  \hline
  $\mathcal{K}_{1}$ & $0.147$ & $0.3$ & $0.0$ &  $0.01$ \\ 
  \hline
  $\mathcal{K}_{2}$ & $0.147$ & $0.3$ & $0.0$ &  $0.1$ \\ 
  \hline
  $\mathcal{K}_{3}$ & $0.147$ & $0.3$ & $0.0$ &  $0.15$ \\ 
  \hline
  $\mathcal{K}_{4}$ & $0.147$ & $0.3$ & $0.0$ &  $0.2$ \\ 
  \hline
  $\mathcal{K}_{5}$ & $0.147$ & $0.3$ & $0.0$ &  $0.3$ \\ 
  \hline
  $\mathcal{K}_{6}$ & $0.147$ & $0.3$ & $0.0$ &  $0.4$ \\ 
  \hline
  $\mathcal{K}_{7}$ & $0.295$ & $0.3$ & $0.0$ &  $0.2$ \\ 
    \hline
  $\mathcal{K}_{8}$ & $0.590$ & $0.3$ & $0.0$ &  $0.2$ \\ 
  \hline
  $\mathcal{K}_{9}$ & $0.0737$ & $0.1$ & $2.0$ &  $0.02$ \\ 
  \hline
  $\mathcal{K}_{10}$ & $0.0737$ & $0.1$ & $2.0$ &  $0.07$ \\ 
  \hline
\end{tabular}}
\caption{Summary of the simulations discussed in the main text. $\mathcal{I}_{i}$ simulations do not include plasma nor SR growth. $\mathcal{J}_{i}$ simulations do include SR growth yet are still without plasma, while $\mathcal{K}_{i}$ simulations do include the plasma. We denote the axionic coupling $k_{\rm a}\Psi_{0}$, the mass coupling $\mu M$, the artificial SR parameter $C$, and the ratio between the initial amplitude of the electric field $E_{0}$~\eqref{eq:InitialElectric} and the scalar field $\Psi_{0}$~\eqref{eq:normalization}. In the simulations that include the plasma ($\mathcal{K}_{i}$), we report the plasma frequency $\omega_{\rm p}$ as well, while the initial EM amplitude is $10^{4}E_{0}M/\Psi_0 = 8.1$, which is not shown in the table due to space limitations. In {\it all} our simulations, we initialize the EM pulse at $r_{0} = 40M$ with $\sigma = 5M$.}
\label{tb:simulations}
\end{table}

To evolve the system, we solve numerically the equations of motion~\eqref{eq:evoleqns} around a Schwarzschild BH with mass $M$. We denote the spatial part of the Maxwell field $\mathcal{A}_{i}$, the electric field $E^{i}$, the magnetic field $B^{i}$, an auxiliary field $\mathcal{Z}$, and finally the conjugate momentum $\Pi$ of the scalar field. Using these variables and applying the 3+1 decomposition to the equations of motion, we obtain the evolution equations for the scalar field, EM field, and the plasma. A detailed account of the formulation of our system as a Cauchy problem can be found in Appendix~\ref{sec:appendDecomp}.

Besides the evolution equations, the 3+1 decomposition also provides us with a set of constraint equations, which are shown explicitly in~\eqref{eq:constrainteqn}. The initial data we construct should satisfy these equations. For the electric field, we assume the following profile:
\begin{equation}
\begin{aligned}
\label{eq:InitialElectric}
E^r &=E^\theta=\mathcal{A}_i=0\,,\\
E^{\varphi} &=E_{0}e^{-\left(\frac{r-r_{0}}{\sigma}\right)^{2}} M\,,
\end{aligned}
\end{equation}
where $E^{i}=F^{i\mu}n_{\mu}\,(i=r,\theta,\varphi)$, with $n_{\mu}$ defined as the normal vector of the spacetime foliation. Here, $E^{\varphi}$ can be an arbitrary function of $r$ and $\theta$. We choose a Gaussian profile with $E_{0}$, $r_{0}$, and $\sigma$ the typical amplitude, radius and width of the Gaussian, respectively. The EM pulse is initialized in {\it all} our simulations at $r_{0} = 40M$ with $\sigma = 5M$. Moreover, we have tested that our results do not depend on these factors, thus confirming their generality. For the initial data of the scalar field, we use a quasi-bound state that is constructed through Leaver's method (see Appendix~\ref{appA:boundstates}). We consider the cloud to occupy the dominant (dipolar) growing mode with an amplitude $\Psi_{0}$, whose normalization is defined in~\eqref{eq:normalization}. Finally, the constraint equation for the plasma is trivially satisfied as we explain in Appendix~\ref{app:evolPlasma}, and for simplicity we take a constant density plasma as initial data.

To keep track of the scalar and EM field during the time evolution, we perform a multipolar decomposition. In the scalar case, we directly project the field $\Psi$ onto spheres of constant coordinate radius using the spherical harmonics with spin weight $s_{\rm w} = 0$ to obtain $\Psi_{\ell m}$~\eqref{eq:scalarextract}. In the EM case, we track the evolution of the field using the Newman-Penrose scalar $\Phi_{2}$, which captures the outgoing EM radiation at infinity (see Appendix~\ref{sec:Waveextraction}). Analogous to the scalar case, we project these using spherical harmonics, yet now using spin weight $s_{\rm w} = -1$ to obtain $(\Phi_{2})_{\ell m}$~\eqref{eq:NPextract}. In our figures, we show $|(\Phi_{2})_{\ell m}| = \sqrt{(\Phi_{2})^{*}_{\ell m}(\Phi_{2})_{\ell m}}$. Since this captures massless waves, $|(\Phi_{2})_{\ell m}| \propto 1/r$ at large spatial distances.

Throughout this work, we will discuss various simulations. In Table~\ref{tb:simulations}, the specific parameters of these simulations are listed. Furthermore, a schematic illustration of our setup can be seen in Fig.~\ref{fig:Evolution}.
%%%%%%%%%%%%%%%%%%%%%%%%%%%%%%%%%%%%%%%%%%%%%%%%%%%%%%%%
\section{Superradiance turned off}\label{sec:withoutSR}
%%%%%%%%%%%%%%%%%%%%%%%%%%%%%%%%%%%%%%%%%%%%%%%%%%%%%%%%
An axion cloud coupled to the Maxwell sector can give rise to a burst of EM radiation. The initial explanation of this phenomenon was outlined in~\cite{RosaKephartStimulated}, while the full numerical exercise followed in~\cite{Boskovic:2018lkj, Ikeda:2018nhb}. In this section, our goal is to carefully perform a further analysis and, as we will see, find some new features of the system. Throughout this section, we assume SR growth to be absent, as in~\cite{Boskovic:2018lkj, Ikeda:2018nhb}. Even though this is clearly an artificial assumption, as it means that the cloud was allowed to grow without being coupled to the Maxwell sector, it allows us to isolate and understand better some of the phenomena. The full case will be dealt with afterwards.

As shown analytically on a flat spacetime, but also numerically in a Kerr background~\cite{Boskovic:2018lkj, Ikeda:2018nhb}, upon growing the cloud to some predetermined value, an EM instability is triggered depending on the quantity $k_{\rm a}\Psi_{0}$. In particular, there exist two regimes, a {\it subcritical} regime and a {\it supercritical} regime.  In the former, no instability is triggered and some initial EM fluctuation does not experience exponential growth. Conversely, in the supercritical regime, an instability is triggered and the axion field ``feeds'' the EM field, which grows exponentially, resulting in a burst of radiation.
The boundary between these regimes is set by two competing effects; the parametric production rate of the photon, $\propto \mu k_{\rm a} \Psi_0$, and their escape rate from the cloud, $\propto \mu^{2} M$. The latter is approximated by the inverse of the cloud size. Similarly to previous works, we find the boundary to be on the order $k_{\rm a} \Psi_0 \sim 0.1-0.4$ for $\mu M \sim 0.2-0.3$.
\begin{figure}
   \hspace*{-0.3cm}
    \includegraphics[width = 0.48\textwidth]{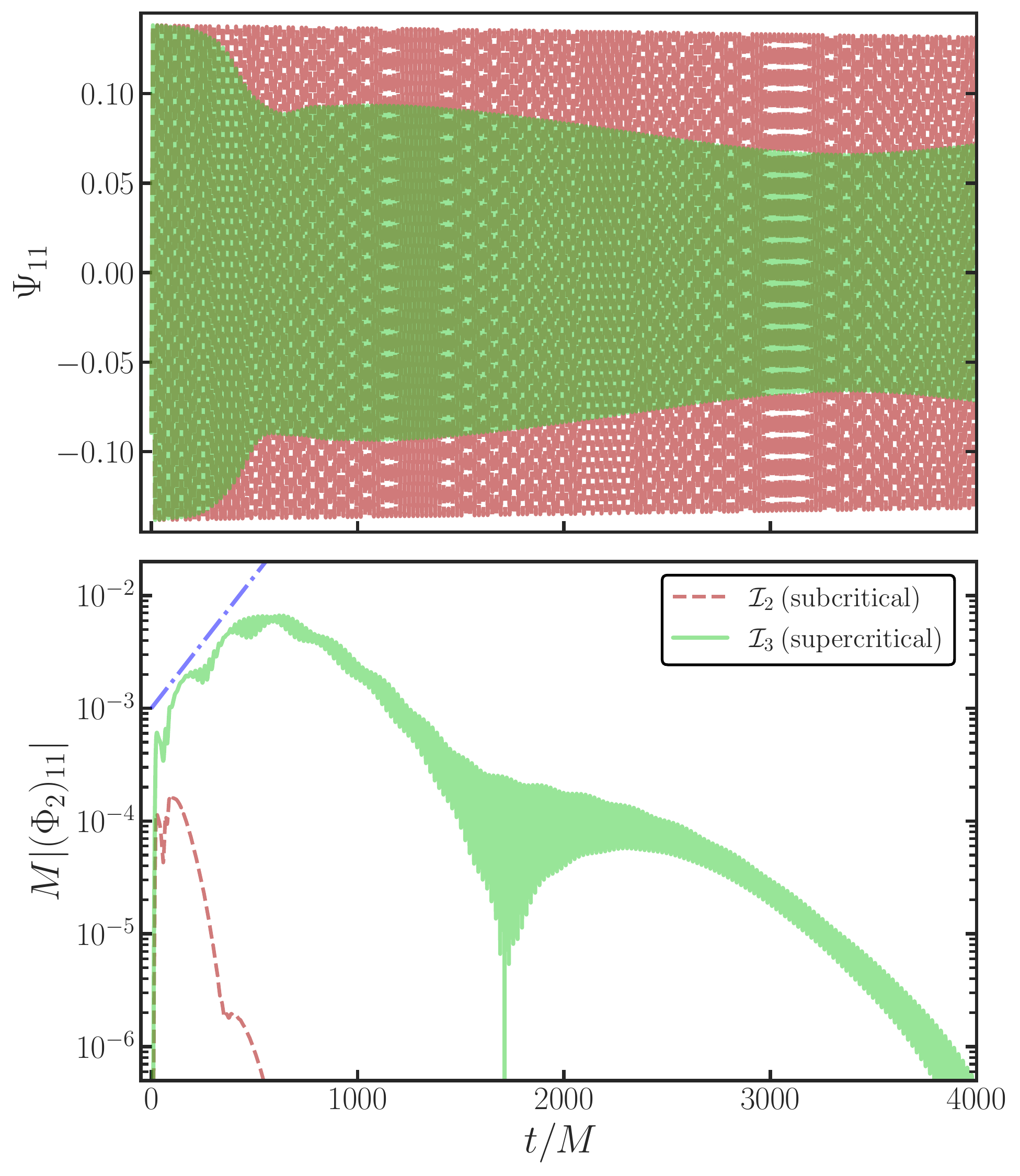}
    \caption{{\bf Top Panel:} The time evolution of the (real part of the) dipolar, $\ell = m = 1$, bound state component of an axion cloud around a Schwarzschild BH in two scenarios: coupling is subcritical (dashed red, $\mathcal{I}_{2}$) or supercritical (green, $\mathcal{I}_{3}$). As a consistency check, we also evolve with a vanishing coupling to the Maxwell sector ($\mathcal{I}_{1}$), which we find to be almost indistinguishable from the subcritical case.
    {\bf Bottom Panel:} The time evolution of the absolute value of the $\ell = m = 1$ component of the Newman-Penrose scalar $\Phi_{2}$ for a subcritical and supercritical coupling. The blue dash-dotted line shows the growth rate, $\lambda = 0.0054$, as estimated from equation~$(77)$ of~\cite{Boskovic:2018lkj}. In both panels, the field is extracted at $r_{\rm ex} = 20M$ and $\mu M = 0.3$.}
    \label{fig:BurstmuM03r20}
\end{figure}
%%%%%%%%%%%%%%%%%%%%%%%%%%%%%%%%%%%%%%%%%%%%%%%%%%%%%%%%%%%%%%%%
\subsection{The process at large}
%%%%%%%%%%%%%%%%%%%%%%%%%%%%%%%%%%%%%%%%%%%%%%%%%%%%%%%%%%%%%%%%
In the following, we explore these two regimes by evolving the coupled system describing a SR cloud of axions coupled to the Maxwell sector, while we initialize it with a small vector fluctuation $E_{0}$. 

Figure~\ref{fig:BurstmuM03r20} summarizes well the possible outcomes, which depend on the strength of the coupling $k_{\rm a} \Psi_0$. For small enough couplings to the Maxwell sector, the axion field is left unaffected, and remains in a bound state of (near) constant amplitude around the BH. For large couplings however, in what we term the {\it supercritical} regime, the amplitude of the axion field decreases. This transition signals a parametric instability whereby axions are quickly converted into photons. 

The bottom panel of Fig.~\ref{fig:BurstmuM03r20} shows the behavior of the EM radiation during this process (we show only the dipolar component $\ell = m = 1$ of the Newman-Penrose scalar, but we find that higher modes are also excited to important amplitudes, see Appendix~\ref{appB:higherorder}). In the subcritical regime, any initial EM fluctuation decays on short timescales. However, in the supercritical regime a burst is initiated; these are the photons that are created by the axion cloud. Furthermore, we find the growth rate to follow estimates from earlier work~\cite{Boskovic:2018lkj}. Specifically, it is approximated by taking the production rate of the photons and subtracting the rate at which photons leave the cloud: $\lambda \sim \lambda_{*} - \lambda_{\rm esc}$, where $\lambda_{*} \sim \frac{1}{2}\mu k_{\rm a} \Psi_{0}$ and $\lambda_{\rm esc} \sim 1/d$, where $d$ is the size of the cloud. This estimate is indicated by the blue dash-dotted line in Fig.~\ref{fig:BurstmuM03r20}.

At late times, the system settles to a final, stationary state. In the subcritical regime, this final state is almost the same as its initial state since the axion cloud is barely affected by the EM perturbation. Conversely, in the supercritical regime, the parametric instability has driven the axion field to decrease to a subcritical value. Therefore, in the absence of SR growth, no further instability can be triggered and the axion cloud settles on a final state with a lower amplitude than its initial value while the created photons travel outwards.
\begin{figure}
    \centering
    \includegraphics[width = 0.5\textwidth]{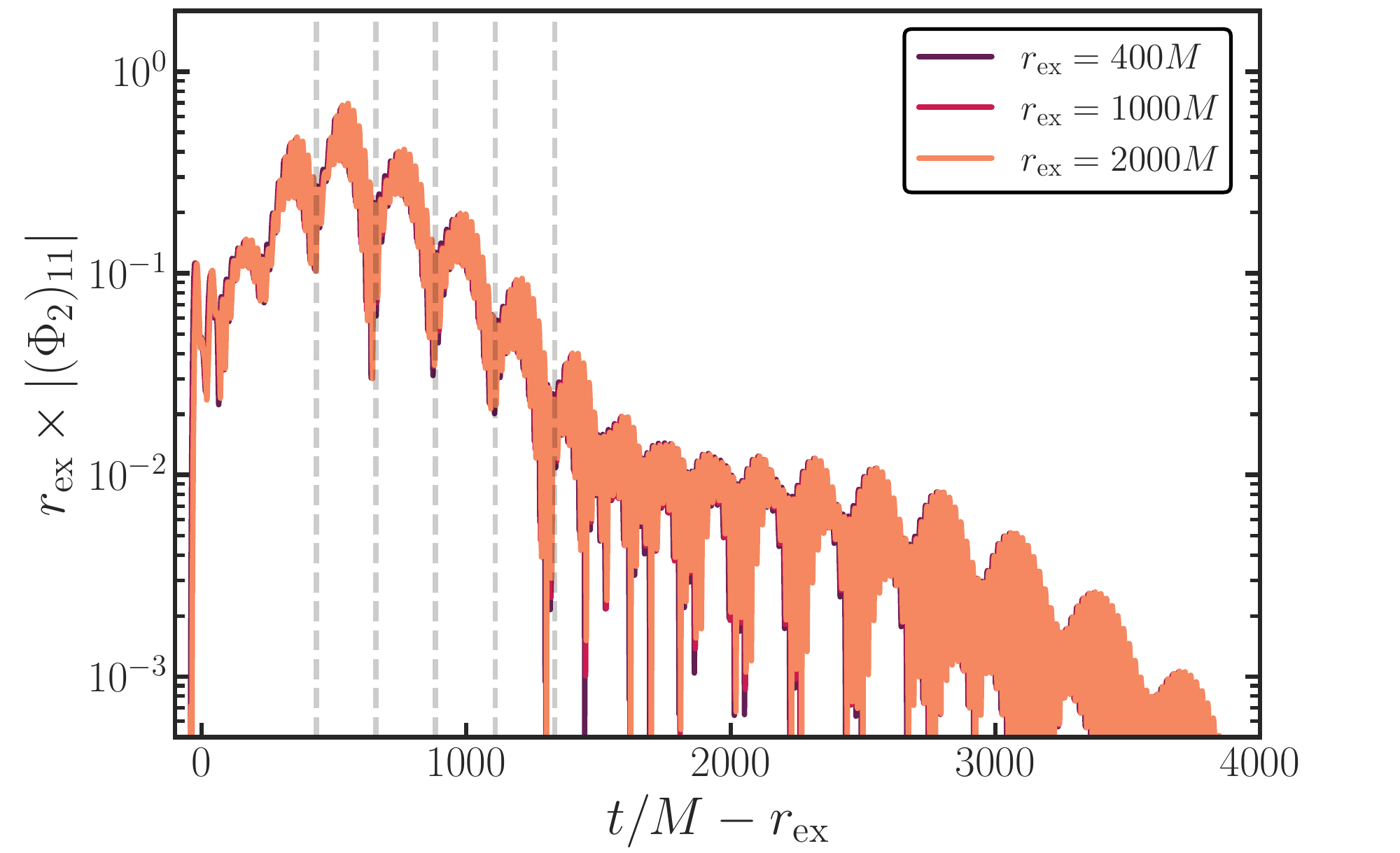}	
    \hspace*{-0.1cm}
    \includegraphics[width = 0.5\textwidth]{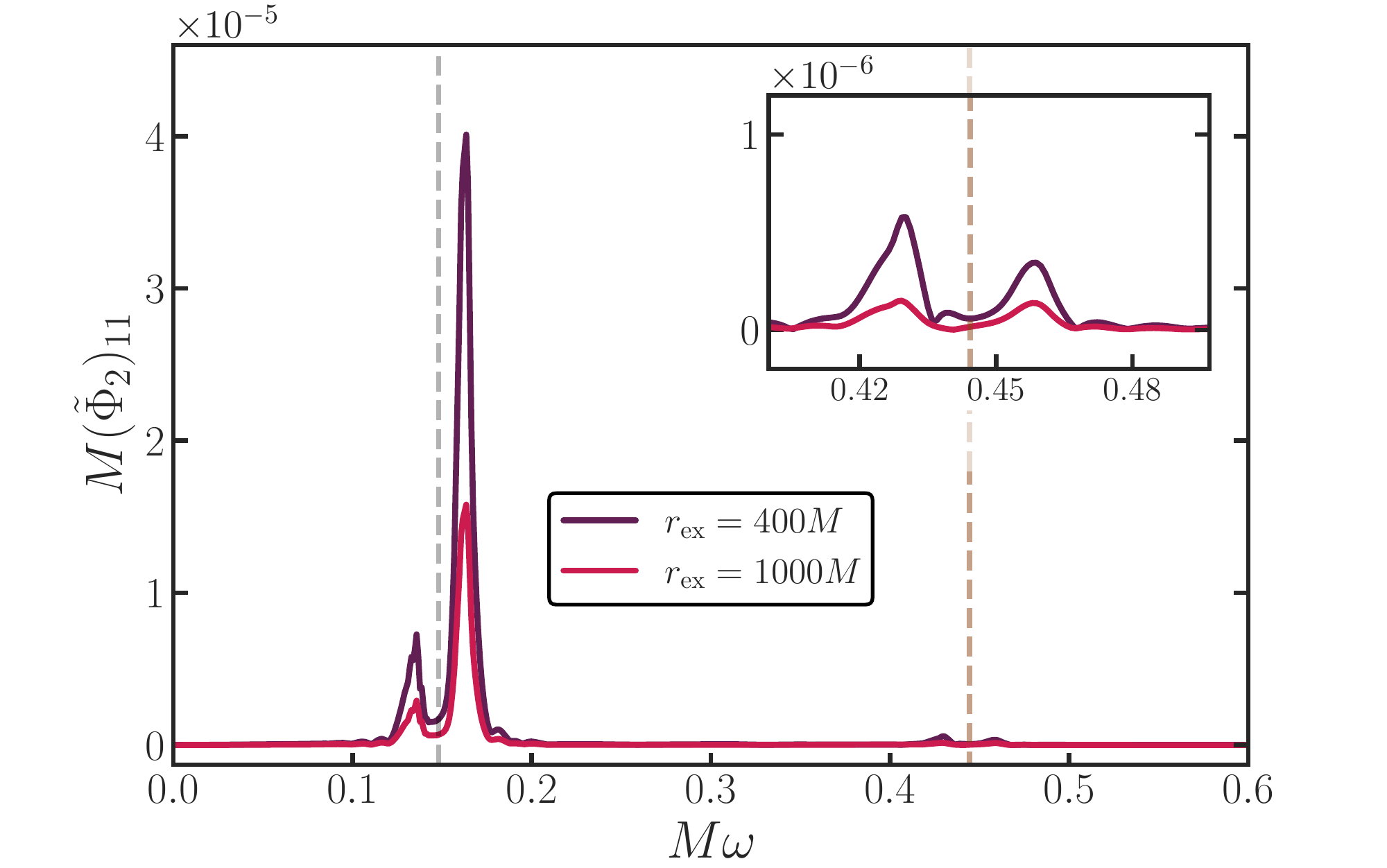}
    \caption{{\bf Top Panel:} The dipolar component $|(\Phi_{2})_{11}|$ of the EM field in the supercritical regime with $\mu M = 0.3$, for simulation $\mathcal{I}_{3}$. The alignment of the waveforms shows that we are dealing with EM radiation. High-frequency oscillations are set by the mass scale $\mu$, whereas ``beatings'' (circumscribed by vertical dashed lines) are controlled by $1/\mu^2$.
    {\bf Bottom Panel:} Fourier transform of the signal, taken on the entire time domain, showing the dominant frequencies in the problem. The gray and brown dashed lines show $N (\omega_0 / 2)$ for $N = 1, 3$, respectively, where $\omega_0 \approx \mu$ is the frequency of the fundamental mode.}    \label{fig:EMfieldBurstmuM03r481020}
\end{figure}
%%%%%%%%%%%%%%%%%%%%%%%%%%%%%%%%%%%%%%%%%%%%%%%%%%%%%%%%%%%%%%
\subsection{Axion and photon emission}
%%%%%%%%%%%%%%%%%%%%%%%%%%%%%%%%%%%%%%%%%%%%%%%%%%%%%%%%%%%%%%
Although the axion is massive, EM waves are massless and allowed to travel freely once outside the cloud. To study EM wave propagation, we monitor the system at large radii. The top panel of Fig.~\ref{fig:EMfieldBurstmuM03r481020} summarizes our findings for EM radiation, where we align waveforms in time. Some features are worth noting:~(i) we find that $\Phi_{2}$ decays like the inverse of the distance to the BH, as might be expected for EM waves;~(ii) the pattern of the waveform is not changing as it propagates, typical of massless fields. The radiation travels at the speed of light, as it should.

Additionally, we observe an interesting morphology in the EM burst. It has a high-frequency component slowly modulated by a beating pattern. While the high-frequency component is set by the boson mass $\mu$, with oscillation period $\sim \!2\pi/(\mu/2)$, the beating frequency scales with $1/\mu^{2}$ as its origin lies with the presence of the cloud. Specifically, when the photons are produced inside the cloud, they travel through it allowing for further interactions. These photon ``echoes'' exhibit a symmetric frequency distribution with respect to the primary photons, lying around $\mu/2$. This can be seen in the Fourier transform in the bottom panel of Fig.~\ref{fig:EMfieldBurstmuM03r481020}. The frequency difference between these peaks, $\Delta \omega$, corresponds to the observed beating timescale, $\sim \!2\pi/\Delta\omega$, indicated by the dashed lines in the top panel of Fig.~\ref{fig:EMfieldBurstmuM03r481020}. In addition to the bulk of photon frequencies near half the axion mass, there are other peaks in the frequency domain, namely two around $M\omega \sim 0.45$. We believe these higher order peaks do not originate from a parametric resonance, as one would expect peaks for each integer $N$ at $N\mu/2$, while we find the peaks at {\it even} $N$ to be absent. Furthermore, the bandwidth of higher order parametric resonances is extremely small, making it hard to trigger those. We rather believe these additional peaks to be a result from photon ``echoes'' as well, generated at later times, i.e.,~photons produced by the parametric mechanism that are up-scattered by the axion cloud. These results are different from a homogeneous axion background, where only echoes with the same frequency are produced. The discrepancy is due to the large momentum tail of the axion cloud.  

\begin{figure}
    \centering
    \includegraphics[width = 0.5\textwidth]{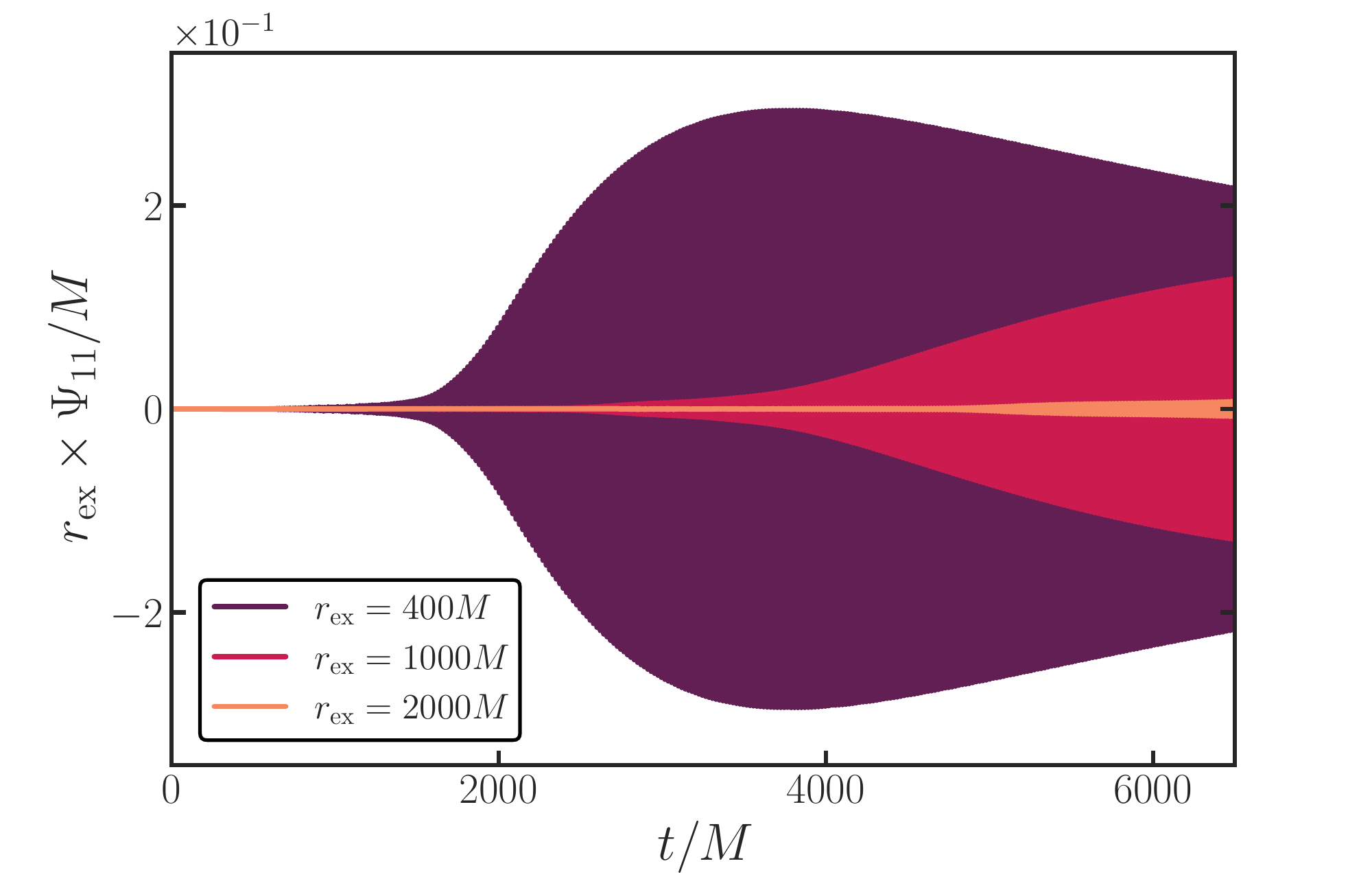}\\
		\includegraphics[width = 0.498\textwidth]{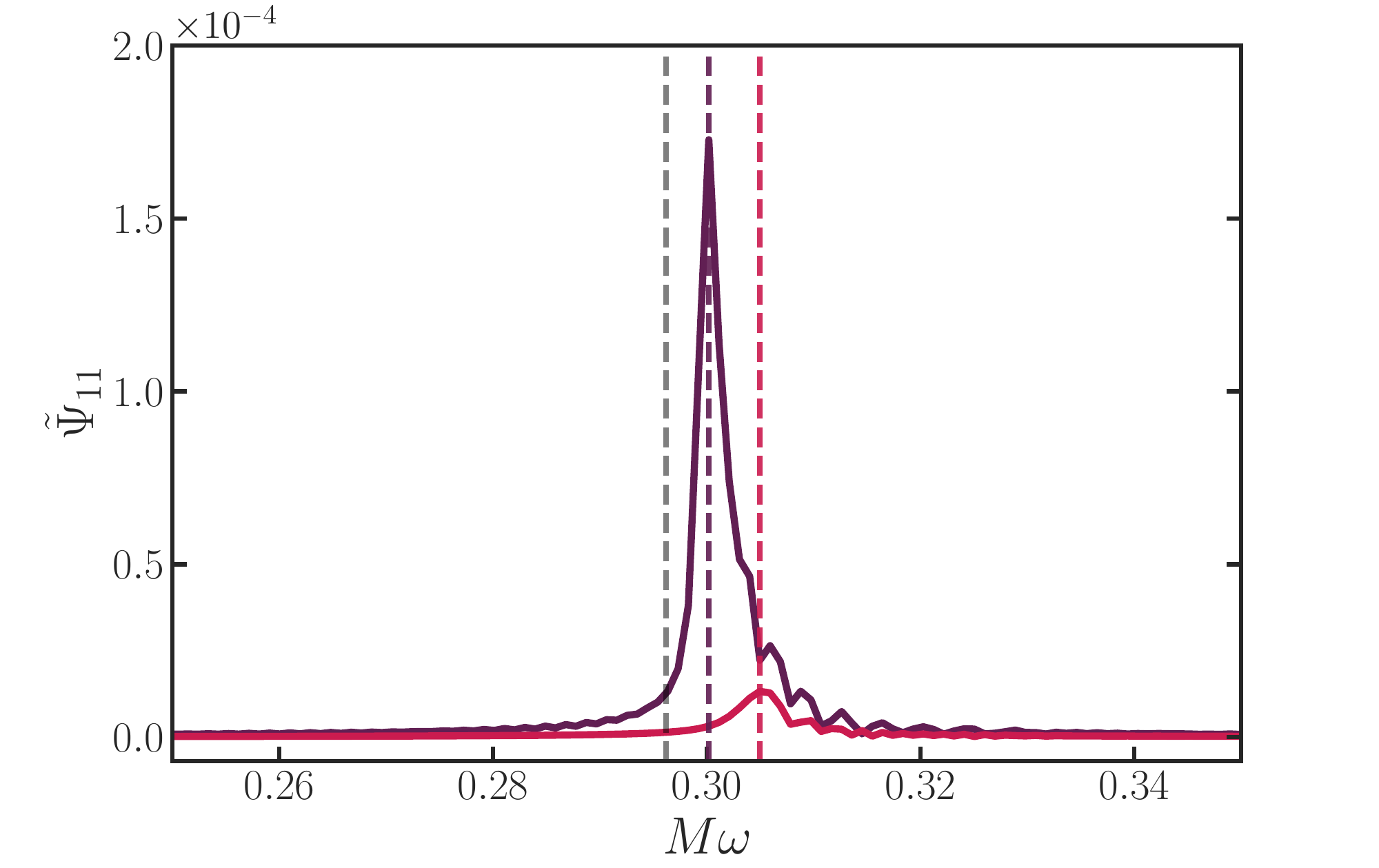}
    \caption{{\bf Top Panel:} The dipolar component of the axion field in the supercritical regime (simulation $\mathcal{I}_{3}$). When extracted at large radii, the non-zero value of $\Psi_{11}$ is only present in the supercritical regime and explained by the production of axion waves. Due to dispersion, the morphology of the wave changes while traveling outwards. 
    {\bf Bottom Panel:} The Fourier transform of the dipolar component, taken on the entire time domain shown in the top panel. The gray dashed line shows the frequency of the fundamental mode of the bound state, while the colored dashed lines indicate the frequency of the peak of each curve situated at $\omega > \mu$, confirming these are axion waves.}
    \label{fig:BurstAxionmuM03r40100}
\end{figure}

Besides the expected EM radiation, the reverse process -- two photons combining to create an axion -- may also provide a non-negligible contribution. In fact, this process has been explored in the context of axion clusters~\cite{Kephart1995}, where energetic axions are created that can not be stimulated anymore, hence they escape the cluster (so-called ``sterile axions''). Projecting this scenario in the context of SR instabilities translates into the creation of unbound axion states, with frequencies $\omega>\mu$, that are thus able to escape to infinity. Such axion waves are indeed produced in our setup, as can be seen in Fig.~\ref{fig:BurstAxionmuM03r40100}. The large scalar field contribution far away from the cloud is {\it only} present in the supercritical regime. Since these are massive waves, the dependence with time and distance from the source is less simple due to dispersion. As components with different frequencies travel with different velocities, the wave changes morphology when traveling to infinity, which is apparent in the top panel of Fig.~\ref{fig:BurstAxionmuM03r40100}. 

The Fourier transform of the axion waves is shown in the bottom panel of Fig.~\ref{fig:BurstAxionmuM03r40100}. It indeed contains components with frequency $\omega > \mu$, showing that the field is energetic enough to travel away from the source. The frequencies of the peaks correspond to a group velocity $v=\sqrt{1-\mu^2/\omega^2} = 0.036$ and $v = 0.18$ for $r_{\rm ex} = 400M$ and $r_{\rm ex} = 1000M$, respectively. Note that this Fourier transform is taken over the full time domain and thus dominated by the late signal of the axion waves, consisting of larger amplitude, non-relativistic waves. This also explains why the peak for the $r_{\rm ex} = 1000M$ curve is at higher frequency; the slower waves did not have time to arrive yet at this larger radius. If we instead calculate the Fourier transform on only the first part of the signal, then we capture the (more) relativistic components. These emitted axion waves can in principle be detected by terrestrial axion detectors if the BH is close enough to Earth.

Besides the dipole component, also higher order scalar multipoles are created by the photons. In fact, from our initial data, only scalar multipoles with odd $\ell$ can be produced. This selection rule is detailed in Appendix~\ref{sec:selectionrules}. The higher multipoles for both the axion and photon radiation are shown in Appendix~\ref{appB:higherorder}, where it can also be seen that excited photons can recombine to create axion waves with twice the axion mass. 
%%%%%%%%%%%%%%%%%%%%%%%%%%%%%%%%%%%%%%%%%%%%%%%%%%%%%%%%%%%%%
\section{Superradiance turned on}\label{sec:withSR}
%%%%%%%%%%%%%%%%%%%%%%%%%%%%%%%%%%%%%%%%%%%%%%%%%%%%%%%%%%%%%
The formation of an EM burst is determined by whether the photon production from the parametric instability is dominant over the escape rate from the cloud or vice versa. The initialization of the system in a supercritical state however, is artificial. Instead, it starts in the subcritical regime and potentially grows supercritical through SR. Previous works claimed that this process developed through a burst-quiet sequence:~a burst of EM and axion waves would deplete the cloud, which would then grow on a SR timescale before another burst occurred~\cite{Ikeda:2018nhb}. We argue that in fact bursts do not occur, and that the process is smoother than thought. As we will show, the presence of SR introduces two important differences:~(i) the growth rate of the EM field is modified and~(ii) the system is forced into a stationary phase.
%%%%%%%%%%%%%%%%%%%%%%%%%%%%%%%%%%%%%%%%%%%%%%%%%
\subsection{Numerical results}\label{subsec:sim}
%%%%%%%%%%%%%%%%%%%%%%%%%%%%%%%%%%%%%%%%%%%%%%%%%%
We numerically evolve the coupled axion-photon system under the influence of a SR growing cloud. In these simulations, we start the system in the  subcritical regime, and let it evolve to supercriticality via (artificial) SR, since now $C\neq 0$. 
\begin{figure}[t!]
    \hspace*{-0.6cm}
    \includegraphics[width = 0.455\textwidth]{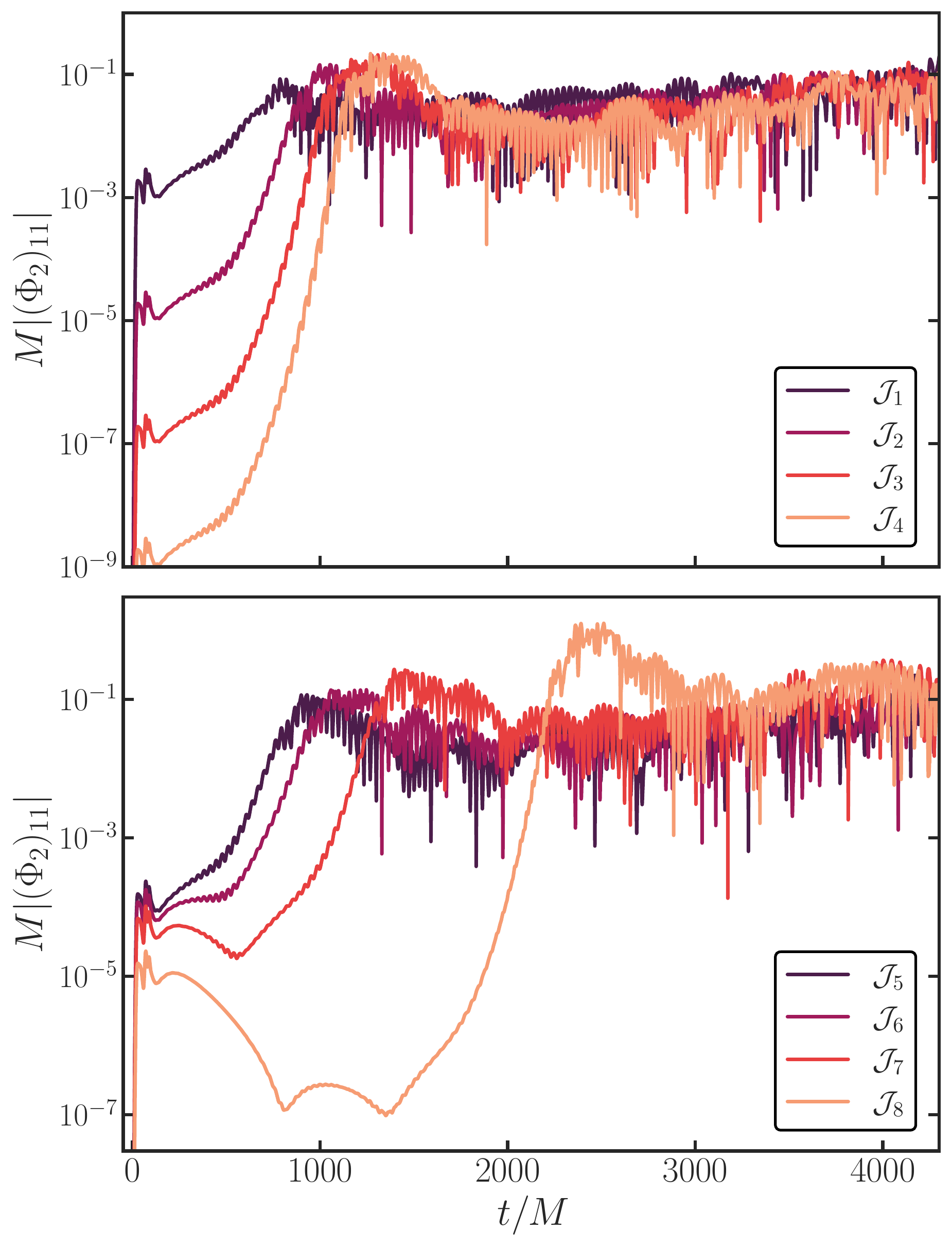}
    \caption{{\bf Top Panel:} The time evolution of the dipolar component of the EM field including SR growth for different strengths of the initial EM pulse (see Table~\ref{tb:simulations}). The field is extracted at $r_{\rm ex} = 20M$ and $\mu M = 0.2$. 
    {\bf Bottom Panel:} Same as above, but while varying the initial coupling strength $k_{\rm a}\Psi_{0}$. Notice how a stationary state is reached instead of a burst of EM waves. Moreover, the final EM value is independent of initial conditions, even though the timescale required to reach saturation does depend on how the fields were initialized.}    \label{fig:Phi2VaryingEMpulse_Phi2VaryingInitialScalar}
\end{figure}

Figure~\ref{fig:Phi2VaryingEMpulse_Phi2VaryingInitialScalar} illustrates the behavior of the system. We evolve different initial conditions, corresponding to different seed EM fields $E_{0}/\Psi_{0}$ and different couplings $k_{\rm a}\Psi_{0}$, and we see a {\it saturation} of the EM field, to a value which is independent on the initial conditions.\footnote{As could be anticipated, the timescale to reach saturation does depend on initial conditions:~lower field values or lower couplings require larger timescales.} This stability is simply achieved by turning on SR growth like in equation~\eqref{eq:ASR}. In contrast to the previous section where the bound state solely loses energy, the supplement to the axion cloud is dominant at first, resulting in exponential growth. As the cloud approaches the critical value, parametric decay to the EM field begins to compensate for the energy gain from the BH, ultimately reaching a phase where energy gain and loss are balanced. As a result, the entire system consisting of the axion cloud and the EM field is constantly pumped by SR growth, with a steady emission of EM waves traveling outwards.
\begin{figure}[t!]
\hspace*{-0.4cm}
\includegraphics[width = 0.455\textwidth]{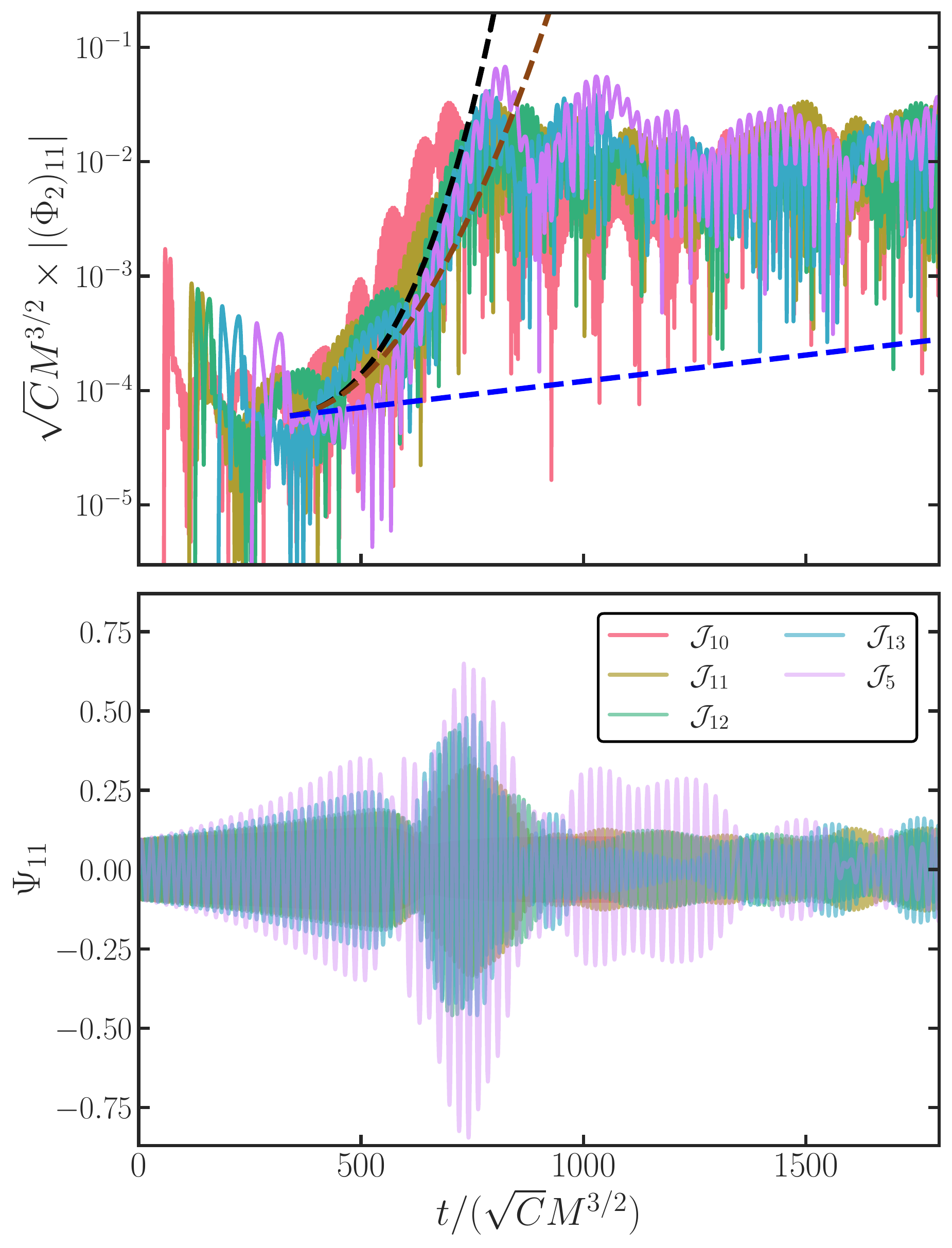}
\caption{{\bf Top Panel:} The dipolar component of EM radiation with SR turned on. The field is extracted at $r_{\rm ex} = 400M$ and $\mu M = 0.2$. The blue, black and brown dashed lines show the growth rate predicted by the standard Mathieu equation, the SR Mathieu equation~\eqref{eq:modifiedgrowthrate} and its first order expansion~\eqref{eq:modifiedgrowthrateFirst}, respectively. The value of $C$ is varied for simulations $\mathcal{J}_{i}$, see Table~\ref{tb:simulations}. Note the rescaling on both the horizontal and vertical axis with $\sqrt{C}$, where we scale all simulations onto $\mathcal{J}_{14}$. Thus, our results indicate a clear, simple dependence on the SR rate, which we investigate analytically below in Section~\ref{subsec:analyticalgrow} and~\ref{subsec:analytical}. 
{\bf Bottom Panel:} Same as above, but for the scalar dipolar component extracted at $r_{\rm ex} = 20M$.}
    \label{fig:Phi2VaryingC}
\end{figure}

The saturation value of the EM field does depend on the SR parameter $C$. This is simply due to the fact that the more axions that are created by SR, the more photons that can be produced through the parametric mechanism. We find the saturation value to be proportional to $\sqrt{C}$, shown in the top panel of Fig.~\ref{fig:Phi2VaryingC}. This result is also supported by analytical estimates in Section~\ref{subsec:analytical}, in particular equation~\eqref{eq:saturationvalue}. Additionally, our results demonstrate that the timescale required to reach saturation (at fixed initial field values), scales with $\sqrt{C}$ as well. This behavior is explained in the following section. 

These simulations provide us with robust evidence that the saturation phase is~(i) independent of the initial data, and~(ii) occurring for all tested values of $C$ that span two orders of magnitude, allowing for universal predictions. In the following, we discuss various features related to the saturation phase. 

%%%%%%%%%%%%%%%%%%%%%%%%%%%%%%%%%%%
{\it Evolution of the cloud's morphology} $-$
%%%%%%%%%%%%%%%%%%%%%%%%%%%%%%%%%%%
\begin{figure*}
    \includegraphics[width = 0.95\textwidth]{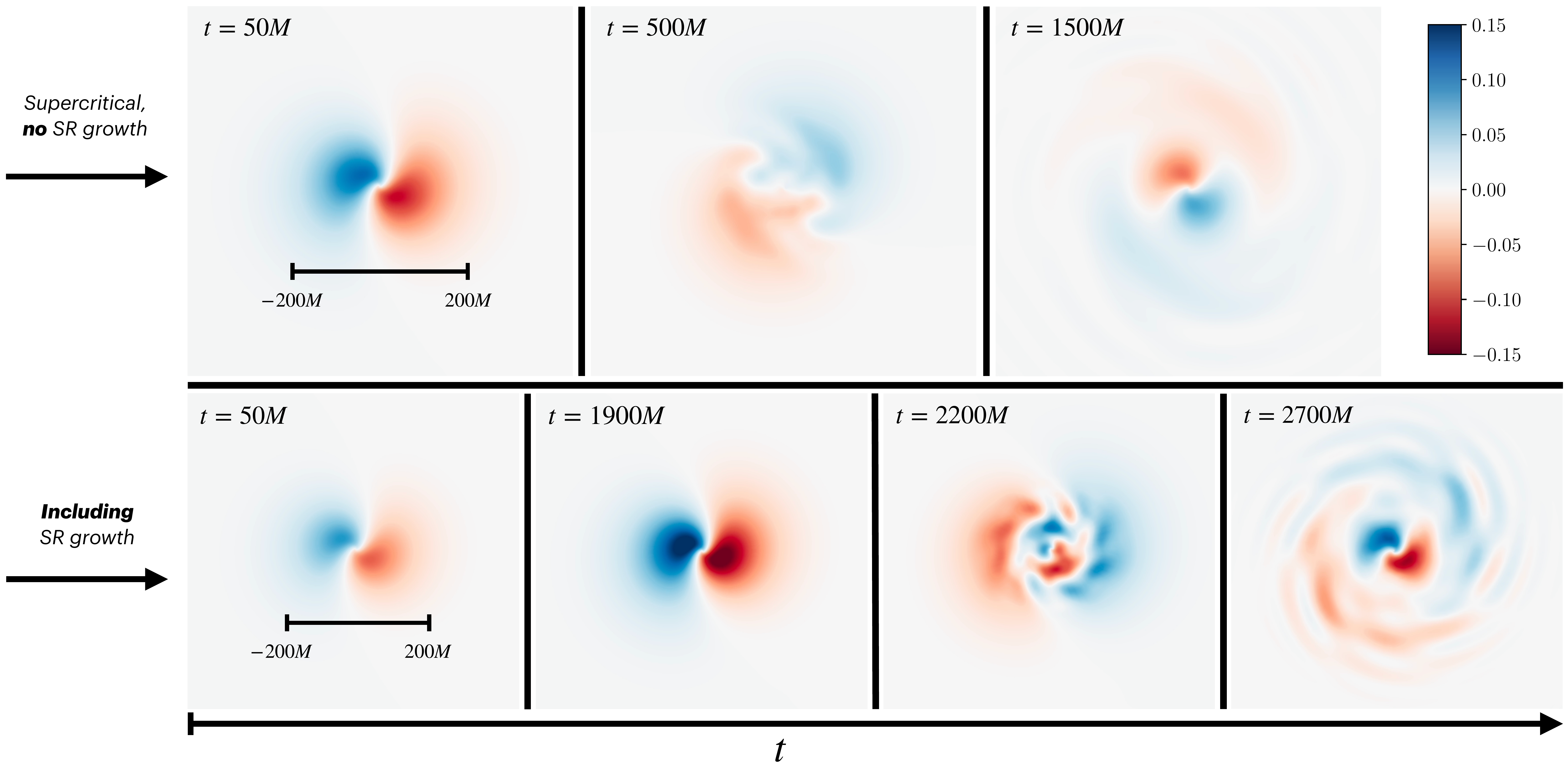}
    \caption{Snapshots of the axion profile during the evolution of the axion-photon system. Upper row shows the system in the supercritical regime (simulation $\mathcal{I}_{3}$), but {\it without} SR growth. The cloud starts in its initial dipole state, and gets disrupted by the parametric instability. Afterwards the configuration settles down while axion waves propagate to infinity. On the bottom row, we show an initially subcritical cloud, yet {\it with} SR growth turned on, $C = 10^{-3} M^{-1}$ (simulation $\mathcal{J}_{12}$). Again, the cloud starts in its dipole mode, yet now it grows in amplitude due to SR. Afterwards, the cloud is disrupted due to the EM instability, eventually settling down to a saturation phase wherein axion waves are continuously produced.}
    \label{fig:Snapshots}
\end{figure*}
When the system has just reached the critical boundary, the EM field starts growing (super-)exponentially until it reaches the saturation value. When this happens, the non-linear backreaction in the Klein-Gordon equation becomes important. In absence of SR, the EM field quickly decays in time after reaching its maximum and with that its backreaction onto the axion field, allowing the cloud to settle back to a stable configuration at late times (see Fig.~\ref{fig:BurstmuM03r20}). Conversely, in presence of SR, the EM field settles to a large and constant value that continuously backreacts onto the axion field. Consequently, it starts to exhibit strong deviations from the initial pure bound state configuration as overtones are triggered, i.e.,~it acquires a beating-like pattern. This can be seen in Fig.~\ref{fig:Phi2VaryingC}, where around $t\sim 700 \sqrt{C}M^{3/2}$ the saturation phase ensues and there is no relaxation to the pure quasi-bound state. We show a series of snapshots from the cloud's evolution in the two distinct scenarios in Fig.~\ref{fig:Snapshots}.

%%%%%%%%%%%%%%%%%%%%%%%%%%%%%%%%%%%%%%%%%%%%%%%%%%
{\it Angular structure of outgoing EM waves} $-$
%%%%%%%%%%%%%%%%%%%%%%%%%%%%%%%%%%%%%%%%%%%%%%%%%%
\begin{figure}
    \hspace*{-0.4cm}
    \includegraphics[width = 0.46\textwidth]{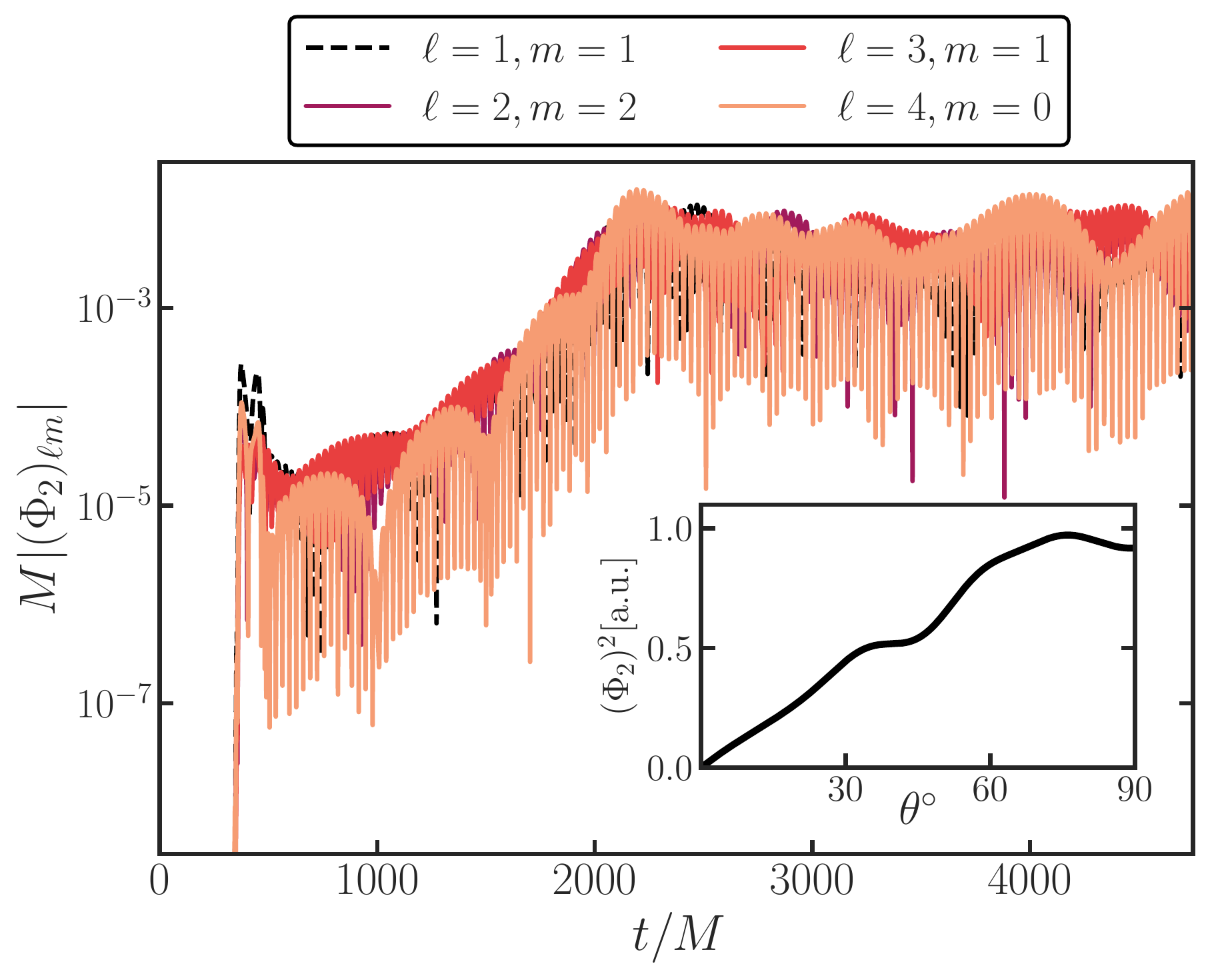}
    \caption{A subset of the multipoles of the EM field for simulation $\mathcal{J}_{12}$, extracted at $r_{\rm ex} = 400M$. Every multipole up to $\ell \leq 6$ has approximately a similar contribution. The inset shows the angular structure of the outgoing EM waves in the saturation phase (where all multipoles up to $\ell \leq 8$ are taken into account). The $y$ axis is reported in arbitrary units (a.u.). The dominant production of the photons is on the equatorial plane where the cloud's density is highest.}
    \label{fig:Phi2MultipolesLargeradii}
\end{figure}
During the saturation phase, there is a nearly constant emission of EM waves. For observational purposes, we probe the angular structure of the outgoing radiation. We do this through the multipole components of $\Phi_{2}$, while up to now only the dipole was considered. A subset of these multipoles is shown Fig.~\ref{fig:Phi2MultipolesLargeradii}. From the differences in amplitude between different modes, we conclude that the radiation is not isotropic. In fact, we find that the dominant radiation is on the equatorial plane (see inset of Fig.~\ref{fig:Phi2MultipolesLargeradii}), where the density of the axion cloud is highest. To strengthen this result, we also compute analytically the excitation coefficients of these multipoles given our initial data. We report them in Appendix~\ref{sec:selectionrules}. 
%%%%%%%%%%%%%%%%%%%%%%%%%%%%%%%%%%%%%%%%%%%%%%%%%%%%%%%%%%%%%%
\subsection{Growth rate}\label{subsec:analyticalgrow}
%%%%%%%%%%%%%%%%%%%%%%%%%%%%%%%%%%%%%%%%%%%%%%%%%%%%%%%%%%%%%%
In previous work~\cite{Boskovic:2018lkj}, it was shown that in absence of SR, the growth rate of the EM field can be approximated by a simple, analytical expression. This is a consequence of the fact that when the background spacetime is Minkowski and the background axion field is a coherently oscillating, homogeneous condensate, the Maxwell equations can be rearranged in the form of a Mathieu equation~\cite{Boskovic:2018lkj, Hertzberg:2018zte}. The growth rate is then found by taking the production rate of the photons ($\sim$~the Floquet exponent of the dominant, unstable mode of the Mathieu equation) and subtracting the escape rate of the photons from the cloud ($\sim$~inverse of the cloud size). However, while this approach yields accurate predictions in absence of SR (see Fig.~\ref{fig:BurstmuM03r20}), it does not in presence of SR (blue dashed line in Fig.~\ref{fig:Phi2VaryingC}). Remarkably, as we will show, a simple adjustment to the Minkowski toy model restores its validity. Additional details are provided in Appendix~\ref{sec:appendMathieu}.

Let us consider the Maxwell equations in flat spacetime. We adopt Cartesian coordinates and  assume the following ansatz for the EM field,
\begin{equation}
\label{eq:MathieuEM}
    A_\mu (t,\vec{x})=\alpha_\mu(t, \vec{p}) e^{i (\vec{p} \cdot \vec{x} - \omega t)}\,,
\end{equation}
where $\vec{p}$ is the wave vector which we assume to be aligned in the $\hat{z}$ direction without loss of generality, i.e.,~$\vec{p}=(0,0,p_z)$. To mimic the amplification of the axion field via SR, we consider a homogeneous condensate that exponentially grows in time as\footnote{We adopt a different notation to distinguish the amplitude of the homogeneous axion field in this toy model, $\psi_0$, with the one of the axion cloud around the BH, $\Psi_0$.}
\begin{equation}
    \Psi=\frac{1}{2}(\psi_0 e^{-i \mu t}+ \psi_0^* e^{i \mu t}) e^{C t}\,.
\end{equation}
Adopting the field redefinition $y_k =e^{i \omega t}\alpha_k$, rescaling the time as $T=\mu t$ and projecting along a circular polarization basis $e_\pm$ such that $y=y_\omega e_\pm$, we obtain a ``superradiant'' Mathieu-like (SM) equation:
\begin{equation}\label{eq:modifiedMathieu}
    \partial_T^2 y_\omega\!+\!\frac{1}{\mu^{2}}\Big(p_z^2+ 2  p_z e^{\frac{CT}{\mu}}\psi_0 k_{\rm a}(C \cos T\!-\!\mu \sin T)\Big)y_\omega=0\,.
\end{equation}
Unsurprisingly, for $C=0$, this equation reduces to the original Mathieu equation.\footnote{Our result includes a sine instead of the cosine found in~\cite{Boskovic:2018lkj}, which originates from a small sign mistake in their derivation, see equation~(19). This has no consequences for the physics, as it only induces a $\pi/2$ phase shift.} From~\eqref{eq:modifiedMathieu}, we find two new features; an extra oscillating term $\sim C \cos T$, and, most importantly, an exponentially growing factor $\sim e^{\frac{CT}{\mu}}$. By solving~\eqref{eq:modifiedMathieu} numerically, we find that, similar to the standard Mathieu equation, this equation admits instability bands, albeit with a larger growth rate. Fitting the exponent of the numerical solution, we conclude that the solution to the superradiant Mathieu equation is well described by a super-exponential expression $y_\omega \sim e^{\lambda_{\scalebox{0.55}{$\mathrm{SM}$}} t}$, with
\begin{equation}
\label{eq:modifiedgrowthrate}
    \lambda_{\scalebox{0.65}{$\mathrm{SM}$}}=\frac{\mu}{2}k_{\rm a }\psi_0  e^{C t/2}\,.
\end{equation}
In Appendix~\ref{sec:appendMathieu}, we show the comparison between this expression and the numerical solutions as well as an analytic derivation of~\eqref{eq:modifiedgrowthrate} using a multiple-scale method. 

We confront the growth rate of the EM field when considering the full axion-photon system in a Schwarzschild background with the growth rate from our toy model~\eqref{eq:modifiedgrowthrate} in Fig.~\ref{fig:Phi2VaryingC}. Here, the standard Mathieu growth rate (blue dashed line), the full solution and first order expansion in $C$ (black and brown dashed lines, respectively) can be seen, where the latter is defined by
\begin{equation}
\label{eq:modifiedgrowthrateFirst}
    \lambda_{\scalebox{0.65}{$\mathrm{SM}$}} t \approx \frac{\mu}{2}k_{\rm a}\psi_0 (t+\frac{C t^2}{2})\,.
\end{equation}
In all of the curves, the time it takes for photons to leave the cloud has been taken into account. Furthermore, we use the critical value for the coupling $k_{\rm a}\Psi_{0}$. Remarkably, the SM growth rate matches the numerical results closely. Moreover, the $C t^2$ term from~\eqref{eq:modifiedgrowthrateFirst} that appears at first order naturally explains why the timescale to reach saturation scales as $\sqrt{C}$. 

Hence, when considering the axion-photon system under the influence of SR, a simple extension to Mathieu equation allows for elegant, analytic predictions of our numerical results. Note that the true value for SR is many orders of magnitude lower than the one considered in this work and thus we expect the correction to be subdominant (unlike here). Finally, this prediction neglected the backreaction onto the axion cloud. Hence, it naturally breaks down when the rate of energy loss due to conversion to photons becomes comparable the SR growth, i.e.,~when the saturation phase ensues.
%%%%%%%%%%%%%%%%%%%%%%%%%%%%%%%%%%%%%%%%%%%%%%%%%%%%%%%%%%%%%%
\subsection{Saturation phase}\label{subsec:analytical}
%%%%%%%%%%%%%%%%%%%%%%%%%%%%%%%%%%%%%%%%%%%%%%%%%%%%%%%%%%%%%%
As can be seen in Figs.~\ref{fig:Phi2VaryingEMpulse_Phi2VaryingInitialScalar} and~\ref{fig:Phi2VaryingC}, turning on SR growth forces the system into a stationary configuration. Here, the energy loss of the cloud due to the parametric instability balances the SR pump sourced by the rotational energy of the BH. A description of this phase is remarkably simple as we will show below. A similar conclusion was found in~\cite{RosaKephartStimulated}. 

For the photons to reach an equilibrium phase, it is required that the parametric decay rate, $\lambda_{\rm pd}$, equals the escape rate of the photon, $\lambda_{\rm esc}$. Assuming that the former can be approximated by the decay rate in the homogeneous condensate case~\cite{Boskovic:2018lkj,Hertzberg:2018zte}, we have
\begin{equation}\label{eq:saturation}
    \lambda_{\rm pd} \approx \frac{k_{\rm a} \Psi_{\rm sat} \mu}{2} \approx \frac{1}{d} \approx \lambda_{\rm esc}\,,
\end{equation}
where $\Psi_{\rm sat}$ is the average amplitude of the scalar field {\it within the cloud} at saturation. In the non-relativistic regime, we can approximate the size of the cloud by the standard deviation of the radius $d = \langle r \rangle \approx 2 \sqrt{3} r_{\rm c} = 4 \sqrt{3} / (\mu^{2}M)$. This yields a relation for which the cloud reaches saturation, namely
\begin{equation}\label{eq:scalarsaturation}
    k_{\rm a} \Psi_{\rm sat}  \approx \frac{\mu M}{2\sqrt{3}}\,.
\end{equation}
This is indeed what we find in the bottom panel of Fig.~\ref{fig:Phi2VaryingC}, since $k_{\rm a}\Psi_{\rm sat} \approx 0.2/(2\sqrt{3}) \approx 0.06$.

Additionally, we consider the equilibrium condition of the axion cloud. It is sourced by the SR rate $\Gamma_{\rm SR}$, yet loses energy due to the parametric instability, $\Gamma_{\rm PI}$. In our setup, these two rates are
\begin{equation}\label{eq:rates}
    \Gamma_{\scalebox{0.65}{$\mathrm{SR}$}} = \frac{C}{2} \quad \mathrm{and} \quad \Gamma_{\scalebox{0.75}{$\mathrm{PI}$}} \approx 2 \Gamma_{{}_{\Psi \rightarrow \gamma \gamma }} f_{\gamma}\,,
\end{equation}
where $\Gamma_{\scalebox{0.65}{$\Psi \rightarrow \gamma \gamma$}} = \hbar k_{\rm a}^2 \mu^3/(16 \pi)$ is the perturbative decay width of the axion-to-photon conversion and $f_\gamma$ is the photon occupation number. When the dominant production is in a narrow band around $p_{\gamma} = \mu/2$, we have~\cite{Carenza:2019vzg}
\begin{equation}
    f_{\gamma} (p_\gamma=\mu/2) = \frac{8\pi^3 n_{\gamma}}{4\pi p^{2}_{\gamma} \Delta p_{\gamma}} \approx \frac{2\pi^2 A^{2}_{\gamma}}{\hbar k_{\rm a} \Psi_{\rm sat} \mu^2}\,,
\end{equation}
where $\Delta p_{\gamma} \approx 2 k_{\rm a} \Psi_{\rm sat} \mu$ is the photon dispersion bandwidth, approximated to be the resonant bandwidth at $p_{\gamma} = \mu/2$. Moreover, $A_{\gamma}$ is the photon amplitude, which relates to our measure of the EM field $\Phi_{2}$ as $A_{\gamma} \sim 2\Phi_{2}/\omega$, where $\omega\sim \mu$ is the frequency of the axion field. Finally, $n_\gamma = 2\rho/(\hbar \omega)$ is the photon number density, with $\rho =  A_{\gamma}^{2}\omega^{2}/2$ the energy density. Substituting these relations into~\eqref{eq:rates}, we find that inside the cloud
\begin{equation}\label{eq:saturationvalue}
    \frac{A^{2}_{\gamma}}{\Psi_{\rm sat}^2} \approx \frac{2C}{\pi \mu k_{\rm a}\Psi_{\rm sat}} \approx \frac{C}{\pi \lambda_{\rm esc}}\,,
\end{equation}
where we used again~\eqref{eq:saturation}. Hence, this simple analytical estimate shows that the EM field stabilizes to a value proportional to $\sqrt{C}$. This result is in excellent agreement with our simulations (see Fig.~\ref{fig:Phi2VaryingC}) and it allows us to consider a case in which $C$ coincides with the SR growth timescale.

The total energy flux of the photons with frequency $\mu/2$ is defined as~\cite{RosaKephartStimulated} 
\begin{equation}
    \frac{\mathrm{d}E}{\mathrm{d}t}=\frac{\hbar\mu}{2}n_{\gamma}\lambda_{\rm esc}\,\chi r_{\rm c}^{3}\,,
\end{equation}
where $\chi r^{3}_{\rm c}$ is the volume of the cloud. Here, $\chi$ is a numerical factor where in the non-relativistic regime, $\chi \approx \mathcal{O}(10^{2})$.\footnote{
To obtain $\chi$, we introduce a threshold value $\epsilon$ for the absolute value of the scalar field, and define
\begin{equation}
    \chi =\int_{0}^{\infty}\mathrm{d}r\,r^{2}\int\mathrm{d}\Omega\,\Theta \left(|\Psi|-\epsilon \right)\nonumber\,,
\end{equation}
where $\Theta$ is the Heaviside step function and $\epsilon\sim 0.5  \max \left(\Psi\right)$. We checked that the order of $\chi$ does not strongly depend on $\epsilon$.}
Assuming the SR rate $\Gamma_{\scalebox{0.65}{$\mathrm{SR}$}}$ to be equal to the decay rate $\Gamma_{\scalebox{0.65}{$\mathrm{PI}$}}$ in the saturation phase~\eqref{eq:rates}, we get that
\begin{equation}
\begin{aligned}\label{eq:analyticL}
    \frac{\mathrm{d}E}{\mathrm{d}t}&\approx 7.6\times 10^{45}\left(\frac{\chi}{100}\right)\left(\frac{2.5\times 10^{2}M}{\tau_{s}}\right)\\
    &\times\left(\frac{0.2}{\mu M}\right)^{2}\left(\frac{10^{-13}{\rm GeV}^{-1}}{k_{\rm a}}\right)^{2}{\rm erg/s}\,,
\end{aligned}
\end{equation}
where $\tau_{s}=C^{-1}$. To probe the SR regime, we tune $C$ to match the well-known SR growth rate in the dominant growing mode~\cite{Detweiler:1980uk}
\begin{equation}\label{eq:SRrate}
    \Gamma_{\scalebox{0.65}{$\mathrm{SR}$}} \approx \frac{a_{\scalebox{0.65}{$\mathrm{J}$}}\,(\mu M)^9}{24 M} \quad \text{when} \quad \mu M \ll 1\,,
\end{equation}
where $a_{\scalebox{0.65}{$\mathrm{J}$}}$ is the spin of the BH. Using this in~\eqref{eq:analyticL}, we obtain
\begin{equation}\label{eq:analyticSRLuminosity}
\begin{aligned}
    \frac{\mathrm{d}E}{\mathrm{d}t}&\approx 8.1\times 10^{40}\left(\frac{\chi}{100}\right)\frac{a_{\scalebox{0.65}{$\mathrm{J}$}}}{M}\\ &\times \left(\frac{\mu M}{0.2}\right)^{7}\left(\frac{10^{-13}{\rm GeV}^{-1}}{k_{\rm a}}\right)^{2}{\rm erg/s}\,.
\end{aligned}
\end{equation}
Note how lower couplings lead to higher fluxes. Although this may sound counter-intuitive, it is explained by the fact that for lower couplings, the axion field saturates at higher values. As the EM flux is proportional to the axion field value, this leads to a higher flux.~\footnote{A similar behavior was found in the context of dark photons with kinetic mixings to the Standard Model photon~\cite{Caputo:2021efm}.} As a consequence, the saturation phase opens a channel to constrain axionic couplings ``from below''. Finally, the divergence of~\eqref{eq:analyticL} and~\eqref{eq:analyticSRLuminosity} at small couplings indicates when our model breaks down. In particular, the equilibrium condition~\eqref{eq:scalarsaturation} suggests a minimum value for the coupling for which the cloud's mass becomes larger than its maximum of $M_{\rm c}  = 0.1M$. For example, for $\mu M = 0.2$, the minimum coupling for which our description holds is $k_{\rm a} > 3.8\times 10^{-18}~{\rm GeV}^{-1}$. Lower couplings then this yield an unphysical situation and thus a breakdown of our description.
%%%%%%%%%%%%%%%%%%%%%%%%%%%%%%%%%%%%%%%%%%%%%%
\subsection{Implications for superradiance}\label{subsec:realSR}
%%%%%%%%%%%%%%%%%%%%%%%%%%%%%%%%%%%%%%%%%%%%%%
Using the scaling relation~\eqref{eq:saturationvalue}, we can probe the system in the regime of SR and thereby test the validity of~\eqref{eq:analyticSRLuminosity} with our numerical simulations. Before doing so, however, we must argue why we are able to extend our results beyond the probed regime for $C$. 

Either end of the tested parameter space for $C$ is accompanied by numerically challenges. For high $C$, there is an extreme growth on short timescales which makes the code diverge. For low $C$, the evolution timescale of the system becomes prohibitively large. Besides the simulations shown in Fig.~\ref{fig:Phi2VaryingC}, we did two additional simulations, $\mathcal{J}_{9}$ (low $C$) and $\mathcal{J}_{14}$ (high $C$), on each end of the spectrum. Regarding the latter, we find indications that the saturation phase is ruined as the EM field starts to grow from its saturation value after spending some time in the saturation phase. Physically, this is to be expected. In the case of extreme SR growth, the balance between the production and escape rate of the photons is distorted; the photons simply do not have time to escape the cloud while plenty of axions are produced. In this scenario, a more burst-like radiation pattern could be possible.\footnote{Such a scenario could be realized in the case of a Bosenova~\cite{Yoshino:2012kn}, where the axion density sharply rises on short timescales.} Since we are interested in SR in this work, we do not probe this regime further. 

The regime of low $C$ is of interest as the SR rate is at significantly lower values than what we can probe numerically~\eqref{eq:SRrate}. From the lowest $C$ we probe, $\mathcal{J}_{9}$, we find that even though there is an apparent decrease after the super-exponential growth, the $\sqrt{C}$ scaling is respected at late times. Physically, the perseverance of the saturation phase makes sense. When the growth rate is small, the system becomes adiabatic; as the axions are slowly produced, the system steadily approaches the critical value at which the system is in equilibrium and a saturation phase ensues.

To extract the energy flux from our simulations, we exploit the properties of the Newman-Penrose scalar. In particular, we can define 
\begin{equation}\label{eq:NPscalarEnergy}
    \frac{\mathrm{d}^{2}E}{\mathrm{d}t \mathrm{d}\Omega} = \lim_{r\rightarrow \infty}\frac{r^{2}}{2\pi}|\Phi_{2}|^{2}\,,
\end{equation}
where $\mathrm{d}\Omega \equiv \sin{\theta}\mathrm{d}\theta\mathrm{d}\varphi$. Decomposing~\eqref{eq:NPscalarEnergy} in terms of spin-weighted spherical harmonics, we obtain
\begin{equation}\label{eq:energyflux}
    \frac{\mathrm{d}E}{\mathrm{d}t} = \sum_{\ell m} \int \mathrm{d}\Omega\,\frac{1}{2\pi} | (\Phi_{2}^{\circ})_{\ell m} \> {}_{{\scalebox{0.65}{$-$}}1}\mkern-2mu Y_{\ell m}|^{2}\,,
\end{equation}
where $\Phi_{2} = \Phi_{2}^{\circ}/r$. From our simulations, we extract $(\Phi_{2})_{\ell m, \rm sim}$ for a certain $C_{\rm sim}$ at large radii ($r = r_{\rm ex}$) by averaging over a sufficient period in the saturation phase. Then, we scale these multipoles to match their saturation value in the case of SR according to 
\begin{equation}\label{eq:Phi2RealSR}
\begin{aligned}
    (\Phi_{2})_{\ell m, \scalebox{0.65}{$\mathrm{SR}$}} &\approx \frac{(\Phi_{2})_{\ell m, \rm sim}}{\sqrt{C_{\rm sim}/(2\Gamma_{\scalebox{0.65}{$\mathrm{SR}$}})} }\,,
\end{aligned}
\end{equation}
where $\Gamma_{\scalebox{0.65}{$\mathrm{SR}$}}$ is defined in~\eqref{eq:SRrate}. We do this for each multipole and sum them according to~\eqref{eq:energyflux} to obtain the total flux. As the contribution to the energy flux becomes smaller for higher multipoles, we sum each multipole until the increment is less than 5\%. In practice, this means summing the first $\sim 8$ values of $\ell$. Following this procedure, we find the following estimate from our simulations for the total, nearly constant, energy flux in the saturation phase:
\begin{equation}
    \frac{\mathrm{d}E}{\mathrm{d}t}  = 9.10 \times 10^{40}\;\left(\frac{10^{-13}~\mathrm{GeV}^{-1}}{k_{\rm a}}\right)^{2}\; \mathrm{erg/s}\,,
    \label{eq:dEdt}
\end{equation}
where we assumed $\mu M = 0.2$ and the BH to be maximally spinning. This matches closely the theoretical prediction from~\eqref{eq:analyticSRLuminosity}.

\begin{figure}
    \centering
    \includegraphics[width = 0.5\textwidth]{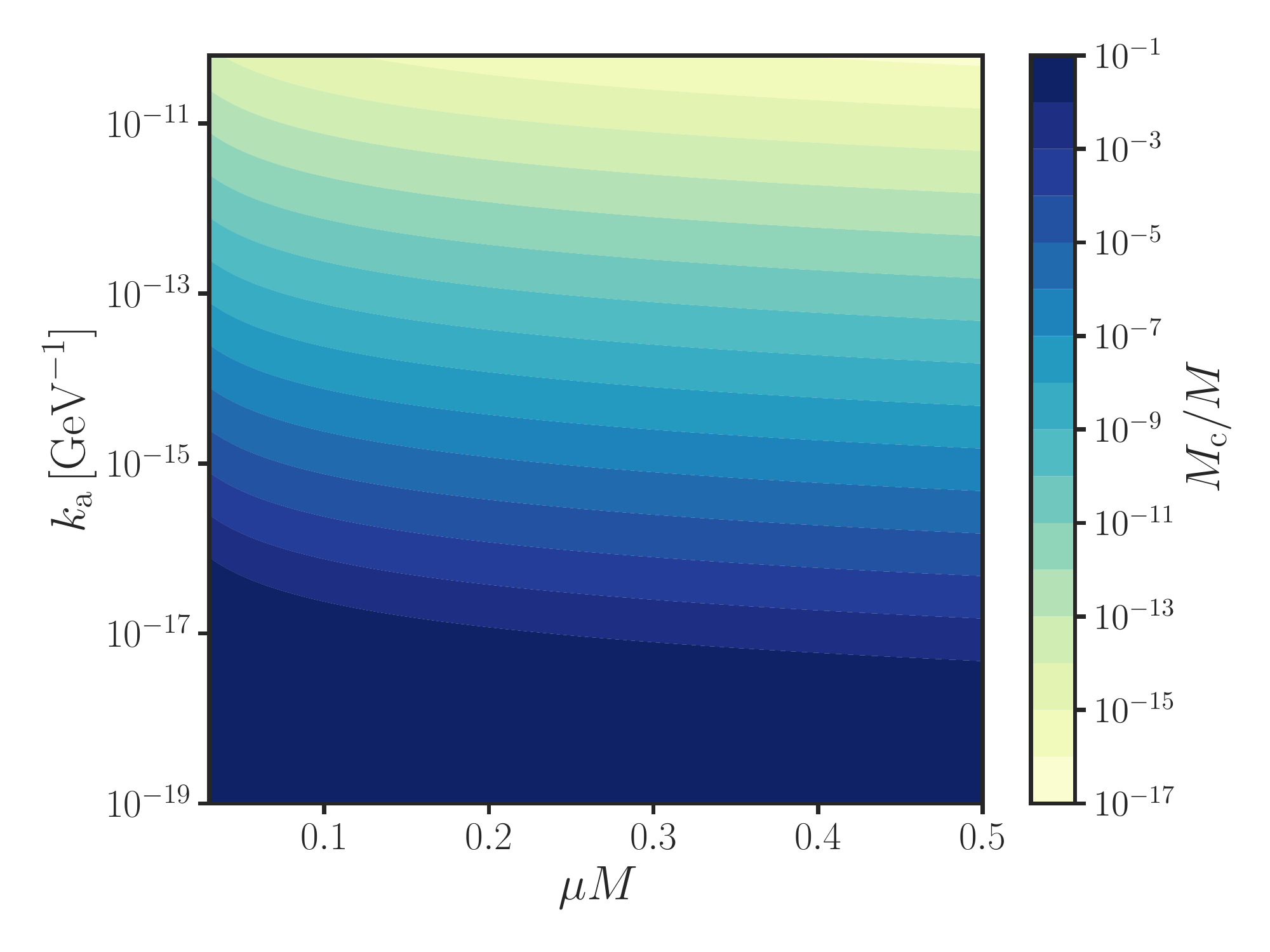}
    \caption{Contour plot of the mass of the cloud ($M_{\rm c}$) {\it at saturation} depending on the axionic and mass coupling, $k_{\rm a}\Psi_{\rm sat}$ and $\mu M$, respectively. The dark blue area at the bottom denotes the maximum mass of $10\%$ that the cloud can achieve in the purely gravitational case.}
    \label{fig:ContouronmuMKaMs}
\end{figure}

Besides the photon production, the parametric instability also affects the axion cloud. As we showed in equation~\eqref{eq:scalarsaturation}, the axion amplitude at saturation is independent of $C$. By translating the amplitude of the axion field to the mass of the cloud, the impact of coupling axions to photons becomes much more apparent~\cite{Brito:2014wla}. It is well-established that, in the purely gravitational case, the cloud is able to obtain a maximum mass of $M_{\rm c} \lesssim 0.1 M$. As can be seen in Fig.~\ref{fig:ContouronmuMKaMs}, through the coupling to the Maxwell sector, the cloud's mass can saturate significantly below this maximum. Note that in this estimate, we assume the profile of the cloud to be hydrogen-like which is only strictly true in the no-coupling case. Consequently, when the cloud is disrupted due to the strong backreaction onto the axion field in the saturation phase, this approximation is not expected to hold. Nevertheless, for the (much) lower SR growth rate, the EM flux is less strong and thus the cloud less disrupted. 

Figure~\ref{fig:ContouronmuMKaMs} has severe implications for current constraints on the mass of ultralight bosons that are set through either GW searches~\cite{LIGOScientific:2021rnv,Tsukada:2018mbp, Palomba:2019vxe, Yuan:2022bem,Ng:2020jqd} or spin measurements of BHs~\cite{Arvanitaki:2010sy,Arvanitaki:2014wva,Brito:2014wla,Brito:2017zvb,Cardoso:2018tly,Stott:2020gjj,Ng:2020ruv,Ng:2019jsx,Davoudiasl:2019nlo,Wen:2021yhz,Fernandez:2019qbj}. Due to the reduced cloud mass, the backreaction to the spin down of the BH could be negligible, which means current constraints no longer apply, as they assume no interactions for the axion. Furthermore, the environmental effect of the SR cloud on the gravitational waveform in BH binaries becomes less relevant~\cite{Baumann:2018vus, Baumann:2019ztm, Baumann:2021fkf,Baumann:2022pkl,Tomaselli:2023ysb,Cole:2022fir,Zhang:2018kib,Zhang:2019eid,Berti:2019wnn}.
%%%%%%%%%%%%%%%%%%%%%%%%%%%%%%%%%%%%%%%%%%%%%%
\section{Surrounding plasma}\label{sec:surroundingplasma}
%%%%%%%%%%%%%%%%%%%%%%%%%%%%%%%%%%%%%%%%%%%%%%
The presence of plasma affects the axion-to-photon conversion in the parametric instability mechanism, as the transverse polarizations of the photon are dressed with an effective mass, i.e., the plasma frequency $\omega_{\rm p}$. Therefore, when $\omega_{\rm p}> \mu/2$, the process $\Psi \rightarrow \gamma + \gamma$ becomes kinematically forbidden. Even though it is common lore to approximate the photon-plasma system with a Proca toy model, the full physics is more involved: already in the simplest case of a cold, collisionless plasma, the longitudinal degrees of freedom are electrostatic, unlike the Proca case (for details see~\cite{Cannizzaro:2020uap, Cannizzaro:2021zbp}). In curved space, these transverse and longitudinal modes are coupled and thus the Proca model cannot assumed to be correct {\it a priori}. Moreover, non-linearities provide additional couplings between the modes, and also the inclusion of collisions or thermal corrections create strong deviations from a Proca theory. Hence, a consistent approach from first principles is imperative. In this section, we take a first step in that direction by studying a linearized axion-photon-plasma system. The underlying physical assumptions of our plasma model can be found in Appendix~\ref{app:theplasma}, while the numerical implementation is detailed in Appendices~\ref{app:evolPlasma}-\ref{sec:DecompMom}.
%%%%%%%%%%%%%%%%%%%%%%%%%%%%%%%%%%%
\subsection{Without superradiance}
%%%%%%%%%%%%%%%%%%%%%%%%%%%%%%%%%%%
We start by studying the axion-photon-plasma system in absence of SR, and initialize the axion cloud in a supercritical state with $k_{\rm a} \Psi_0 \ll 1$. We evolve the system on a BH background for different values of the plasma frequency (see Table~\ref{tb:simulations}). Note that there is no backreaction onto the axion field in our linearized setup.
\begin{figure}
    \hspace*{-0.2cm}
    \includegraphics[width = 0.47\textwidth]{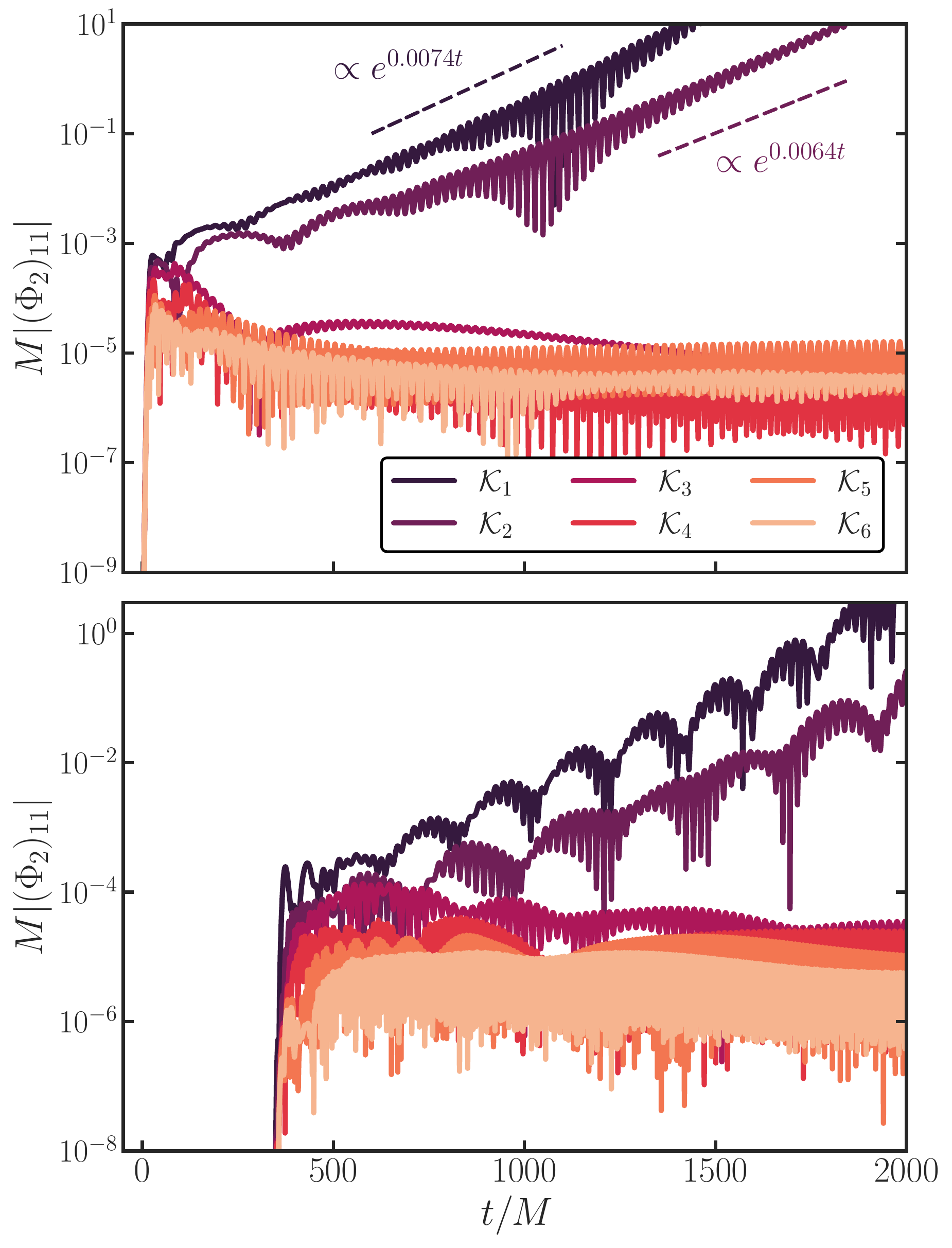}
    \caption{{\bf Top Panel:} Time evolution of the dipolar component $|(\Phi_{2})_{11}|$ of the EM field, extracted at $r_{\rm ex} = 20M$ for $\mu M = 0.3$, in the presence of plasma. The plasma frequency $\omega_{\rm p}$ is progressively larger for simulations $\mathcal{K}_{1}\!-\!\mathcal{K}_{6}$, see Table~\ref{tb:simulations}. The exponential growth rate (dashed lines) is determined from~\eqref{eq:sengrowth}.
    {\bf Bottom Panel:} Same as above but now the field is extracted at $r_{\rm ex} = 400M$. The modulations arise from scattering of photons with the axion cloud, similar to Fig.~\ref{fig:EMfieldBurstmuM03r481020}.}
    \label{fig:BurstPlasmamuM03r20}
\end{figure}

Figure~\ref{fig:BurstPlasmamuM03r20} summarizes the main results. When $\omega_{\rm p} < \mu/2$, the plasma has little impact on the system and the parametric instability ensues. When $\omega_{\rm p} \geq \mu/2$ instead, a suppression of the photon production is seen. We find the growth rate estimated in equation~(19) of~\cite{SenPlasma} to fit our simulations well, when taking into account the finite-size effect of the cloud as $\lambda_{\rm esc} \sim 1/d$, i.e.,
\begin{equation}\label{eq:sengrowth}
    \lambda \approx \frac{\mu^{2}\sqrt{\mu^{2}-4\omega_{\rm p}^{2}}}{2\mu^{2}-4\omega_{\rm p}^{2}}k_{\rm a} \Psi_0 - \lambda_{\rm esc}\,.
\end{equation}
Finally, the beating pattern in the EM radiation at larger radii (see bottom panel of Fig.~\ref{fig:BurstPlasmamuM03r20}) is explained by the photons having to travel through the cloud, thereby scattering of the axions.

\begin{figure}
    \centering
    \includegraphics[width = 0.5\textwidth]{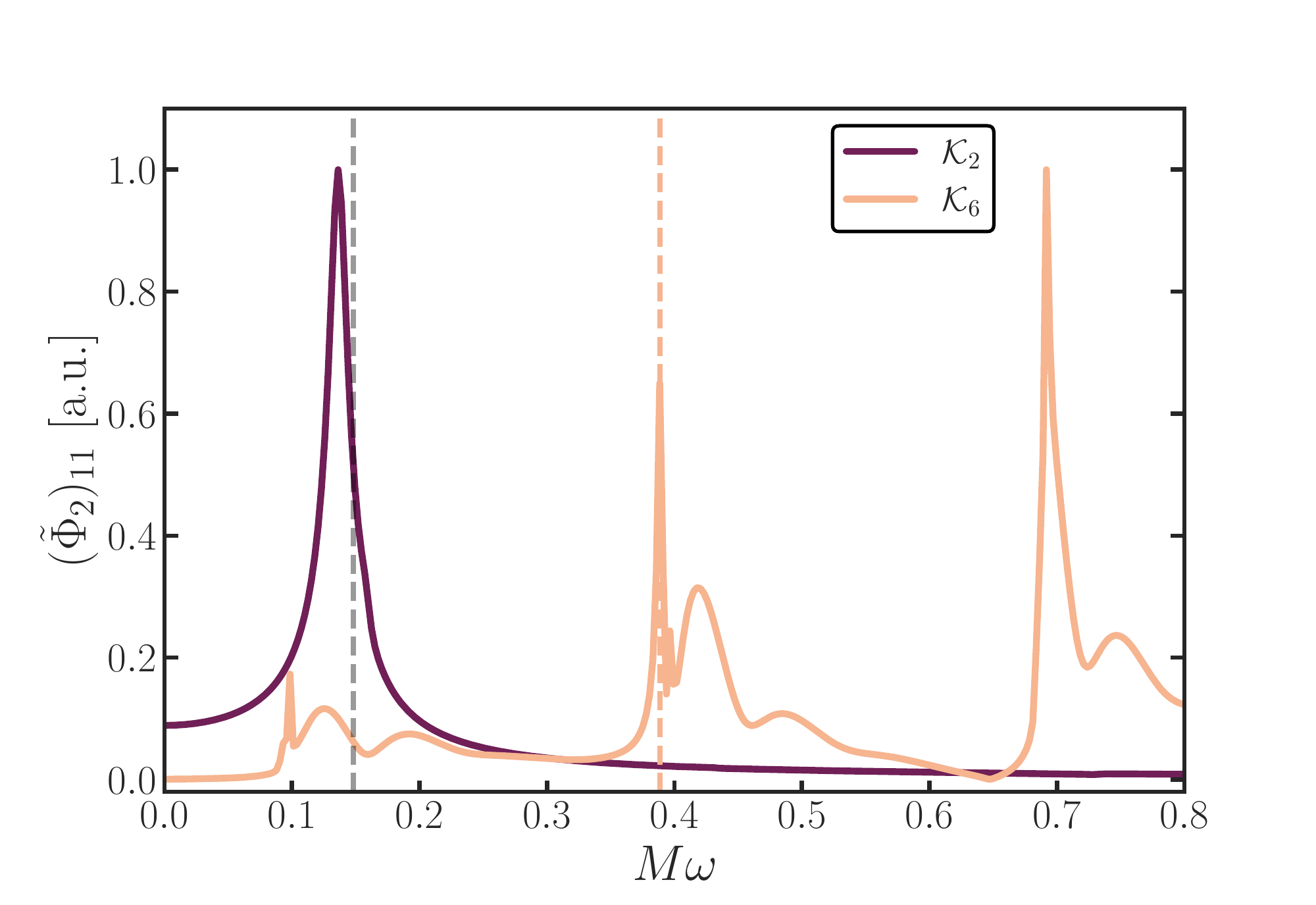}
    \caption{Fourier transform of $(\Phi_{2})_{11}$ extracted at $r_{\mathrm{ex}} = 20M$ for simulations $\mathcal{K}_{2}$ ($\omega_{\rm p} < \mu/2$) and $\mathcal{K}_{6}$ ($\omega_{\rm p} > \mu/2$) with $\mu M = 0.3$. The $y$ axis is shown in arbitrary units such that both are visible on the same figure. The gray dashed line denotes half the boson mass, while the other dashed line shows the plasma-driven quasi-bound state frequency for $\omega_{\rm p} = 0.4$, located at $M\omega=0.3887-0.0016i$.}
    \label{fig:FourierPlasma}
\end{figure}

Additionally, we show the Fourier decomposition of the signal in Fig.~\ref{fig:FourierPlasma}. As we concluded from Fig.~\ref{fig:BurstPlasmamuM03r20}, for low $\omega_{\rm p}$, the parametric instability is barely hindered and a clear peak arises at half the boson mass. However, when $\omega_{\rm p}>\mu/2$, we observe the presence of modes with a frequency very close to $\omega_{\rm p}$. We find good agreement between these peaks and the plasma-driven quasi-bound states computed in a similar setup~\cite{Cannizzaro:2020uap}. Note however, that these bound states are extremely fragile and geometry dependent, and may disappear if more realistic plasma models are considered~\cite{Dima:2020rzg}. We conjecture the origin of the two additional peaks at $\omega_{\rm p}\pm \mu$ to be up and down-scattering from the quasi-bound state photons with the axion cloud. Due to the fact that modes with frequency $\omega_{\rm p}- \mu$ are decaying, their amplitude is highly suppressed compared to the up-converted ones. 

\begin{figure}
    \centering    
    \includegraphics[width = 0.5\textwidth]{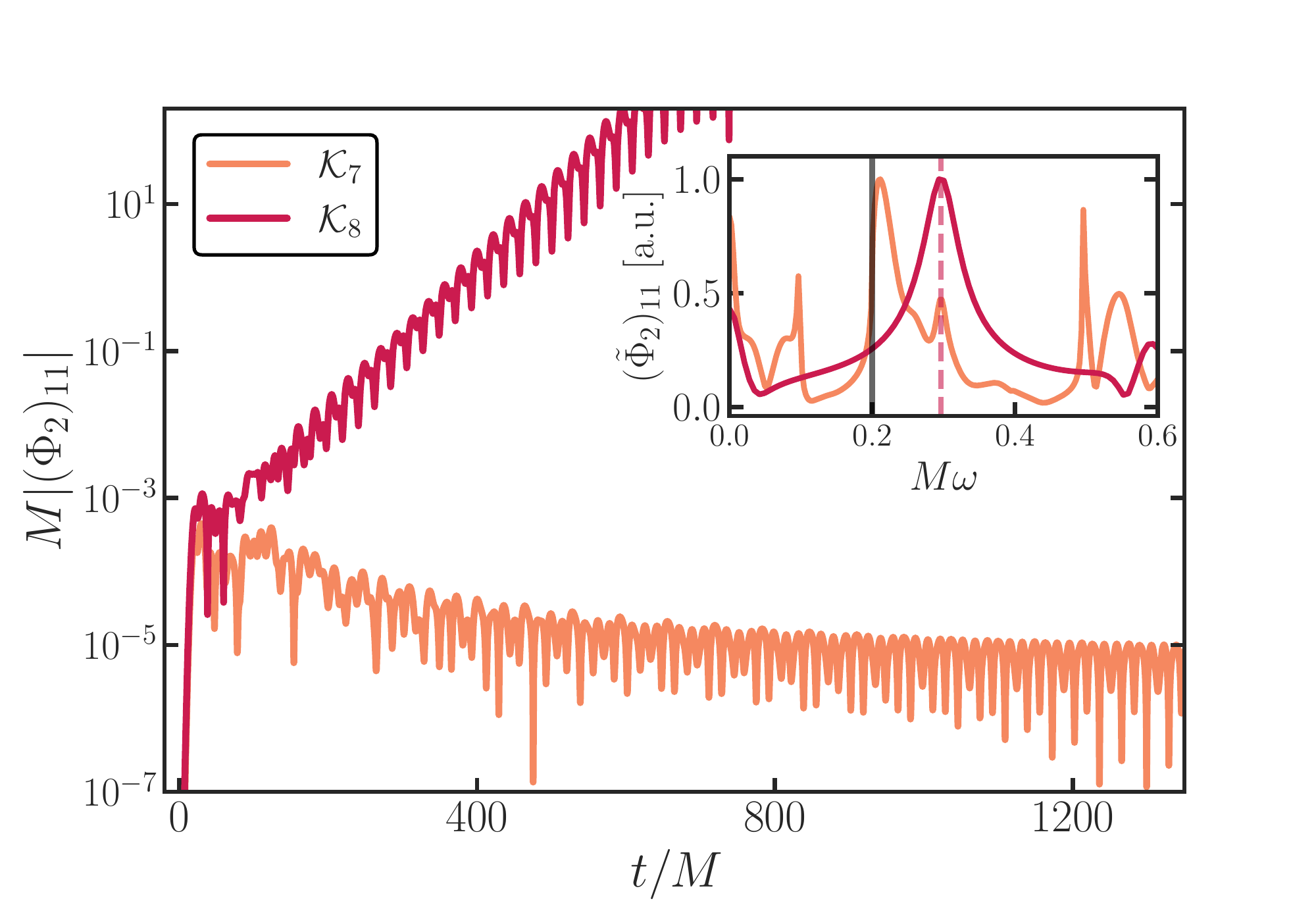}
    \caption{Time evolution of $|(\Phi_{2})_{11}|$ extracted at $r_{\mathrm{ex}} = 20M$ for simulations $\mathcal{K}_{7}$ and $\mathcal{K}_{8}$ with $\mu M = 0.3$. Even though in both simulations $\omega_{\rm p} = 0.2 > 0.15 = \mu/2$, the instability can be restored when $k_{\rm a}\Psi_0$ is high enough. The inset shows the Fourier transform of both curves, with the gray line indicating the plasma frequency. The red dashed line shows the frequency of the peak for $\mathcal{K}_8$, which is at $M\omega = 0.3$, indicating that the second instability band of the Mathieu solution is triggered.}
    \label{fig:FourierPlasmaSen}
\end{figure}

We now focus on the high axionic coupling regime. In the toy model considered in~\cite{SenPlasma}, it was shown that even when $\omega_{\rm p} \geq \mu/2$, an EM instability could be triggered for high enough $k_{\rm a}\Psi_0$. In Fig.~\ref{fig:FourierPlasmaSen}, we confirm this prediction numerically and show, for the first time, the presence of an instability in dense plasmas. This might seem in tension with the kinematic argument that for $\omega_{\rm p}>\mu/2$ the axion decay into two photons is forbidden. However, as we show in the inset of Fig.~\ref{fig:FourierPlasmaSen}, the frequency centers at $\omega=\mu$ instead of the usual $\omega=\mu/2$. This suggests the photon production to be dominated by a different process, namely $\Psi + \Psi \rightarrow \gamma + \gamma$.

To support this hypothesis, we study again the connection with the Mathieu equation (see e.g.~\cite{Kovacic2018MathieusEA}). As we detail in Appendix~\ref{sec:appendMathieu}, in flat spacetime, the Maxwell equations in presence of a plasma can indeed be recasted into a Mathieu equation which admits instability bands whenever $\omega^2 = p_z^2+\omega_{\rm p}^2 = n^2\,\mu^2/4$, with $n \in \mathbb{N}$. Therefore, when $\omega_{\rm p}>\mu/2$, the first instability band ($n=1$) at $\omega=\mu/2$ can indeed not be triggered, yet it is still possible to trigger the second band at $n=2$, where $\omega=\mu$. This matches exactly the phenomenology observed in Fig.~\ref{fig:FourierPlasmaSen} and thus we conclude the EM instability to correspond to the second instability band of the Mathieu equation, which indeed is triggered by the process $\Psi + \Psi \rightarrow \gamma + \gamma$ (and kinematically viable even for $\omega_{\rm p}>\mu/2$)~\cite{Hertzberg:2018zte}. This analysis can be continued for even higher branches. However, since these get progressively narrower, (extremely) high values of the axionic coupling could be necessary to trigger instabilities in higher bands. 
%%%%%%%%%%%%%%%%%%%%%%%%%%%%%%%%
\subsection{With superradiance}
%%%%%%%%%%%%%%%%%%%%%%%%%%%%%%%%
We now probe the axion-photon-plasma system starting from a subcritical regime, yet letting it evolve to supercritical values via SR. Based on the previous section, we expect the system to turn unstable at some point, as the axionic coupling $k_{\rm a} \Psi_0$ grows indefinitely. Due the longer timescales associated with this process, we anticipate assumption~(v), regarding the neglecting of the gravitational term, to be violated for the same parameter choices as before. Therefore, we evolve the system with $\mu M = 0.1$, such that all the assumptions are still justified. 

\begin{figure}
    \centering
    \includegraphics[width = 0.5\textwidth]{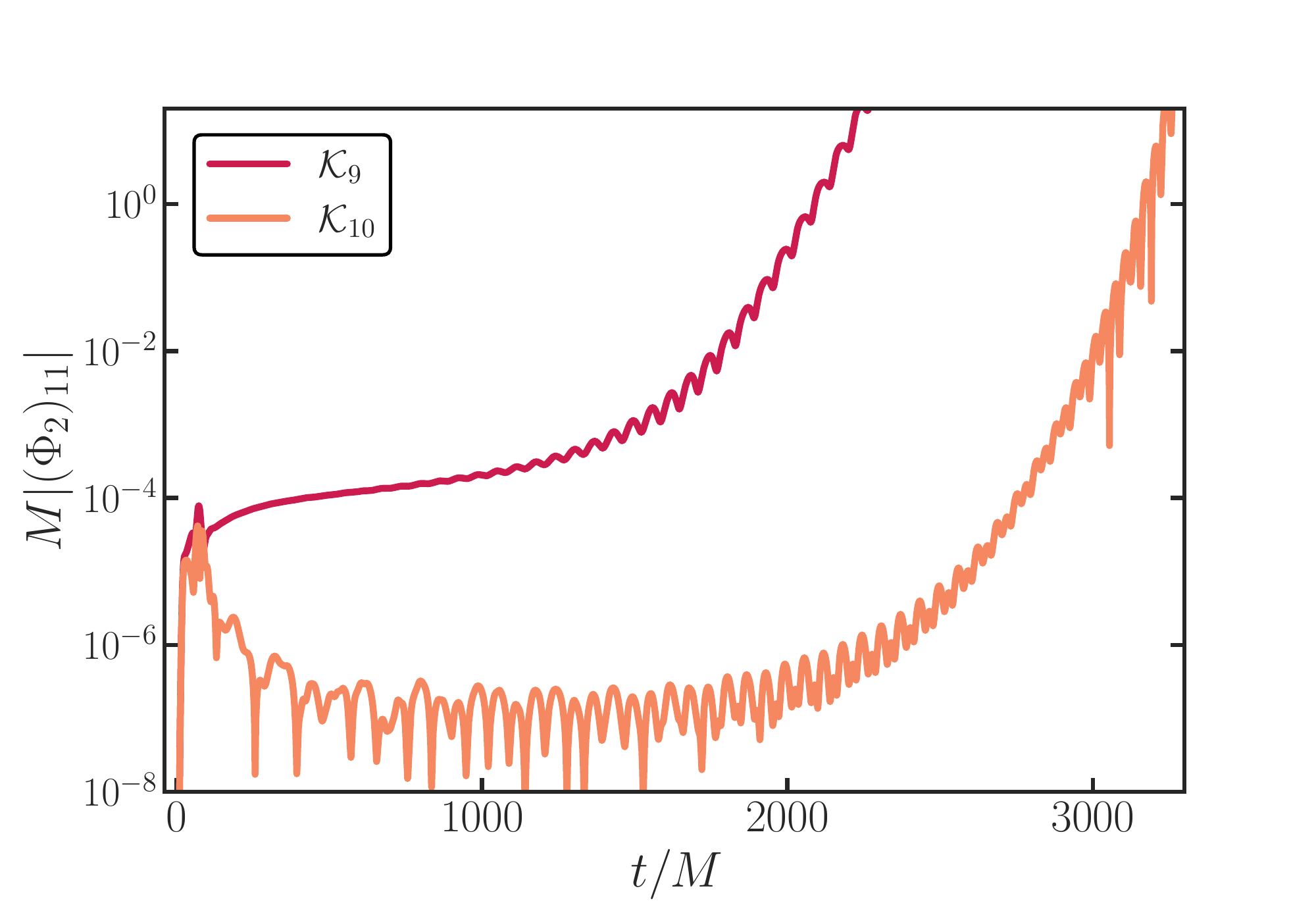}
    \caption{The dipolar component of EM radiation extracted at $r_{\mathrm{ex}} = 20M$ for simulations $\mathcal{K}_{9}$ ($\omega_{\rm p} = 0.02$) and $\mathcal{K}_{10}$ ($\omega_{\rm p} = 0.07$) with $\mu M = 0.1$. Although initially subcritical, SR drives the axionic coupling high enough such that an instability can occur, even when  $\omega_{\rm p} = 0.07 > 0.05 = \mu/2$.}
    \label{fig:FourierPlasmawithC}
\end{figure}

In Fig.~\ref{fig:FourierPlasmawithC}, we show two simulations, $\mathcal{K}_{9}$ ($\omega_{\rm p} < \mu /2$) and $\mathcal{K}_{10}$ ($\omega_{\rm p} > \mu /2$), which capture well the two distinct outcomes. In the former, the usual instability with $\omega = \mu /2$ ensues when the system has reached the supercritical threshold, yet in the latter, the time to reach this threshold is longer as the axionic coupling must grow sufficiently to trigger the second instability band. Note that, similar to the previous section, there is no backreaction onto the axion field, which therefore merely acts as a big reservoir for the EM field. This naturally explains the absence of a saturation phase. Should the backreaction be included however, there is no physical reason to expect that the saturation phase is ruined by the presence of plasma as it does not interfere with the balance between the energy inflow from the BH and energy outflow from the emitted photons. We therefore expect that the general outcome from the analysis in Section~\ref{subsec:analytical} still holds, aside from minor modifications.\footnote{For example, the presence of a plasma affects the escape rate, as photons will travel more slowly through it, i.e.,~$v_{\gamma} < c$.}

By allowing the axionic coupling to take on arbitrarily high values, an instability is thus always triggered, regardless of the plasma frequency. In practice however, it is bounded by constraints on the coupling constant $k_{\rm a}$ and the mass of the cloud (which relates to $\Psi_0$) when SR growth is saturated. We can estimate this maximum axionic coupling, and therefore the maximum plasma frequency (= electron density) for which an instability occurs. We do this using the flat space toy model detailed in Appendix~\ref{app:plasmaMathieu}. From~\eqref{eq:PlasmaMathieu}, it is immediate to see that the critical value to trigger an instability in the presence of an overdense plasma (when $\omega_{\rm p} \gtrsim p_z$) is given by
\begin{equation}
    k_{\rm a} \psi_0 \gtrsim \frac{\omega_{\rm p}^2+ p_z^2}{2\mu p_z}\approx \frac{\omega_{\rm p}^2}{\mu^2}\,.
\end{equation}
This condition corresponds to the requirement that the harmonic term in the Mathieu-like equation dominates over the non-oscillatory one [cf.~\eqref{eq:PlasmaMathieu}]. We have confirmed that this flat spacetime model closely matches the simulations in curved spacetime. Therefore, we can safely use the (flat spacetime) relation between the axion amplitude and the mass of the cloud~\cite{Brito:2015oca} to obtain
\begin{equation}
    k_{\rm a} \gtrsim 8 \times 10^{2} \biggl(\frac{\omega_{\rm p}^2}{\mu^2}\biggr)\biggl(\frac{0.1}{M_{\rm c}/M}\biggr)^{1/2}\biggl(\frac{0.2}{\mu M}\biggr)^{2}\,.
\end{equation}
Note that when $\omega_{\rm p} \approx \mu$, this condition reduces to the one derived in~\cite{Ikeda:2018nhb}, while in the case $\omega_{\rm p}\gg\mu$, stronger constrains on the coupling are imposed. We can translate this into the following condition for when an instability is triggered
\begin{equation}\label{eq:criticalcoupling}
\begin{aligned}
\frac{10^{-13}~\mathrm{GeV}^{-1}}{k_{\rm a}} &\lesssim 8 \times 10^{5} \biggl(\frac{10^{-3}~\mathrm{cm}^{-3}}{n_{\rm e}}\biggr)
\biggl(\frac{M_{\rm c}/M}{0.1}\biggr)^{1/2}\\&\times \biggl(\frac{1M_{\odot}}{M}\biggr)^{2}\biggl(\frac{\mu M}{0.2}\biggr)^{4}\,.
\end{aligned}
\end{equation}
Since current constraints on the coupling constant are around $10^{-13}~\mathrm{GeV}^{-1}$~\cite{AxionLimits}, this means a plasmic environment can at least be a few orders of magnitude higher than the interstellar medium ($n_{\rm e} \in (10^{-3}, 1)~\mathrm{cm}^{-3} $) and still an EM instability would be triggered. 
%%%%%%%%%%%%%%%%%%%%%
\section{Observational prospects}\label{sec:observational}
%%%%%%%%%%%%%%%%%%%%%
Based on the previous section, there are two distinct outcomes for parametric photon production in presence of plasma; (i) the dominant instability for $\omega_{\rm p} < \mu/2$, and (ii) higher band instabilities in the regime of large axionic couplings, for $\omega_{\rm p} > \mu/2$. In situation (i), the plasma frequency establishes a threshold for the frequency of the emitted photons. In the case of the interstellar medium, characterized by an electron density of approximately $1~\mathrm{cm}^{-3}$~\cite{Ferriere:2001rg}, the value of $\omega_{\rm p}$ is estimated to be around $10^{-11}~\mathrm{eV}/\hbar$~\eqref{eq:plasmafreq}, corresponding to a frequency of $7.6~\mathrm{kHz}$. This should be compared to e.g.~a BH with mass $5 M_\odot$, which can effectively ($\mu M = 0.4)$ accumulate an axion cloud with the same frequency, i.e.,~$\mu \approx 10^{-11}~\mathrm{ eV}$. In this case, the axion cloud would decay into pairs of photons with a frequency of approximately $3.8~\mathrm{kHz}$, which is close to the threshold value required for observation. For higher $\mu$, the mass of the plasma can be considered negligible, and we anticipate that the primary photon flux will exhibit a nearly monochromatic energy of $\mu/2$ within the radio-frequency band. Note however, that for higher $\mu$, we need to invoke subsolar-mass BHs to grow the cloud on astrophysically relevant timescales. Besides the total photon flux derived analytically~\eqref{eq:analyticSRLuminosity} and extracted from our simulations~\eqref{eq:dEdt}, we also demonstrate, for the first time, the anisotropic emission morphology in the frame of the BH, see Fig.~\ref{fig:Phi2MultipolesLargeradii}. Consequently, one expects varying observer inclination angles to result in quantitatively distinct signals.

Situation (ii) presents an opportunity to observe photons produced by axion clouds beyond stellar-mass BHs. Still, the typical frequency of these photons fall below the $\mathrm{MHz}$ band, and thus poses a challenge for current Earth-based radio observations. However, the forthcoming moon-based radio observatories can potentially detect these signals~\cite{burns2019farside}. Moreover, in the case of a rapidly spinning BHs resulting from binary mergers, one can anticipate that the radio signals will follow strong gravitational wave emissions with a delay determined by the SR timescale. Consequently, by employing multimessenger observations between gravitational wave detectors and lunar radio telescopes, constraints can be imposed on the axion-photon coupling. 

\begin{figure}
    \centering
    \includegraphics[width = 0.5\textwidth]{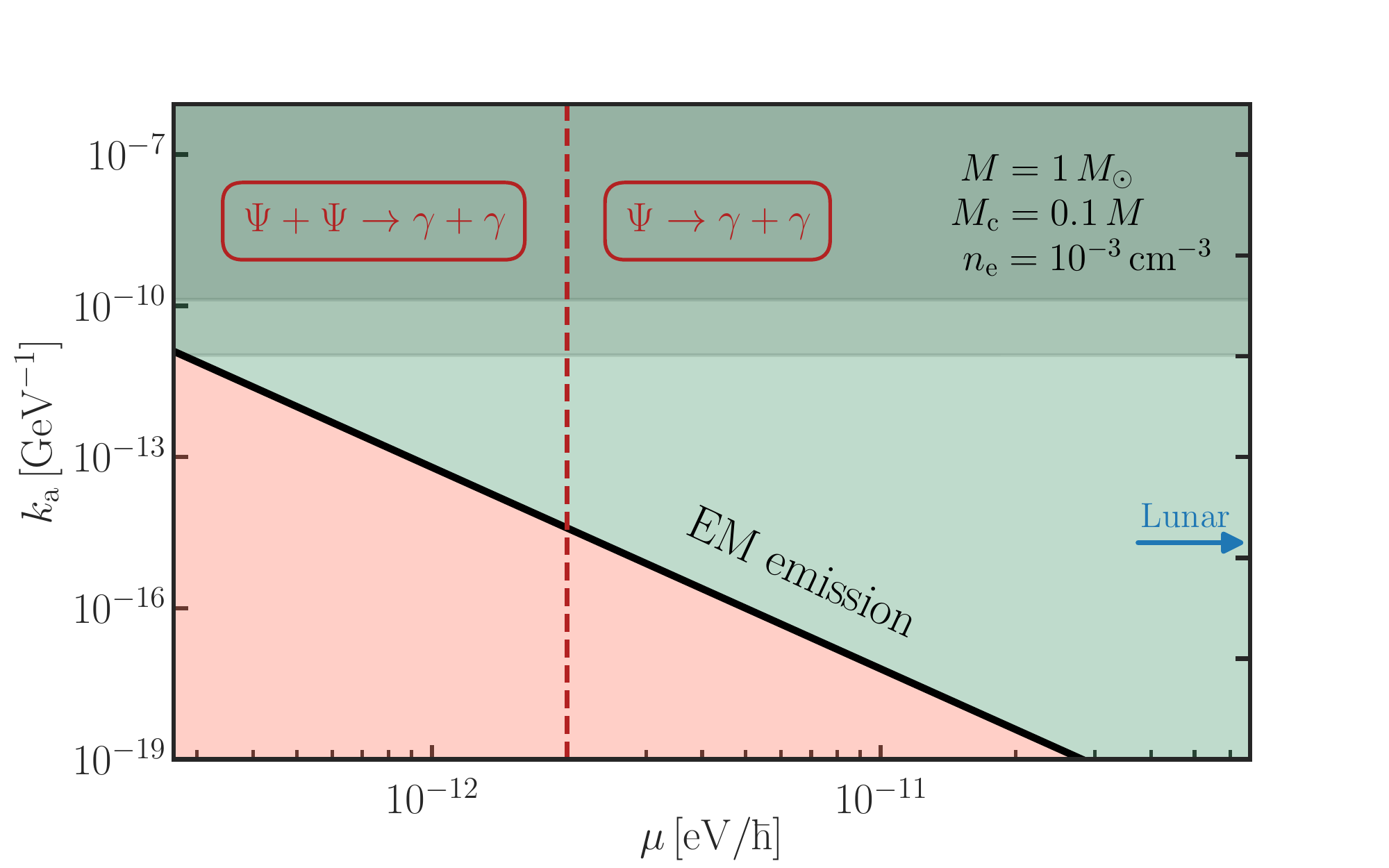}
    \caption{Possible outcomes of the axion-photon-plasma system depending on the boson’s mass and the axion-photon coupling. From~\eqref{eq:criticalcoupling}, we find that above the threshold, EM emission is triggered (green area), while below it, the system never turns supercritical. When the boson frequency is low, i.e.,~$\mu \lesssim 2\omega_{\rm p}$, the EM instability can still be triggered in the presence of plasma, given a high enough axionic coupling. This happens when the system enters the second instability band, indicated by the red dashed line. In the future, Lunar-based radio observatories could probe a part of this parameter space, denoted by the blue arrow. Finally, we show two of the most robust constraints on the axion-photon coupling, namely from the solar axion experiment CAST~\cite{CAST:2017uph} (darker region) and from measurements on supernova 1987A (lighter region)~\cite{Hoof:2022xbe}. This data was collected from~\cite{AxionLimits}.}
    \label{fig:SummaryBurst}
\end{figure}

Finally, the projected saturated value of $k_{\rm a} \Psi_{\rm sat}$ can induce a rotation in the linear polarization emitted in the vicinity of BHs~\cite{Carroll:1989vb,Harari:1992ea}. This phenomenon has been investigated in the context of supermassive BHs~\cite{Chen:2019fsq,Chen:2021lvo,Chen:2022oad}. It should be noted however that due to the significant hierarchy between the ultra-low SR mass window and the plasma mass generated by a dense environment, higher-order instabilities are not expected to occur in supermassive BHs. Consequently, the axion cloud outside an supermassive BH remains robust against axion-photon couplings. A summary of our findings in the presence of plasma can be found in Fig.~\ref{fig:SummaryBurst}

%%%%%%%%%%%%%%%%%%%%%%%%%%%%%%
\section{Summary and 
conclusions}\label{sec:conclusions}
%%%%%%%%%%%%%%%%%%%%%%%%%%%%%%
The presence of ultralight bosons is ubiquitous in beyond-the-Standard Model theories. Proposed as a solution to the dark matter or strong {\it CP} problem, they are of interest across various areas of physics. Detecting them however, is notoriously hard especially when their coupling to the Standard Model is weak. Through SR, a new channel for detection opens up, turning BHs into powerful ``particle detectors'' in the cosmos. 

In this work, we performed a detailed numerical study of the dynamics of boson clouds around BHs with axionic couplings to the Maxwell sector. We demonstrate the existence of an EM instability, thereby confirming previous studies. However, while those works assumed that the cloud is allowed to grow {\it before} turning on the coupling, we relax this assumption and study the axionic coupling {\it simultaneously} with the growth of the cloud, conform to the SR mechanism. We find that, in this setup, a stationary state emerges, wherein every produced axion by SR is converted into photons that escape the cloud at a steady rate. This leads to strong observational signatures, as the nearly monochromatic and constant EM signal could have a luminosity comparable with some of the brightest sources in our universe. Moreover, the depletion of the axion cloud impacts current constraints on the boson mass. 

Additionally, we study the influence of a surrounding plasma on the EM instability. In the regime of small axionic couplings, we find the expected suppression of the instability when the plasma frequency exceeds half the boson mass. Surprisingly however, the instability can be restored for high enough couplings. We show how the Maxwell equations in presence of plasma reduce to a Mathieu equation, which naturally explains how the restoring of the instability is associated with higher-order instability bands. With this interpretation in mind, we conclude that (very) dense plasmas do not necessarily quench the parametric instability; it solely depends on the value of the axionic coupling. As higher instability bands correspond to higher frequencies of the emitted photons, it might be possible to detect the parametric mechanism beyond the stellar-mass BH regime.

In order to make clear observational predictions, at least two issues need to be resolved;~(i) the geometry of realistic plasmic environments and (ii) the non-linear dynamics of the axion-photon-plasma system. While in this work, we assumed a constant plasma density throughout space, astrophysical environments can be nearly planar, for example in the case of thin disks (see e.g.~\cite{Abramowicz2013}), which will impact the strength of the resulting EM flux. Additionally, even though our linearized model captures well the impact of the plasma on the EM instability, the backreaction onto the axion field is essential for the late-time behavior. Moreover, as we showed, high axionic couplings could be necessary to trigger the EM instability in dense plasmas. Therefore, a natural extension of this work would be to study non-linear dynamics of the coupled axion-photon-plasma system. Indeed, the propagation of large amplitude EM waves in dense plasmas in the non-linear regime can carry a plethora of interesting features (see e.g.~\cite{1970PhFl...13..472K,1971PhRvL..27.1342M,Cardoso:2020nst,Cannizzaro:2023ltu}). We intend to treat these points in a future work and with that fully understand EM instabilities of axion clouds in realistic astrophysical environments.
%%%%%%%%%%%%%%
\begin{acknowledgments}
%%%%%%%%%%%%%%
We thank Giovanni Maria Tomaselli for comments on the manuscript, and David Hilditch, Zhen Zhong, and Miguel Zilhão for useful discussions. V.C.\ is a Villum Investigator and a DNRF Chair, supported by VILLUM Foundation (grant no.~VIL37766) and the DNRF Chair program (grant no.~DNRF162) by the Danish National Research Foundation. V.C.\ acknowledges financial support provided under the European Union's H2020 ERC Advanced Grant ``Black holes: gravitational engines of discovery'' grant agreement
no.~Gravitas--101052587. Views and opinions expressed are however those of the author only and do not necessarily reflect those of the European Union or the European Research Council. Neither the European Union nor the granting authority can be held responsible for them.
This project has received funding from the European Union's Horizon 2020 research and innovation programme under the Marie Sklodowska-Curie grant agreement No.~101007855. 
E.C.~acknowledges additional financial support provided by Sapienza, ``Progetti per Avvio alla Ricerca,'' protocol number AR1221816BB60BDE.
\end{acknowledgments}
\appendix
%%%%%%%%%%%%%%%%%%%%%%%%%%%%%%%%%%%%%%%%%%%%%%%%%%%%%%%%%%%%%%%%
\section{Benchmarks for evolution of scalar fields}\label{app:freescalars}
%%%%%%%%%%%%%%%%%%%%%%%%%%%%%%%%%%%%%%%%%%%%%%%%%%%%%%%%%%%%%%%%
The purpose of this appendix is to study in some detail the time evolution of free massive scalar fields in the vicinity of a Schwarzschild BH. Even though SR requires a spinning BH and thus the use of the Kerr metric, timescales are prohibitively large. Nevertheless, the main focus of our work is on physics related to the {\it existence} of scalar clouds, more than to what caused them in the first place.

As such, we mimic SR growth without the need of a spinning BH (see Section~\ref{sec:app_artificial_super} below) and therefore we consider a Schwarzschild spacetime for simplicity. We still need to guarantee that, on the required timescales, a bound state exists, so that it can mimic well the true SR clouds. Fortunately, massive scalars around non-spinning BHs do settle on quasi-bound states which, while not unstable, have extremely large lifetimes. Thus, we want to show first of all that our numerical framework reproduces well such states.
%%%%%%%%%%%%%%%%%%%%%%%%%%%%%%%%%%%%%%%%%%%%%%%%%%%%%%%%%%%%%%%%
\subsection{Bound states}\label{appA:boundstates}
%%%%%%%%%%%%%%%%%%%%%%%%%%%%%%%%%%%%%%%%%%%%%%%%%%%%%%%%%%%%%%%%
The initial data whose time evolution we will study, are the quasi-bound states of a massive scalar field, which are solutions localized in the vicinity of the BH and prone to become unstable in the SR regime (if the BH is allowed to spin). There exist various methods to find such quasi-bound solutions, either by direct numerical integration or using continued fractions~\cite{Leaver:1985ax,Cardoso:2005vk,Dolan:2007mj,Berti:2009kk}. In this work, we use Leaver's continued fraction approach~\cite{Leaver:1985ax}. It is crucial to have accurate solutions describing pure quasi-bound states, as deviations from such a pure state may trigger excitations of overtones, resulting in a beating pattern~\cite{Witek:2012tr}.

In Boyer-Lindquist (BL) coordinates ($t_{\scalebox{0.55}{$\mathrm{BL}$}}$, $r_{\scalebox{0.55}{$\mathrm{BL}$}}$, $\theta_{\scalebox{0.55}{$\mathrm{BL}$}}$, $\varphi_{\scalebox{0.55}{$\mathrm{BL}$}}$), the scalar field bound state is given by\footnote{We include spin here for generality, although we evolve the scalar field in a Schwarzschild background}
\begin{equation}\label{eq:scalarfieldBL}
\Psi_{\ell m}=e^{-i \omega t_{\scalebox{0.55}{$\mathrm{BL}$}}} e^{-i m \varphi_{\scalebox{0.55}{$\mathrm{BL}$}}} S_{\ell m}\left(\theta_{\scalebox{0.55}{$\mathrm{BL}$}}\right) R_{\ell m}\left(r_{\scalebox{0.55}{$\mathrm{BL}$}}\right)\,,
\end{equation}
where $S_{\ell m}(\theta_{\scalebox{0.55}{$\mathrm{BL}$}})$ are the spheroidal harmonics. In a Schwarzschild geometry, the angular dependence is fully captured by the familiar spherical harmonics 
$S_{\ell m}(\theta_{\scalebox{0.55}{$\mathrm{BL}$}})e^{i m\varphi_{\scalebox{0.55}{$\mathrm{BL}$}}} = Y_{\ell m}(\theta_{\scalebox{0.55}{$\mathrm{BL}$}}, \varphi_{\scalebox{0.55}{$\mathrm{BL}$}})$. The radial dependence is given by
\begin{equation}\label{eq:radialdep}
    \begin{aligned}
       R_{\ell m}\left(r_{\scalebox{0.55}{$\mathrm{BL}$}}\right)=&\left(r_{\scalebox{0.55}{$\mathrm{BL}$}}-r_{\scalebox{0.55}{$\mathrm{BL, +}$}}\right)^{-i \sigma}\left(r_{\scalebox{0.55}{$\mathrm{BL}$}}-r_{\scalebox{0.55}{$\mathrm{BL, -}$}}\right)^{i \sigma+\chi-1}\times \\&e^{r_{\scalebox{0.55}{$\mathrm{BL}$}} q} \sum_{n=0}^{\infty} a_n\left(\frac{r_{\scalebox{0.55}{$\mathrm{BL}$}}-r_{\scalebox{0.55}{$\mathrm{BL, +}$}}}{r_{\scalebox{0.55}{$\mathrm{BL}$}}-r_{\scalebox{0.55}{$\mathrm{BL, -}$}}}\right)^n\,,
    \end{aligned}
\end{equation}
where
\begin{equation}
\begin{aligned}
    \sigma&=\frac{2 M r_{\scalebox{0.55}{$\mathrm{BL, +}$}}\left(\omega-\omega_{\rm c}\right)}{r_{\scalebox{0.55}{$\mathrm{BL, +}$}}-r_{\scalebox{0.55}{$\mathrm{BL, -}$}}}, \quad q=\pm \sqrt{\mu^2-\omega^2}\,,\\
    \chi&=M\frac{\mu^2-2 \omega^2}{q}\,.
\end{aligned}
\end{equation}
Here, $r_{\scalebox{0.55}{$\mathrm{BL, \pm}$}} = M \pm \sqrt{M^{2}-a_{\scalebox{0.60}{$\mathrm{J}$}}^{2}}$ are the inner~$({\scalebox{0.85}{$\mathrm{-}$}})$ and outer~$({\scalebox{0.85}{$\mathrm{+}$}})$ horizon, $\omega_{\rm c} = m \Omega_{H} = m a_{\scalebox{0.60}{$\mathrm{J}$}}/(2Mr_{\scalebox{0.55}{$\mathrm{BL, +}$}})$ is the critical SR frequency, $a_{\scalebox{0.65}{$\mathrm{J}$}}$ is the spin of the BH and to obtain quasi-bound states, one should consider the minus sign in the expression for $q$. Since all the terms in these expressions are known in closed form, we only need to solve for the frequency of the mode of interest, $\omega$. This is found by solving the following condition for $\omega$:
\begin{equation}\label{eq:Leaver}  
\beta_{0}-\frac{\alpha_{0}\gamma_{1}}{\beta_{1}-}\frac{\alpha_{1}\gamma_{2}}{\beta_{2}-}\cdots=0\,,
\end{equation}
where all the coefficients can be found in e.g.~\cite{Dolan:2007mj}. In~\eqref{eq:radialdep}, the amplitude of the scalar field is defined arbitrarily (as long as one neglects the backreaction of the field on the background geometry). Hence, we must choose a suitable normalization. We will normalize the field by assigning a predetermined value to the maximum of the radial wave function. In previous works~\cite{Boskovic:2018lkj, Ikeda:2018nhb}, the hydrogenic approximation was used instead, where the wave function is defined as $\Psi=\Psi_0 r_{\scalebox{0.55}{$\mathrm{BL}$}} M \mu^2 e^{-r_{\scalebox{0.55}{$\mathrm{BL}$}} M \mu^2 / 2} \cos \left(\varphi_{\scalebox{0.55}{$\mathrm{BL}$}}-\omega_{\scalebox{0.65}{$\mathrm{R}$}} t\right) \sin \theta_{\scalebox{0.55}{$\mathrm{BL}$}}$, where $\omega_{\scalebox{0.65}{$\mathrm{R}$}}$ is the real part of the eigenfrequency. In order to allow for a direct comparison with those works, we relate our normalization, the maximum value of the real part of the field, $\left(R_{\ell m}\right)_{\rm max}$, to this parameter $\Psi_0$. They are related by
\begin{equation}\label{eq:normalization}
    \left(R_{\ell m}\right)_{\rm max} = \frac{4 \Psi_0 \sqrt{2\pi/3}}{e}\,,
\end{equation}
where the factor $\sqrt{2\pi/3}$ comes from the normalization of the spherical harmonics $\ell = 1$ modes, and should be adapted accordingly for higher multipoles. We will introduce relevant quantities in terms of $\Psi_0$.

For numerical purposes, BL coordinates are not ideal due to the coordinate singularity at the horizon. Therefore, we employ Kerr-Schild coordinates, which are horizon penetrating coordinates~\cite{Witek:2012tr}. The coordinate transformation from BL to Kerr-Schild (KS) coordinates is given by
\begin{equation}\label{eq:coordBLKS}
\begin{aligned}
&\mathrm{d} t_{\scalebox{0.55}{$\mathrm{KS}$}}=\mathrm{d} t_{\scalebox{0.55}{$\mathrm{BL}$}}+\frac{2 M r_{\scalebox{0.55}{$\mathrm{BL}$}}}{\Delta} \mathrm{d} r_{\scalebox{0.55}{$\mathrm{BL}$}}, \quad \mathrm{d} r_{\scalebox{0.55}{$\mathrm{KS}$}}=\mathrm{d} r_{\scalebox{0.55}{$\mathrm{BL}$}},\\  &\mathrm{d} \theta_{\scalebox{0.55}{$\mathrm{KS}$}}=\mathrm{d} \theta_{\scalebox{0.55}{$\mathrm{BL}$}}, \quad \mathrm{d} \varphi_{\scalebox{0.55}{$\mathrm{KS}$}}=\mathrm{d} \varphi_{\scalebox{0.55}{$\mathrm{BL}$}}+\frac{a_{\scalebox{0.60}{$\mathrm{J}$}}}{\Delta} \mathrm{d} r_{\scalebox{0.55}{$\mathrm{BL}$}}\,,
\end{aligned}
\end{equation}
where $\Delta \equiv r^{2} -2 M r + a^{2}_{\scalebox{0.60}{$\mathrm{J}$}}$. Using this coordinate transformation in~\eqref{eq:scalarfieldBL}, we can construct the bound state scalar field as
\begin{equation}\label{eq:scalarfieldKS}
\begin{aligned}
\Psi_{\ell m}&=e^{-i \omega t_{\scalebox{0.55}{$\mathrm{KS}$}}}\left(r_{\scalebox{0.55}{$\mathrm{KS}$}}-r_{\scalebox{0.55}{$\mathrm{KS, +}$}}\right)^P\left(r_{\scalebox{0.55}{$\mathrm{KS}$}}-r_{\scalebox{0.55}{$\mathrm{KS, -}$}}\right)^Q \times \\ 
&\left(\frac{r_{\scalebox{0.55}{$\mathrm{KS}$}}-r_{\scalebox{0.55}{$\mathrm{KS, +}$}}}{r_{{}_{\rm KS}}-r_{\scalebox{0.55}{$\mathrm{KS, -}$}}}\right)^R Y_{\ell m}\left(\theta_{\scalebox{0.55}{$\mathrm{KS}$}}, \varphi_{\scalebox{0.55}{$\mathrm{KS}$}}\right) R_{\ell m}(r_{\scalebox{0.55}{$\mathrm{KS}$}})\,,
\end{aligned}
\end{equation}
where $P=\frac{2 i \omega M r_{\scalebox{0.45}{$\mathrm{KS, +}$}}}{r_{\scalebox{0.45}{$\mathrm{KS, +}$}}- r_{\scalebox{0.45}{$\mathrm{KS, -}$}}}, Q=-\frac{2 i \omega M r_{\scalebox{0.45}{$\mathrm{KS, -}$}}}{r_{\scalebox{0.45}{$\mathrm{KS, +}$}}-r_{\scalebox{0.45}{$\mathrm{KS, -}$}}},\linebreak 
R=\frac{i m a_{\scalebox{0.45}{$\mathrm{J}$}}}{r_{\scalebox{0.45}{$\mathrm{KS, -}$}}-r_{\scalebox{0.45}{$\mathrm{KS, +}$}}}$. Unless otherwise stated, we use KS coordinates without the subscript. In our non-spinning BH case, $Q = R = 0$. The remaining extra term instead exactly cancels the divergence of the field at the BH horizon. We test our numerical setup by constructing the bound state initial configurations for scalar fields with mass couplings $\mu M = 0.1$ and $\mu M = 0.3$ and evolving them in a Schwarzschild background. 

\begin{figure}
    \hspace*{-0.6cm}
    \includegraphics[width = 0.455\textwidth]{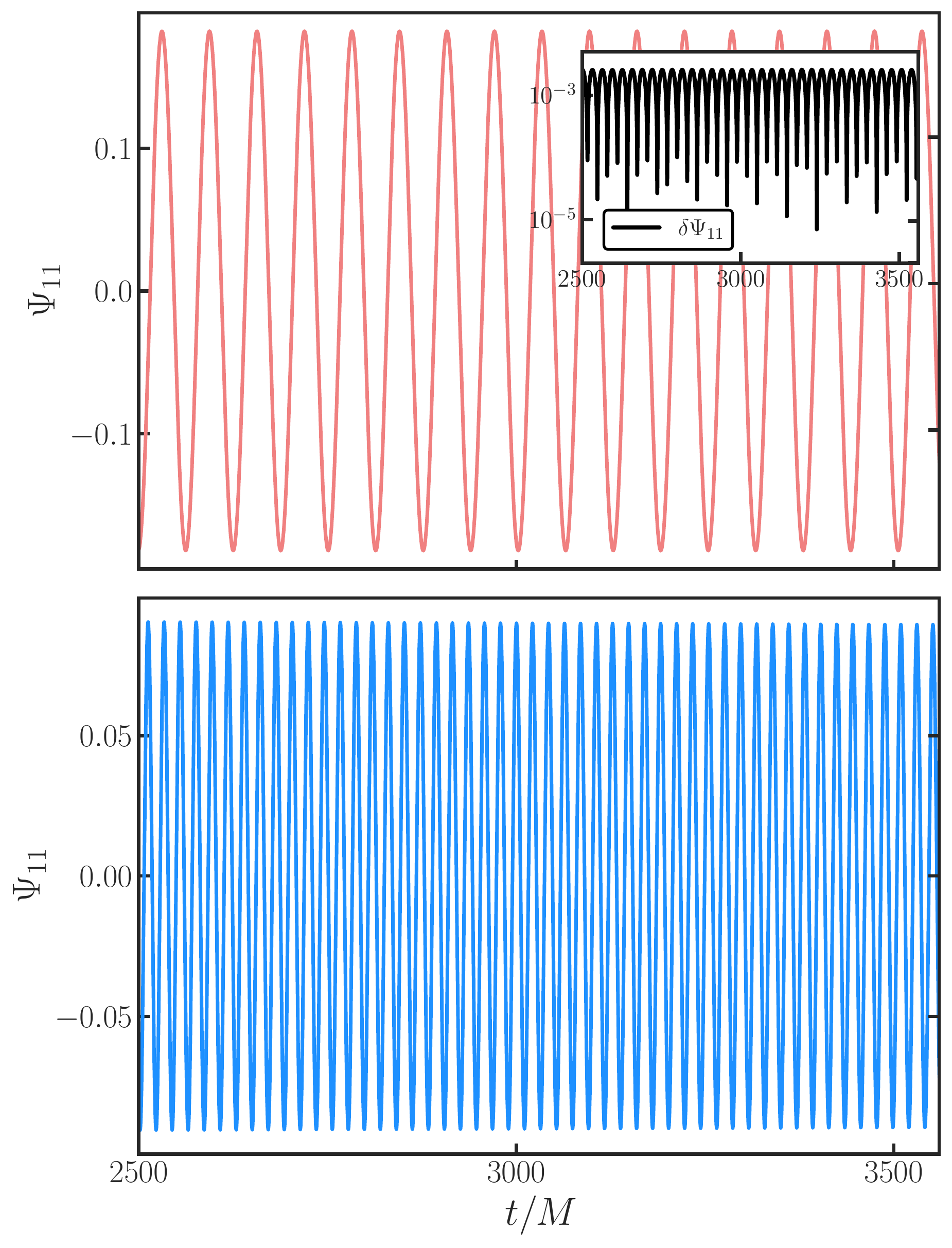}
    \caption{{\bf Top Panel:} The $\ell = m = 1$ component of a scalar field around a Schwarzschild BH. Figure shows fraction of the time evolution of the initial conditions from~\eqref{eq:scalarfieldKS}. The field is extracted at $r_{\rm ex} = 100 M$ and $\mu M = 0.1$. Inset shows $\delta \Psi_{11}$, which is the difference between the numerical output and the theoretically predicted fundamental mode $\Psi^{\mathrm{Fund}}_{11} \sim \cos{(\omega_{\scalebox{0.55}{$\mathrm{R}$}} t)}e^{-\omega_{\scalebox{0.55}{$\mathrm{I}$}} t}$, where $\omega_{\scalebox{0.65}{$\mathrm{R}$}}, \omega_{\scalebox{0.55}{$\mathrm{I}$}} $ are the real and imaginary part of the eigenfrequency, respectively. These were independently computed using Leaver's method.
    {\bf Bottom Panel:} Same for $\mu M=0.3$ and extraction radius $r_{\rm ex} = 40 M$. There is an apparent decay of the field on timescales shorter than those implied by the quasi-bound state decay. This effect is due to finite resolution, and its magnitude is small enough such that we can ignore it in our study.}
    \label{fig:scalarfieldmuM01}
\end{figure}

In Fig.~\ref{fig:scalarfieldmuM01}, we show the non-vanishing multipolar component of the field for $\mu M = 0.1$ and $\mu M = 0.3$, where we only display a fraction of the time evolution such that individual oscillations are visible. For $\mu M = 0.1$, the scalar field is exceptionally stable on timescales longer than $5000M$. For $\mu M = 0.3$, there is a decrease in the amplitude of a few percent on those timescales, which does not have severe consequences. In fact, this problem can be resolved by increasing the spatial resolution. 

\begin{figure}
    \centering
    \includegraphics[width = 0.43\textwidth]{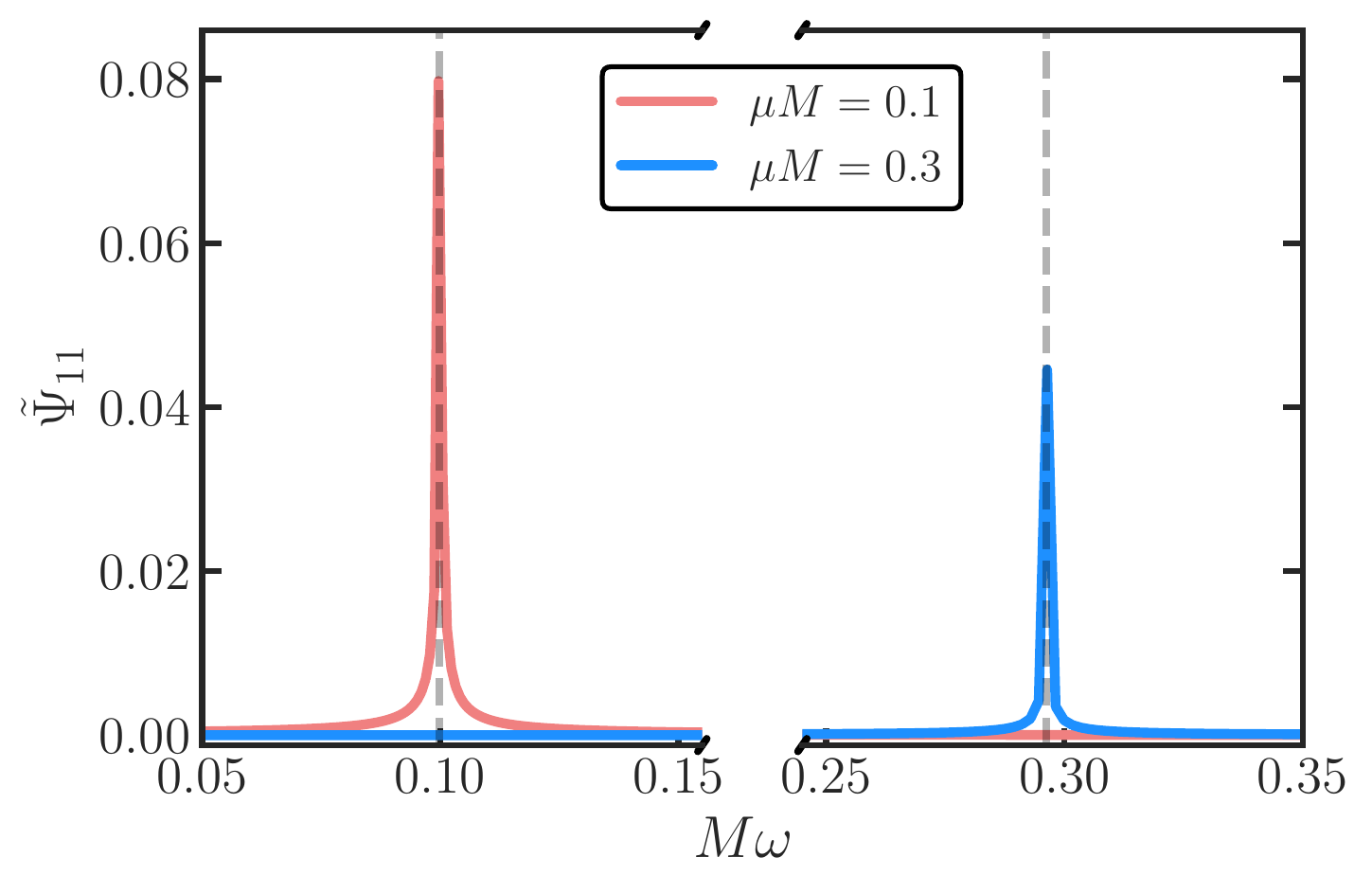}
    \caption{Fourier transform of the dipole component of the scalar field for $\mu M = 0.1$ and $\mu M = 0.3$ when the field is extracted at $r_{\rm ex} = 100 M$ and $r_{\rm ex} = 40 M$, respectively. Fourier transform is taken on the entire time evolution of Figure~\ref{fig:scalarfieldmuM01}. Dashed lines indicate the (real part of the) frequency of the fundamental mode for $\mu M = 0.1$ and $\mu M = 0.3$. Clearly, we are not triggering any overtones.}
    \label{fig:Fourierplots}
\end{figure}

As a last check, we show in Fig.~\ref{fig:Fourierplots} the Fourier transform for both $\mu M = 0.1$ and $\mu M = 0.3$ and compare it with the real part of the eigenfrequency of the fundamental mode. We find an excellent comparison, showing that we are not triggering any overtones.
%%%%%%%%%%%%%%%%%%%%%%%%%%%%%%%%%%%%%%%%%%%%%%%%%%%%%%
\subsection{Artificial superradiance}\label{sec:app_artificial_super}
%%%%%%%%%%%%%%%%%%%%%%%%%%%%%%%%%%%%%%%%%%%%%%%%%%%%%%
Studying SR for scalars is numerically challenging, since timescales for SR growth are very large. Fortunately, an effective SR-like instability can be introduced by adding a simple $C \partial \Psi/\partial t $ term to the KG equation as shown in~\eqref{eq:ASR}. This ``trick'' was first used by Zel'dovich~\cite{ZelDovich1971, ZelDovich1972, Cardoso:2015zqa} and it can mimic the correct description of many SR systems. The addition of this Lorentz-invariance-violating term causes an instability on a timescale of the order $1/C$, where we can tune $C$ to be within our numerical limits. For reference, let us report the timescales in our problem.

\noindent ``{\it Normal SR}'':
\begin{equation}
    t_{\scalebox{0.65}{$\mathrm{SR}$}} \sim 48 \left(\frac{a_{\scalebox{0.65}{$\mathrm{J}$}}}{M}(\mu M)^{-9} \right) M \quad \text{when} \quad \mu M \ll 1\,, 
\end{equation}
``{\it Artificial SR}'':
\begin{equation}
    t_{\scalebox{0.65}{$\mathrm{ASR}$}} \sim \frac{1}{C} M\,, 
\end{equation}
``{\it EM instability}'':
\begin{equation}
    t_{\scalebox{0.65}{$\mathrm{EM}$}} \sim 10 k_{\rm a}^{-1}\left(\frac{M}{M_{\rm c}}\right)^{1/2}(\mu M)^{-3} = 5 (\mu M)^{-1} M\,,
\end{equation}
which is the EM instability timescale that was found in~\cite{Ikeda:2018nhb} and where we used $k_{\rm a} \geq 2 (\frac{M}{M_{\rm c}})^{1/2}(\mu M)^{-2} M^{-1}$. 

Accordingly, for a reasonable mass coupling of $\mu M = 0.1$, and while optimizing SR growth with a maximally spinning BH, normal SR timescales are on 
the order of $t_{\scalebox{0.65}{$\mathrm{SR}$}} \sim 10^{10} M$. This should be compared to the EM instability, which is on $t_{\scalebox{0.65}{$\mathrm{EM}$}} \sim 50 M$. 

\begin{figure}[h!]
    \centering
    \includegraphics[width=0.5\textwidth]{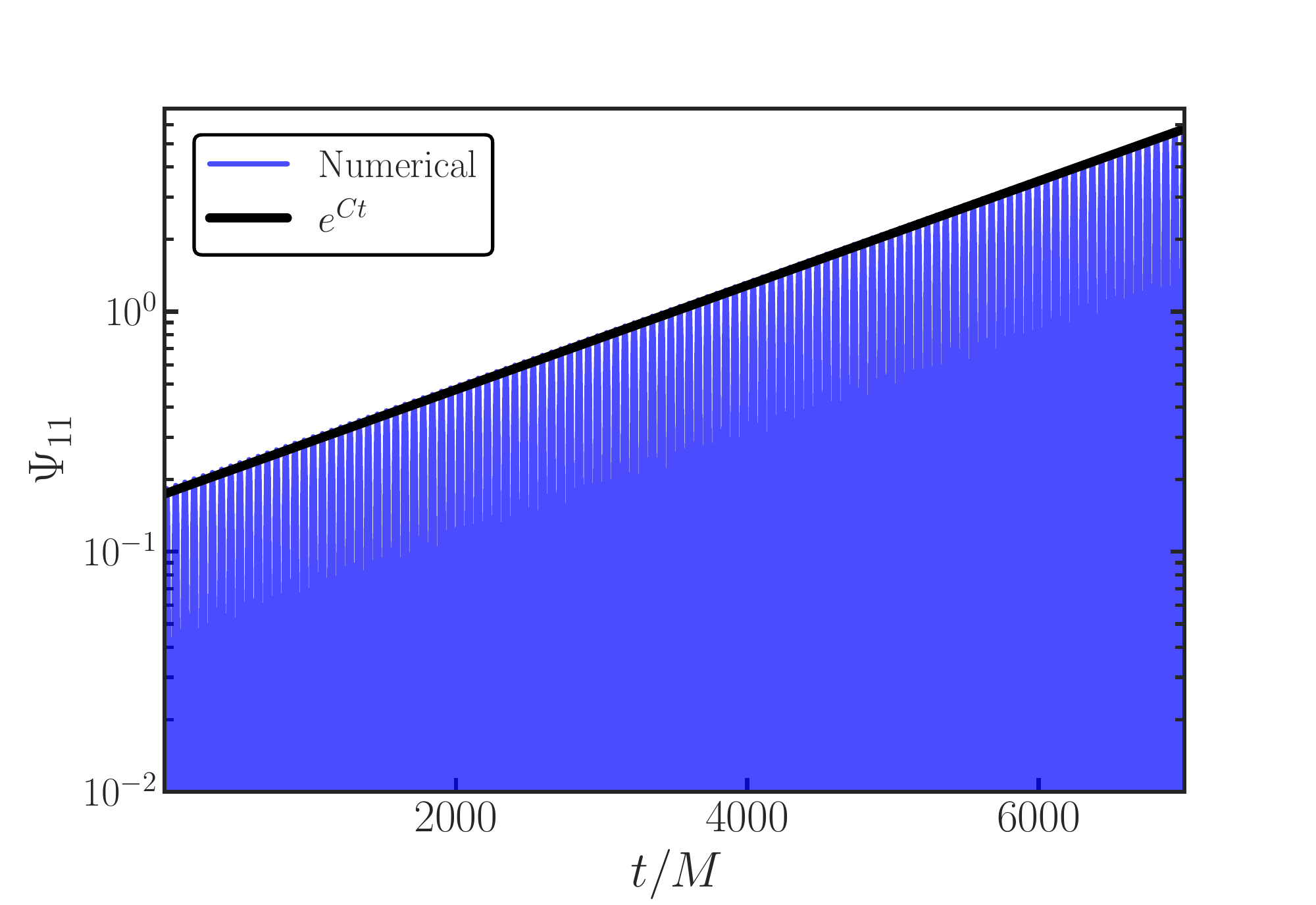}
    \caption{The time evolution of $\Psi_{11}$ extracted at $r_{\rm ex} = 100 M$ with $C = 5 \times 10^{-4} M^{-1}$ and $\mu M = 0.1$. We show the growth of the scalar field from the initial conditions~\eqref{eq:scalarfieldKS}, using Zel'Dovich trick described in the main text.}
    \label{fig:ASR}
\end{figure}

To test whether we implemented the artificial SR growth in the correct way, we set $C = 5 \times 10^{-4} M^{-1}$ and evolve the scalar field. From Fig.~\ref{fig:ASR}, we can see that the artificial SR is correctly implemented in the code, as it leads to the desired exponential evolution of the field. 
%%%%%%%%%%%%%%%%%%%%%%%%%%%%%%%%%%%%%%
\section{Wave extraction}\label{sec:Waveextraction}
%%%%%%%%%%%%%%%%%%%%%%%%%%%%%%%%%%%%%%
From the simulations, we extract the radiated scalar and vector waves at some radius $r = r_{\rm ex}$. For the scalar waves, we project the field $\Psi$ and its conjugated momentum $\Pi$ onto spheres of constant coordinate radius using the spherical harmonics with spin weight $s_{\rm w} = 0$:
\begin{equation}\label{eq:scalarextract}
\begin{aligned}
    \Psi_{\ell m}(t)&= \int \mathrm{d}\Omega\,\Psi(t, \theta, \varphi)\, {}_{0}\mkern-2mu Y_{\ell m}^{*}(\theta, \varphi), \\ \Pi_{\ell m}(t)&= \int \mathrm{d}\Omega\, \Pi(t, \theta, \varphi)\, {}_{0}\mkern-2mu Y_{\ell m}^{*}(\theta, \varphi)\,.
\end{aligned}
\end{equation}
To monitor the emitted EM (vector) waves, we use the Newman-Penrose formalism~\cite{NewmanPenrose}. In this formalism, the radiative degrees of freedom are given by complex scalars. For EM, these are defined as contractions between the Maxwell tensor and vectors of a null tetrad ($k^{\mu}, \ell^{\mu}, m^{\mu}, \Bar{m}^{\mu}$), where $k^\mu \ell_\mu=-m^\mu \bar{m}_\mu = -1$. The null tetrad itself is constructed from the orthonormal timelike vector $n^\mu$ and a Cartesian orthonormal basis $\left\{u^i, v^i, w^i\right\}$ on the spatial hypersurface. Asymptotically, the basis
vectors $\left\{u^i, v^i, w^i\right\}$ behave as the unit radial, polar and azimuthal vectors, respectively. For our purposes, the quantity of interest is the gauge-invariant Newman-Penrose scalar $\Phi_{2}$, which captures the outgoing EM radiation at infinity and is defined as
\begin{equation}\label{eq:NPscalar2}
    \Phi_{2} = F_{\mu \nu}\ell^{\mu}\Bar{m}^{\nu}\,,
\end{equation}
where $\ell^{\mu} = \frac{1}{\sqrt{2}}(n^{\mu}-u^{\mu})$ and $\Bar{m}^{\mu} = \frac{1}{\sqrt{2}}(v^{\mu}-iw^{\mu})$. Decomposing the Maxwell tensor gives
\begin{equation}
F_{\mu \nu}=n_\mu E_\nu-n_\nu E_\mu+D_\mu \mathcal{A}_\nu-D_\nu \mathcal{A}_\mu\,,
\end{equation}
where $E_{\mu} = F_{\mu\nu}n^{\nu}$ and $\mathcal{A}_\mu$ is the spatial part of the vector field $A_{\mu}$. The real and imaginary components of $\Phi_2$ are then given by
\begin{equation}
\begin{aligned}
\Phi_2^{\rm R} =&-\frac{1}{2}\biggl[E_i^{\rm R} v^i+u^i v^j\left(D_i \mathcal{A}_j^{\rm R}-D_j \mathcal{A}_i^{\rm R}\right)\\ 
& +E_i^{\rm I} w^i+u^i w^j\left(D_i \mathcal{A}_j^{\rm I}-D_j \mathcal{A}_i^{\rm I}\right)\biggr]\,, \\ 
\Phi_2^{\rm I} =&\,\frac{1}{2}\biggl[E_i^{\rm R} w^i+u^i w^j\left(D_i \mathcal{A}_j^{\rm R}-D_j \mathcal{A}_i^{\rm R}\right)  \\
& -E_i^{\rm I} v^i-u^i v^j\left(D_i \mathcal{A}_j^{\rm I}-D_j \mathcal{A}_i^{\rm I}\right)\biggr]\,.
\end{aligned}
\end{equation}
Similar to the scalar case, we obtain the multipoles of $\Phi_{2}$ at a certain extraction radius $r_{\rm ex}$, by projecting $\Phi_2$ onto the $s_{\rm w}=-1$ spin-weighted spherical harmonics.
\begin{equation}\label{eq:NPextract}
\begin{aligned}
(\Phi^{\rm R}_{2})_{\ell m}(t) = \int 
\mathrm{d} \Omega&\biggl[\Phi_2^{\rm R}(t, \theta, \varphi)\> {}_{{\scalebox{0.65}{$-$}}1}\mkern-2mu Y_{\ell m}^{\rm R}(\theta, \varphi)\\ & +\Phi_2^{\rm I}(t, \theta, \varphi)\> {}_{{\scalebox{0.65}{$-$}}1}\mkern-2mu Y_{\ell m}^{\rm I}(\theta, \varphi)\biggr]\,, \\
(\Phi^{\rm I}_{2})_{\ell m}(t) = \int 
\mathrm{d} \Omega&\biggl[\Phi_2^{\rm I}(t, \theta, \varphi)\> {}_{{\scalebox{0.65}{$-$}}1}\mkern-2mu Y_{\ell m}^{\rm R}(\theta, \varphi)\\ & -\Phi_2^{\rm R}(t, \theta, \varphi)\> {}_{{\scalebox{0.65}{$-$}}1}\mkern-2mu Y_{\ell m}^{\rm I}(\theta, \varphi)\biggr]\,.
\end{aligned}
\end{equation}
Throughout this work, we will often show $|(\Phi_{2})_{\ell m}| = \sqrt{(\Phi_{2})^{*}_{\ell m}(\Phi_{2})_{\ell m}}$.
%%%%%%%%%%%%%%%%%%%%%%%%%%%%%%%%
\section{Plasma framework}\label{app:theplasma}
%%%%%%%%%%%%%%%%%%%%%%%%%%%%%%%%
In this appendix, we elaborate on the assumptions regarding our plasma model, listed in Section~\ref{sec:Plasma}. 
\begin{enumerate}[(i)]
    \item We drop non-linear terms in the axion-photon-plasma system. That is to say, as long as the EM field is small, it is sufficient to consider only the linear response of the medium. Consequently, the backreaction of the EM field onto the axion field is not included.
    \item We ignore the oscillations of the ions due to the EM field. Whenever the plasma is non-relativistic, this assumption is justified due to the larger inertia of the ions compared to the electrons and we can treat them as a neutralizing background. In the relativistic regime however, there is a critical threshold for which this approximation is no longer valid, defined as $\Gamma_{\rm e} \gg m_{\rm ion}/m_{\rm e}$~\cite{KrallTrivelpiece1973}, where $\Gamma_{\rm e}$ is the Lorentz factor of the electrons. 
    \item For simplicity, we consider as initial data a locally quasi-neutral plasma, i.e.,~$Zn_{\rm ion}\approx n_{\rm e}$, where $Z$ is the atomic number, such that we have a vanishing charge density. As plasma is globally neutral, this approximation is valid at large enough length scales for our setup. In particular, it holds in systems where typical length scales are much larger than the Debye length, which is the length at which free charges are efficiently screened in the plasma. 
    \item We assume a cold and collisionless plasma. In principle, our formalism can be extended straightforwardly to include both thermal and collisional effects by adding a few terms in the momentum equation~\cite{Cannizzaro:2021zbp}. In particular, thermal effects can be included by considering a non-negligible pressure for the fluid and an equation of state, while electron-ion collisions can be modeled using a term proportional to the relative velocity of the two fluid species.
    \item We neglect the evolution of the fluid's four velocity with respect to an {\it Eulerian observer} due to gravity. The 3+1 decomposition of the momentum equation~\eqref{eq:momeqnlinear}, shows how after linearization the evolution is dictated both by a gravitational term $a_i$ and an EM term $E_i$. In a Schwarzschild BH background, the only non-vanishing component of the gravitational term is $a^r=M/r^2$, which assumes the well known form of gravitational acceleration by a spherical object.\footnote{In fact, this term corresponds to the surface gravity when evaluated at the horizon.} Since we are interested in the effect of plasma in a localized region of spacetime, i.e.,~the axion cloud, which is situated around the Bohr radius, an easy estimate shows that the effect of gravity is sufficiently small in the timescales of interest. For example, let us consider a cloud with $\mu M = 0.1$, located around $r=200 M$. In this case, we have that $a^r=2.5 \times 10^{-5} M^{-1}$. Thus for an initially zero velocity fluid, this term only gives significant modifications on a timescale $t \sim 10^5 M$, which is much longer than the growth timescales of the EM field. Furthermore, note that the gravitational term is suppressed with respect to the EM term by a factor of $m_{\rm e}/q_{\rm e} \sim 10^{-22}$. Neglecting the evolution of the velocity due to gravity has two consequences for our system of evolution equations. 

    First of all, while the momentum equation for the electrons still has the EM term, we neglect this term for the ions according to assumption (ii). Therefore, the ionic momentum equation becomes trivial. In fact, it means that ions with initially zero velocity, do not evolve. Consequently, we can consider them as a stationary, neutralizing background without evolving the fluid.

    Second, by neglecting the gravitational term, the constraint of the electron momentum equation~\eqref{eq:constrainMomeqn} reduces at the linear level to $u^\nu \partial_\nu \Gamma=0$. Since $\Gamma = 1$ linearly due to its quadratic dependence on the velocity, the constraint equation is trivially satisfied. Hence, dropping the gravitational term allows for a great simplification of our evolution scheme.
\end{enumerate}
%%%%%%%%%%%%%%%%%%%%%%%%%%%%%%%%%%%%%%%%%%%%%%
\section{Formulation as a Cauchy problem}\label{sec:appendDecomp}
%%%%%%%%%%%%%%%%%%%%%%%%%%%%%%%%%%%%%%%%%%%%%%
In this appendix, we formalize our equations of motion~\eqref{eq:evoleqns} as an (initial value) Cauchy problem and we discuss the initial data.
%%%%%%%%%%%%%%%%%%%%%%%%%%%%%%%%%%%%%%%%%%%%%%
\subsection{3+1 Decomposition}\label{sec:DecompGeneral}
%%%%%%%%%%%%%%%%%%%%%%%%%%%%%%%%%%%%%%%%%%%%%%
The equations of motion of our axion-photon-plasma system are given by~\eqref{eq:evoleqns}. In this work, we will ignore the dynamics of gravity and solve the Klein-Gordon, Maxwell and plasma equations on a fixed spacetime background. In order to evolve the system in time, we use the standard 3+1 decomposition of the spacetime (see e.g.~\cite{Alcubierre}). The metric then takes the following generic form:
\begin{equation}
    \mathrm{d} s^2=-\alpha^2 \mathrm{d} t^2+\gamma_{i j}\left(\mathrm{d} x^i+\beta^i \mathrm{d} t\right)\left(\mathrm{d} x^j+\beta^j \mathrm{d} t\right)\,, 
\end{equation}
where $\alpha$ is the lapse function, $\beta^{i}$ is the shift vector and $\gamma_{ij}$ is the 3-metric on the spatial hypersurface. 
Furthermore, we introduce the scalar momentum as
\begin{equation}
    \Pi = -n^{\mu}\nabla_{\mu}\Psi\,, 
\end{equation}
where $n^{\mu}$ is the unit normal vector to the spatial hypersurface, which takes on the form $n^{\mu} = (1/\alpha, -\beta^{i}/\alpha)$. The vector field $A_{\mu}$ can be decomposed as
\begin{equation}
    A_{\mu} = \mathcal{A}_{\mu} + n_{\mu}A_{\varphi}\,,
\end{equation}
where
\begin{equation}  
\mathcal{A}_i=\gamma^j{ }_i A_j \quad \text{and} \quad A_\varphi=-n^\mu A_\mu\,.
\end{equation}
We also introduce the EM fields
\begin{equation} 
E^i=\gamma^i{ }_j F^{j \nu} n_\nu \quad \text{and} \quad B^i=\gamma^i{ }_j{ }^* \!F^{j \nu} n_\nu\,,
\end{equation}
which are defined with respect to an Eulerian observer.\footref{ft:Eulerian}
As for the plasma quantities, we decompose the fluids' four velocities as~\cite{Gourgoulhon:2007ue}
\begin{equation}\label{eq:fluidvel}
    u_{\rm e}^\mu=\Gamma_{\rm e}(n^\mu + \mathcal{U}^\mu) , \quad  u_{\rm ion}^\mu=\Gamma_{\rm ion}(n^\mu + \mathcal{V}^\mu)\,.
\end{equation}
where $\mathcal{U}^\mu, \mathcal{V}^\mu$ are again defined with respect to an Eulerian observer. From the normalization of the four velocities, the Lorentz factor is then:
\begin{equation}\label{eq:lorentzfactor}
    \Gamma_{\rm e}=- n_\mu u^\mu=\frac{1}{\sqrt{1- \mathcal{U}_\mu \mathcal{U}^\mu}}, \quad \Gamma_{\rm ion}=\frac{1}{\sqrt{1- \mathcal{V}_\mu \mathcal{V}^\mu}}\,.
\end{equation}
Note that even though we include the ion quantities in~\eqref{eq:fluidvel} and~\eqref{eq:lorentzfactor} for generality, we do not actually use them in this work as we ignore the oscillations of the ions [assumption (ii)]. Finally, we introduce the charge density as seen by an Eulerian observer as
\begin{equation}
\label{eq:rho}
    \rho=- n_\mu j^\mu\,.
\end{equation}
Since $j^\mu$ is the sum of the currents of the two fluids, we can express~\eqref{eq:rho} also as $\rho=\rho_{\rm e}+\rho_{\rm ion}$. 

Using the above definitions, we obtain the following evolution equations for the full axion-photon-plasma system (for the decomposition of the momentum equation, we refer to Appendix~\ref{sec:DecompMom}; for the EM part, see e.g.,~\cite{Alcubierre:2009ij}):
\begin{equation}
\begin{aligned}
\label{eq:3+1set}
\partial_t \Psi&= -\alpha \Pi+\mathcal{L}_\beta \Psi\,, \\
\partial_t \Pi&=\alpha\left(-D^2 \Psi+\mu^2 \Psi+K \Pi-2 k_{\mathrm{a}} E^i B_i\right) 
\\&-D^i \alpha D_i \Psi+\mathcal{L}_\beta \Pi\,, \\
\partial_t \mathcal{A}_i&=-\alpha\left(E_i+D_i A_\varphi\right)-A_\varphi D_i \alpha+\mathcal{L}_\beta \mathcal{A}_i\,, \\
\partial_t E^i&= \alpha K E^i-\alpha D_j\left(D^j \mathcal{A}^i-D^i \mathcal{A}^j\right)-\\ &\left(D^i \mathcal{A}^j-D^j \mathcal{A}^i\right) D_j \alpha + \alpha\left(D^{i}\mathcal{Z}-j^{i}\right) \\ &+2 k_{\mathrm{a}}\alpha\left(B^i \Pi+\epsilon^{i j k} E_k D_j \Psi\right)+ \mathcal{L}_\beta E^i\,,\\
\partial_t A_\varphi&=-\mathcal{A}^i D_i \alpha+\alpha\left(K A_\varphi-D_i \mathcal{A}^i-\mathcal{Z}\right)+\mathcal{L}_\beta A_\varphi\,,
&\\
\partial_{t}(\Gamma_{\rm e}\mathcal{U}_{i}) &= \alpha\left(\frac{q_{\rm e}}{m_{\rm e}}E_{i} + \epsilon_{ijk}\mathcal{U}^{j}B^{k} -\Gamma_{\rm e} a_{i} -\mathcal{U}^{j} D_{j} \left(\Gamma_{\rm e}\mathcal{U}_{i}\right)\right) \\
&+\mathcal{L}_\beta \Gamma_{\rm e}\mathcal{U}_{i}\,,  \\
\partial_{t}\rho_{\rm e} &= -D_{i}(\alpha j^{i}) +\alpha \rho_{\rm e} K + \mathcal{L}_{\beta}\rho_{\rm e}\,, \\
\partial_t \mathcal{Z}&= \alpha\left(D_iE^i+2 k_{\mathrm{a}} B^i D_i \Psi -\rho \right)-\kappa \alpha \mathcal{Z}+\mathcal{L}_\beta \mathcal{Z}\,,
\end{aligned}
\end{equation}
where we have introduced a constraint damping variable $\mathcal{Z}$ to stabilize the numerical time evolution. Furthermore, we define $D_i$ as the covariant derivative with respect to $\gamma_{i j}$, the extrinsic curvature as $K_{i j}=\frac{1}{2 \alpha}\left[-\partial_t \gamma_{i j}+D_i \beta_j+D_j \beta_i\right]$ and $K$ as its trace. Note that the absence of the evolution equations for the ions due to assumption (ii).

Finally, we get the following constraints:
\begin{equation}\label{eq:constrainteqn}
    \begin{aligned}
        D_{i}B^{i} &= 0\,, \\
        D_{i}E^{i} &= \rho -2k_{\rm a}B_{i}D^{i}\Psi\,,\\
        (n^{\mu}+\mathcal{U}^{\mu})\nabla_{\mu}\Gamma_{\rm e}&=\Gamma_{\rm e} \mathcal{U}^{i}\mathcal{U}^{j}K_{ij}-\Gamma_{\rm e} \mathcal{U}^{i}a_{i}-\frac{q_{\rm e}}{m_{\rm e}}E^{i}\mathcal{U}_{i}\,.
    \end{aligned}
\end{equation}
Upon ignoring the gravitational term in the momentum evolution equation [assumption (v)], this last constraint is trivially satisfied on the linear level.
%%%%%%%%%%%%%%%%%%%%%%%%%%%%%%%%%%%%%%
\subsection{Background metric}
%%%%%%%%%%%%%%%%%%%%%%%%%%%%%%%%%%%%%%
As discussed in Appendix~\ref{appA:boundstates}, we employ Kerr-Schild coordinates in our numerical setup to avoid the coordinate singularity at the horizon. These are related to Cartesian coordinates by
\begin{equation}
\begin{aligned}
&x=r \cos \varphi \sin \theta-a_{\scalebox{0.60}{$\mathrm{J}$}} \sin \varphi \sin \theta\,,\\
&y=r \sin \varphi \sin \theta+a_{\scalebox{0.60}{$\mathrm{J}$}} \cos \varphi \sin \theta\,,\\
&z=r \cos \theta\,.
\end{aligned}
\end{equation}
In these coordinates, the metric takes on the following form:
\begin{equation}
\mathrm{d}s^{2}=(\eta_{\mu\nu}+2Hl_{\mu}l_{\nu})\mathrm{d}x^{\mu}\mathrm{d}x^{\nu}\,,
\end{equation}
where
\begin{equation}
\begin{aligned}
H&=\frac{r^{3}M}{r^{4}+a_{\scalebox{0.60}{$\mathrm{J}$}}^{2}z^{2}}\,,\\
l_{\mu}&=\left(
1,\frac{rx+a_{\scalebox{0.60}{$\mathrm{J}$}}y}{r^{2}+a_{\scalebox{0.60}{$\mathrm{J}$}}^{2}},\frac{-a_{\scalebox{0.60}{$\mathrm{J}$}}x+ry}{r^{2}+a_{\scalebox{0.60}{$\mathrm{J}$}}^{2}},\frac{z}{r}
\right)\,,\\
r&=\biggl[\frac{1}{2}\biggl(x^{2}+y^{2}+z^{2}-a_{\scalebox{0.60}{$\mathrm{J}$}}^{2}\\&+\sqrt{(x^{2}+y^{2}+z^{2})^{2}+4a_{\scalebox{0.60}{$\mathrm{J}$}}^{2}z^{2}}\,\biggr)\biggr]^{1/2}\,,
\end{aligned}
\end{equation}
where we consider Schwarzschild, i.e.,~$a_{\scalebox{0.65}{$\mathrm{J}$}} = 0$. Furthermore, we define
\begin{equation}
\begin{gathered}
\alpha=\frac{1}{\sqrt{1+2 H}}\,, \\
\beta_i=2 H l_i\,, \\
\gamma_{i j}=\delta_{i j}+2 H l_i l_j, \\
K_{i j}=\frac{\partial_i\left(H l_j\right)+\partial_j\left(H l_i\right)+2 H (l^{*})^k \partial_k\left(H l_i l_j\right)}{\sqrt{1+2 H}}\,,
\end{gathered}
\end{equation}
which are the lapse function, shift vector, spatial metric, and the extrinsic curvature, respectively.
%%%%%%%%%%%%%%%%%%%%%%%%%%%%%%%%%%%%%%%%%%%%%%%%%%
\subsection{Evolution without plasma}\label{app:evolwithoutplasma}
%%%%%%%%%%%%%%%%%%%%%%%%%%%%%%%%%%%%%%%%%%%%%%%%%
Since the simulations {\it with} and {\it without} plasma have a slightly different structure, we separate these clearly in the following sections. First, we consider the full set of equations in the absence of plasma. These belong to the simulations $\mathcal{I}_{i}$ and $\mathcal{J}_{i}$ from Sections~\ref{sec:withoutSR} and~\ref{sec:withSR}. They are
\begin{equation}
\begin{aligned}
\partial_t \Psi&= -\alpha \Pi+\mathcal{L}_\beta \Psi\,, \\
\partial_t \Pi&=\alpha\left(-D^2 \Psi+\mu^2 \Psi+K \Pi-2 k_{\mathrm{a}} E^i B_i\right) 
\\&-D^i \alpha D_i \Psi+\mathcal{L}_\beta \Pi\,, \\
\partial_t \mathcal{A}_i&=-\alpha\left(E_i+D_i A_\varphi\right)-A_\varphi D_i \alpha+\mathcal{L}_\beta \mathcal{A}_i\,, \\
\partial_t E^i&= \alpha K E^i-\alpha D_j\left(D^j \mathcal{A}^i-D^i \mathcal{A}^j\right)-\\ &\left(D^i \mathcal{A}^j-D^j \mathcal{A}^i\right) D_j \alpha + \alpha D^i \mathcal{Z}\\ &+ 2 k_{\mathrm{a}}\alpha\left(B^i \Pi+\epsilon^{i j k} E_k D_j \Psi\right) + \mathcal{L}_\beta E^i\,,\\
\partial_t A_\varphi&=-\mathcal{A}^i D_i \alpha+\alpha\left(K A_\varphi-D_i \mathcal{A}^i-\mathcal{Z}\right)+\mathcal{L}_\beta A_\varphi\,,
&\\
\partial_t \mathcal{Z}&= \alpha\left(D_iE^i+2 k_{\mathrm{a}} B^i D_i \Psi \right)-\kappa \alpha \mathcal{Z}+\mathcal{L}_\beta \mathcal{Z}\,.
\end{aligned}
\end{equation}
Note that these are the same as considered in~\cite{Boskovic:2018lkj, Ikeda:2018nhb}.\\

%%%%%%%%%%%%%%%%%%%%%%%%%%%%%%%%%%%%%%%%%%%%%%%%%%
{\it Initial Data} $-$
%%%%%%%%%%%%%%%%%%%%%%%%%%%%%%%%%%%%%%%%%%%%%%%%%%
To construct the initial data for our simulations, we must solve the constraint equations~\eqref{eq:constrainteqn}. By doing so on the initial time-slice, the Bianchi identity will ensure they are satisfied throughout the evolution. As explained in Appendix~\ref{appA:boundstates}, we use Leaver's method to construct the scalar field bound state. For the electric field, we use initial data analogous to~\cite{Boskovic:2018lkj,Ikeda:2018nhb}. In particular, we choose a Gaussian profile defined in~\eqref{eq:InitialElectric}. 
%%%%%%%%%%%%%%%%%%%%%%%%%%%%%%%%%%%%%%%%%%%%
\subsection{Evolution with plasma}\label{app:evolPlasma}
%%%%%%%%%%%%%%%%%%%%%%%%%%%%%%%%%%%%%%%%%%%%
In the simulations {\it with} plasma, we linearize the axion-photon-plasma system due to the complexity of the problem, and we neglect ion perturbations. We express the perturbed quantities with a tilde, such that
\begin{equation}
\begin{aligned}
\mathcal{A}^{i}&=\mathcal{A}^{i}_{\rm b}+\varepsilon\tilde{\mathcal{A}}_{i}, \quad E^{i}=E^{i}_{\rm b}+\varepsilon \tilde{E}_{i}, \quad A_{\varphi}=A_{\mathrm{b},\varphi}+\varepsilon \tilde{A}_{\varphi}\,, \\ 
\mathcal{U}^{i}&=U^{i}_{\mathrm{b}}+\varepsilon \tilde{\mathcal{U}}_{i}, \quad \Gamma_{\rm e}=1\,, \quad \rho=\rho_{\mathrm{b, e}}+\rho_{\mathrm{b, ion}}+\varepsilon\tilde{\rho}_{\rm e}\,. 
\end{aligned}
\end{equation}
where we denote background quantities with a subscript $\mathrm{b}$ and where $\varepsilon$ is the arbitrarily small parameter in the perturbation scheme. For simplicity, we consider a quasi-neutral, field-free background plasma, i.e.,~$E^{i}_{\rm b}=A^{i}_{\mathrm{b}}=A_{\mathrm{b},\varphi}=0$, and $\rho_{\mathrm{b, e}}=-\rho_{\mathrm{b, ion}}$. The problem at hand naturally introduces two distinct reference frames: the Eulerian observer rest frame and the plasma rest frame. The relative velocity between the two is the background quantity $\mathcal{U}^i$. We consider a plasma co-moving with the Eulerian observer, such that the plasma is static in the spacetime foliation.
Since the background field of the electron charge density does not vanish, according to~\eqref{eq:3+1set} it should evolve as 
\begin{equation}
\label{eq:bkgdensity}  \partial_{t}\rho_{\mathrm{b}}=\alpha\rho_{\mathrm{b, e}}K+ \mathcal{L}_\beta \rho_{\mathrm{b, e}}\,.
\end{equation}
We are mainly interested in the evolution of this variable in a localized region of spacetime far away from the BH, i.e.,~the axion cloud, and thus the evolution of~\eqref{eq:bkgdensity} due to strong gravity terms is extremely slow compared to the linear system. Therefore, we neglect its evolution similarly to the gravitational influence on the evolution of the background velocity [assumption (v)].

Before proceeding, there is one other subtlety. The plasma response to the perturbing EM field is proportional to the electron charge-to-mass ratio, which is extremely large $q_{\rm e}/m_{\rm e}\approx 10^{22}$.
Nevertheless, as we are linearizing the system, and therefore neglecting the backreaction of the EM field onto the axion field, the amplitude of the former is arbitrary in our scheme. If we were to consider the full problem including backreaction instead, the amplitude of the axion field would clearly introduce a scale. Due to this freedom, we rescale the EM variables as
\begin{equation}
\begin{aligned}
\bar{E}^{i}&=\frac{q_{\rm e}}{m_{\rm e}}\tilde{E}^{i}\,,\\
    \bar{A}_{i}&=\frac{q_{\rm e}}{m_{\rm e}}\tilde{A}_{i}\,,\\
    \bar{\mathcal{Z}}&=\frac{q_{\rm e}}{m_{\rm e}}\mathcal{Z}\,.
\end{aligned}
\end{equation}
Then, we can write down the full set of equations {\it including} the plasma as
\begin{equation}\label{eq:evoleqnplasma}
\begin{aligned}
\partial_t \Psi&= -\alpha \Pi+\mathcal{L}_\beta \Psi\,, \\
\partial_t \Pi&=\alpha\left(-D^2 \Psi+\mu^2 \Psi+K \Pi \right) -D^i \alpha D_i \Psi+\mathcal{L}_\beta \Pi\,, \\
\partial_t \bar{\mathcal{A}}_i&=-\alpha\left(\bar{E}_i+D_i \bar{A}_\varphi\right)-\bar{A}_\varphi D_i \alpha+\mathcal{L}_\beta \bar{\mathcal{A}}_i\,, \\
\partial_t \bar{E}^i&= \alpha K \bar{E}^i-\alpha D_j\left(D^j \bar{\mathcal{A}}^i-D^i \bar{\mathcal{A}}^j\right)-\\ &\left(D^i \bar{\mathcal{A}}^j-D^j \bar{\mathcal{A}}^i\right) D_j \alpha + \alpha\left(D^i \bar{\mathcal{Z}} - \omega_{\rm p}^{2}\tilde{\mathcal{U}}^{i}\right)\\ &+2 k_{\mathrm{a}}\alpha\left(\bar{B}^i \Pi+\epsilon^{i j k} \bar{E}_k D_j \Psi\right) + \mathcal{L}_\beta \bar{E}^i\,,\\
\partial_t \bar{A}_{\varphi}&=-\bar{\mathcal{A}}^{i}D_{i}\alpha+\alpha(K\bar{\mathcal{A}}_{\varphi}-D_{i}\bar{\mathcal{A}}^i-\bar{\mathcal{Z}})+\mathcal{L}_{\beta}\bar{A}_{\varphi}\,,\\
\partial_t\tilde{U}_i&=\alpha\bar{E}_i+\mathcal{L}_{\beta}\tilde{U}_{i}\,,\\
\partial_t\tilde{\omega}_{\rm p}^{2}&=-D_{i}(\alpha\omega_{\rm p}^{2}\tilde{U}^{i})+\alpha\tilde{\omega}^{2}_{\rm p}K+\mathcal{L}_{\beta}\tilde{\omega}^{2}_{\rm p}\,, \\
\partial_t \bar{\mathcal{Z}}&= \alpha\left(D_i \bar{E}^i+2 k_{\mathrm{a}} \bar{B}^i D_i \Psi -\tilde{\omega}^{2}_{\rm p} \right)-\kappa \alpha \bar{\mathcal{Z}}+\mathcal{L}_\beta \bar{\mathcal{Z}}\,, 
\end{aligned}
\end{equation}
where $\omega_{\rm p}^{2}$ is the plasma frequency, and $\tilde{\omega}_{\rm p}^{2}$ its perturbation:
\begin{equation}
    \begin{aligned}
        \omega_{\rm p}^{2}&=\frac{q_{\rm e}}{m_{\rm e}}\rho_{\mathrm{b,e}}\,,\\
        \tilde{\omega}_{\rm p}^{2}&=\frac{q_{\rm e}}{m_{\rm e}}\tilde{\rho}_{\mathrm{e}}\,.
    \end{aligned}
\end{equation}
Note that due to the rescaling, there is no charge-to-mass ratio of the electrons and the field equations are written only in terms of the plasma frequency, which is $\mathcal{O}(1/M)$. 

As we detail in the following subsection, including a linearized fluid model in the equations of motion, causes the system~\eqref{eq:evoleqnplasma} to become ill-posed upon using a damping variable.\footnote{The ill-posedness originates from the linearization of the fluid equation, whereas the full non-linear system of equations is strongly hyperbolic and thus well-posed~\cite{MUNZ2000484,Abgrall:2014rfb}} However, this damping variable is essential in constraining Gauss' law and without it, the simulations diverge for large EM values. As a resolution, we slightly adjust our equations by not including the perturbed plasma frequency, $\tilde{\omega}_{\rm p}$, in the evolution equation of the damping variable $\mathcal{Z}$, i.e.,
\begin{equation}
    \partial_t \bar{\mathcal{Z}}= \alpha\left(D_i \bar{E}^i+2 k_{\mathrm{a}} \bar{B}^i D_i \Psi \right)-\kappa \alpha \bar{\mathcal{Z}}+\mathcal{L}_\beta \bar{\mathcal{Z}}\,.
\end{equation}
This is a minimal change as the perturbed plasma frequency does not enter any other evolution equation of the system in the linearized regime, yet it does restore the well-posedness of our setup. We justify this approach in two ways;~(i) we evolve the system {\it with} and {\it without} the damping variable for large plasma frequencies (where the EM values remain small) and we find excellent agreement between the two, and (ii) for small plasma frequencies, where the EM field is allowed to grow, the effects of the plasma are negligible, and therefore ignoring $\tilde{\omega}_{\rm p}$ leads to a subleading error compared to the EM values. 

%%%%%%%%%%%%%%%%%%%%%%%%%%%%%%%%%%%%%%%%%%%%%%%%%%
{\it Initial Data} $-$
%%%%%%%%%%%%%%%%%%%%%%%%%%%%%%%%%%%%%%%%%%%%%%%%%%
For the plasma part, we assume quasi-neutrality [cf.~assumption (iii)], i.e., $\rho = -n_{\mu}(e n_{\rm e} u_{\rm e}^{\mu} - Z e n_{\rm ion} u_{\rm ion}^{\mu}) = e n_{\rm e} - Z e n_{\rm ion} = 0$. As shown in~\eqref{eq:constrainteqn}, the constraint equation for the plasma is trivially satisfied on the linear level, and thus the initial data listed in the previous section solves all of our constraints. As for the electronic density, in principle, depending on the specific environment we are interested in, we can assume different spatial profiles~\cite{Dima:2020rzg, Wang:2022hra, Abramowicz2013}, which correspond to a space-dependent effective mass for the photon. However, the length scale of interest to us, i.e.,~the size of the axion cloud, is typically much shorter than length scale on which the effective mass varies. Hence, for simplicity, we assume a constant density plasma.
%%%%%%%%%%%%%%%%%%%%%%%%%%%%%%%%%%%%%%%%%%%%%%%%%%%%%%%%%%%%%%%%
\subsection{Hyperbolicity of fluid model}\label{sec:hyperbolicity}
%%%%%%%%%%%%%%%%%%%%%%%%%%%%%%%%%%%%%%%%%%%%%%%%%%%%%%%%%%%%%%%%
The evolution equations~\eqref{eq:evoleqnplasma} are not strongly hyperbolic and therefore do not form a well-posed system. Consequently, the existence of a unique solution that depends continuously on the initial data is not guaranteed and any numerical approach is bound to fail. In this appendix, we proof that our system is not strongly hyperbolic.

Based on~\cite{Hilditch:2013sba}, we introduce an arbitrary unit vector $s^{i}$ and consider the principal part of the system, i.e.,~we consider only the highest derivative terms from~\eqref{eq:evoleqnplasma}:
\begin{equation}
\begin{aligned}
    \partial_{t}\lbrack\partial_{s}^{2}\psi\rbrack&\sim -\alpha\partial_{s}\left(\partial_{s}\bar{A}_{\varphi}\right)+\beta^{s}\partial_{s}\lbrack\partial_{t}^{2}\psi\rbrack\,,\\
    \partial_{t}(\partial_{s}\bar{\mathcal{A}}_{A})&\sim -\alpha \partial_{s}\bar{E}_{A}+\beta^{s}\partial_{s}(\partial_{s}\bar{\mathcal{A}}_{A})\,,\\
    \partial_{t}(\partial_{s}\bar{A}_{\varphi})&\sim \beta^{s}(\partial_{s}\bar{A}_{\varphi})-\alpha\partial_{s}\lbrack\partial_{s}^{2}\psi\rbrack\,,\\
    \partial_{t}\bar{E}_{A}&\sim -\alpha\partial_{s}(\partial_{s}\bar{\mathcal{A}}_{A})+\beta^{s}\partial_{s}\bar{E}_{A}\,,\\
    \partial_{t}\bar{E}_{s}&\sim \alpha\partial_{s}\bar{\mathcal{Z}}+\beta^{s}\partial_{s}\bar{E}_{s}\,,\\
    \partial_{t}\bar{\mathcal{Z}}&\sim \alpha\partial_{s}\bar{E}^{s}+\beta^{s}\partial_{s}\bar{\mathcal{Z}}\,,\\
    \partial_{t}\tilde{U}_{s}&\sim\beta^{s}\partial_{s}\tilde{U}_{s}\,,\\
    \partial_{t}\tilde{U}_{A}&\sim\beta^{s}\partial_{s}\tilde{U}_{A}\,,\\
    \partial_{t}\tilde{\omega}^{2}_{\rm p}&\sim -\alpha\omega_{\rm p}^{2}\partial_{s}\tilde{U}_{s}+\beta^{s}\partial_{s}\tilde{\omega}_{\rm p}^{2}\,,
\end{aligned}
\label{highest derivative terms}
\end{equation}
where the index $A$ denotes the component projected into the surface orthogonal to $s^{i}$, and $\lbrack\partial_{s}^{2}\psi\rbrack$ can be written as
\begin{equation}
\lbrack\partial_{s}^{2}\psi\rbrack=\partial_{s}\bar{\mathcal{A}}_{s}+\bar{\mathcal{Z}}\,.
\end{equation}
Defining the principal symbol of 
$(\lbrack\partial_{s}^{2}\psi\rbrack,\bar{A}_{\varphi})$
$(\bar{\mathcal{Z}},E_{s})$, 
$(\partial_{s}\bar{\mathcal{A}}_{A},\bar{E}_{A})$,
and $(\tilde{U}_{s},\tilde{\omega}_{\rm p}^{2})$ as $P_{\mathcal{G}}$, $P_{\mathcal{C}}$, $P_{\mathcal{P}}$, and $P_{\mathcal{F}}$, respectively, we get:
\begin{equation}
\begin{aligned}
P_{\mathcal{G}}&=\begin{pmatrix}
\beta^{s}&-\alpha\\
-\alpha&\beta^{s}
\end{pmatrix}\,, \quad
&&P_{\mathcal{C}}=\begin{pmatrix}
\beta^{s}&\alpha\\
\alpha&\beta^{s}
\end{pmatrix}\,, \\
P_{\mathcal{P}}&=\begin{pmatrix}
\beta^{s}&-\alpha\\
-\alpha&\beta^{s}
\end{pmatrix}\,, \quad
&&P_{\mathcal{F}}=\begin{pmatrix}
\beta^{s}&0\\
-\alpha\omega_{\rm p}^{2}&\beta^{s}
\end{pmatrix}\,.
\end{aligned}
\end{equation}
We see that the eigenvalue for $P_{\mathcal{F}}$ is degenerate with $\beta^{s}$ {\it and} the eigenvector is $(0,b)^{T}$, where $b$ is an arbitrary value. Therefore, the principal symbol does not have a complete set of eigenvectors, and the system is not strongly hyperbolic.
If we ignore $\tilde{\omega}^{2}_{\rm p}$ in $\mathcal{Z}$, it becomes decoupled from Maxwell's equations, and the only relevant variable for the fluid part that remains, is the linearized four velocity $\tilde{U}_{i}$. As is shown in~\eqref{highest derivative terms}, the principal part for $\tilde{U}_{i}$ is just canonical the advection term.
%%%%%%%%%%%%%%%%%%%%%%%%%%%%%%%%%%%%%%%%%%%%%%%%%%%%%%%%%%%%%%%%
\subsection{Momentum equation}\label{sec:DecompMom}
%%%%%%%%%%%%%%%%%%%%%%%%%%%%%%%%%%%%%%%%%%%%%%%%%%%%%%%%%%%%%%%%
In order to include a plasma in our numerical setup, we need to apply the 3+1 decomposition to the momentum equation~\eqref{eq:evoleqns}. Even though this has been done before, e.g.~in~\cite{Thorne1982}, it is not part of standard literature. Therefore, we do the decomposition explicitly here. Our starting point is the momentum equation for the electrons, given by (we drop the subscript ``$\mathrm{e}$'' here, since we only consider the momentum equation for the electrons)
\begin{equation}\label{eq:momeqn}
    u^{\nu}\nabla_{\nu}u^{\mu} = \frac{q}{m}F^{\mu\nu}u_{\nu}\,.
\end{equation}
The four-velocity can be written as $u^{\mu} = \Gamma(n^{\mu} + \mathcal{U}^{\mu})$, where $\mathcal{U}_{i}$ is tangent to $\Sigma_{t}$, the spatial hypersurface, such that $n^{\mu}\mathcal{U}_{\mu} = 0$ and $u^{\mu}u_{\mu} =-1$. Although we are interested in linear effects and therefore $\Gamma = 1$, we will derive the 3+1 equations in full generality and thus including this factor. 

If we project~\eqref{eq:momeqn} using the projector operator $h^a_{\ b} = \delta^a_{\ b} + n^{a}n_{b}$ we obtain the evolution equation, if we project it onto $n^{a}$ we get the constraint equation. Let us start with the former.

%%%%%%%%%%%%%%%%%%%%%%%%%%%%%%%%%%%%%
{\it Evolution equation} $-$
%%%%%%%%%%%%%%%%%%%%%%%%%%%%%%%%
We start with the left-hand side (LHS) of~\eqref{eq:momeqn}:\footnote{We avoid writing the overall $\Gamma$ coming from $u^{\mu}$ on both sides of~\eqref{eq:momeqn}.}
\begin{equation}\label{eq:lhsmomeq}
\begin{aligned}             
    & h^\rho_{\  \mu}\left(n^{\nu}+\mathcal{U}^{\nu}\right)\nabla_{\nu}\left[\Gamma\left(n^{\mu}+\mathcal{U}^{\mu}\right)\right] =\\
    &\Gamma a^{\rho} +  \underbrace{n^{\nu}\nabla_{\nu}\left(\Gamma \mathcal{U}^{\rho}\right)  - \Gamma \mathcal{U}^{\mu}a_{\mu}n^{\rho}}_\text{I} + \underbrace{\Gamma\mathcal{U}^{\nu}\nabla_{\nu}n^{\rho}}_\text{II} +  \\& \underbrace{h^{\rho}_{\mu}\mathcal{U}^{\nu}\nabla_{\nu}\left(\Gamma \mathcal{U}^{\mu}\right)}_\text{III}\,,
\end{aligned}
\end{equation}
where $a_{\mu} = n^{\nu}\nabla_{\nu}n_{\mu}$ is the acceleration of the Eulerian observer. Since this a projection onto the hypersurface and thus orthogonal to $n^{\mu}$, we focus on $\rho = i$, i.e.,~the spatial part. To write~\eqref{eq:lhsmomeq} in a more convenient form, we work out parts I, II and III separately in~\eqref{eq:partI},~\eqref{eq:partII},~\eqref{eq:partIII}. First, we define the useful relations~\cite{Thorne1982}:
\begin{equation}
\begin{aligned}
    D_{\tau}\mathcal{U}^{i} &= n^{\mu}\nabla_{\mu}\mathcal{U}^{i} - n^{i}a_{\mu}\mathcal{U}^{\mu}\,,\\
     \mathcal{L}_{t}\mathcal{U}^{i} &= \alpha (D_{\tau}\mathcal{U}^{i} + K^{i}_{j}\mathcal{U}^{j}) + \mathcal{L}_{\beta}\mathcal{U}^{i}\,,
\end{aligned}
\end{equation}
where $\mathcal{L}_t=\mathcal{L}_{\alpha n \scalebox{0.55}{$\mathrm{+}$} \beta}$.
For part I, we then have
\begin{equation}\label{eq:partI}
\begin{aligned}
    n^{\mu} \nabla_{\mu}\left(\Gamma\mathcal{U}^{i}\right)- \Gamma n^{i}a_{\mu}\mathcal{U}^{\mu} &= \frac{1}{\alpha}\Big[\mathcal{L}_t\left(\Gamma\mathcal{U}^{i}\right) \\ & -\mathcal{L}_{\beta}\left(\Gamma\mathcal{U}^{i}\right)\Big] - \Gamma K^{i}_{j}\mathcal{U}^{j}\,,
\end{aligned}
\end{equation}
for term II, we find
\begin{equation}\label{eq:partII}
\begin{aligned}
   \Gamma\mathcal{U}^{\nu}\nabla_{\nu}n^{i} = 
   \Gamma\mathcal{U}^{\nu}\left(-a^{i}n_{\nu} - K^{i}_{\nu}\right) = -\Gamma K^{i}_{j}\mathcal{U}^{j}\,,
\end{aligned}
\end{equation}
and finally, for term III, we have
\begin{equation}\label{eq:partIII}       h^{i}_{\mu}\mathcal{U}^{\nu}\nabla_{\nu}\left(\Gamma\mathcal{U}^{\mu}\right) = h^{i}_{\mu}h^{\nu}_{ \alpha}\mathcal{U}^{\alpha}\nabla_{\nu}\left(\Gamma\mathcal{U}^{\mu}\right) =\mathcal{U}^{\alpha}D_{\alpha}\left(\Gamma\mathcal{U}^{i}\right)\,.
\end{equation}
Combining these, we can write the LHS as
\begin{equation}\label{eq:evolLHS}
\begin{aligned}
\Gamma a^{i}&+\frac{1}{\alpha}\left[\mathcal{L}_t\left(\Gamma \mathcal{U}^{i}\right) -\mathcal{L}_{\beta}\left(\Gamma\mathcal{U}^{i}\right)\right]\\ &- \Gamma K^{i}_{j}\mathcal{U}^{j} - \Gamma K^{i}_{j}\mathcal{U}^{j}+ \mathcal{U}^{\alpha}D_{\alpha}\left(\Gamma\mathcal{U}^{i}\right)\,.
\end{aligned}
\end{equation}
For the right-hand side (RHS) of the momentum equation~\eqref{eq:momeqn}, we first write out the standard form of the decomposition of the Maxwell tensor (see, e.g.,~\cite{Alcubierre:2009ij}) and then project it onto the spatial hypersurface using the projector operator: 
\begin{equation}\label{eq:evolRHS}
\begin{aligned}
    &h^\rho_{\ \mu}\frac{q}{m}F^{\mu\nu}(n_{\nu}+\mathcal{U}_{\nu}) \\
    &=h^\rho_{\ \mu}\frac{q}{m}(\epsilon^{\alpha\mu\nu\sigma}n_{\alpha}B_{\sigma}\mathcal{U}_{\nu} + n^{\mu}E^{\nu}\mathcal{U}_{\nu} + E^{\mu}) \\ &=
    \frac{q}{m}\left( ^{(3)}\epsilon^{\rho\nu\sigma}\mathcal{U}_{\nu}B_{\sigma} + E^{\rho} \right)\,,
\end{aligned}
\end{equation}
where we used $E^{\nu}n_{\nu} = 0$, $\epsilon^{\alpha\mu\nu\sigma}n_{\alpha}n_{\nu} = 0$, and $\epsilon^{\alpha\mu\nu\sigma}n_{\alpha} =^{(3)}\epsilon^{\mu\nu\sigma}$.

We can then piece together~\eqref{eq:evolLHS} and~\eqref{eq:evolRHS} to obtain
\begin{equation}
\begin{aligned}
    &\mathcal{L}_t\left(\Gamma \mathcal{U}^{i}\right) -\mathcal{L}_\beta \left(\Gamma\mathcal{U}^{i}\right) + \alpha \mathcal{U}^{j} D_{j} \left(\Gamma\mathcal{U}^{i}\right)\\&= \alpha\left(\frac{q}{m}\left(E^{i} + \epsilon^{ijk}\mathcal{U}_{j}B_{k}\right) - \Gamma a^{i}+2 \Gamma K^{ij}\mathcal{U}_j\right)\,.
\end{aligned}
\end{equation}
Finally, we apply the 3-metric tensor $\gamma_{ij}$ to the above equation to lower the index. To do so, we use the following identities~\cite{Thorne1982}:
\begin{equation}
\begin{aligned}
   \gamma_{ij} \mathcal{L}_t\left(\Gamma \mathcal{U}^{j}\right) &= \mathcal{L}_t \left(\Gamma\mathcal{U}_i \right)- \Gamma \mathcal{U}^j \mathcal{L}_t \gamma_{ij}\,, \\
    \Gamma \mathcal{U}^j \mathcal{L}_t \gamma_{ij} &= - 2 \alpha \Gamma \mathcal{U}^j  K_{ij}+ \Gamma \mathcal{U}^j \mathcal{L}_\beta \gamma_{ij}\,,   
\end{aligned}
\end{equation} 
from which we obtain:
\begin{equation}
\begin{aligned}
\label{eq:Uupieq}
    &\partial_t\left(\Gamma \mathcal{U}_{i}\right) -\mathcal{L}_\beta \left(\Gamma\mathcal{U}_{i}\right) + \alpha \mathcal{U}^{j} D_{j} \left(\Gamma\mathcal{U}_{i}\right)\\&= \alpha\left(\frac{q}{m}\left(E_{i} + \epsilon_{ijk}\mathcal{U}^{j}B^{k}\right) - \Gamma a_{i}\right)\,.
\end{aligned}
\end{equation}
As explained in Appendix~\ref{app:theplasma}, we simplify the plasma to make it more suitable to our numerical setup. By linearizing this equation, which also implies $\Gamma\sim 1$, we are left with
\begin{equation}\label{eq:momeqnlinear}
    \partial_{t}\mathcal{U}_{i} = \alpha\left(\frac{q}{m}E_{i} - a_{i} \right) +\mathcal{L}_\beta \mathcal{U}_{i}\,.
\end{equation}
Finally, we also neglect gravity, which brings us to our final equation:
\begin{equation}
    \partial_{t}\mathcal{U}_{i} = \alpha\frac{q}{m}E_{i}+\mathcal{L}_\beta \mathcal{U}_{i}\,.
\end{equation}

%%%%%%%%%%%%%%%%%%%%%%%%%%%%%%%%%%%%%%%%%%%%%%
{\it Constraint equation} $-$
%%%%%%%%%%%%%%%%%%%%%%%%%%%%%%%%%%%%%%%%%%%%%%
By projecting the momentum equation~\eqref{eq:momeqn} onto the timelike unit vector, $n^{\mu}$, we obtain the constraint equation. Again, we first show the LHS: 
\begin{equation}
    \begin{aligned}
        & n_\mu\left(n^{\nu}+\mathcal{U}^{\nu}\right)\nabla_{\nu}\left[\Gamma\left(n^{\mu}+\mathcal{U}^{\mu}\right)\right] = \\
        & - \Gamma a_{\mu}\mathcal{U}^{\mu} - \Gamma \mathcal{U}^{\nu}\mathcal{U}^{\mu}\nabla_{\nu}n_{\mu}-n^\nu \partial_\nu \Gamma- \mathcal{U}^\nu \partial_\nu \Gamma\,,
    \end{aligned}
\end{equation}
where we used that $n_{\mu}\nabla_{\nu}n^{\mu} = 0$. For the RHS, we have
\begin{equation}
    \begin{aligned}
        &\frac{q}{m}n_\mu(\epsilon^{\alpha\mu\nu\sigma}n_{\alpha}B_{\sigma}\mathcal{U}_{\nu} + n^{\mu}E^{\nu}\mathcal{U}_{\nu} + E^{\mu}) = \\
        &-\frac{q}{m}E^{\nu}\mathcal{U}_{\nu}\,.
    \end{aligned}
\end{equation}
Thus we end up with the following constraint equation:
\begin{equation}\label{eq:constrainMomeqn}
    \begin{aligned}
        n^\nu \partial_\nu \Gamma+ \mathcal{U}^\nu \partial_\nu \Gamma +\Gamma a_{\mu}\mathcal{U}^{\mu} + \Gamma \mathcal{U}^{\nu}\mathcal{U}^{\mu}K_{\mu \nu} = \frac{q}{m}E^{\nu}\mathcal{U}_{\nu}\,.
    \end{aligned}
\end{equation}
The fourth and fifth term are second order and thus drop out in our linearized setup [assumption~(i)]. Furthermore, we can neglect the third term since we ignore the gravity term [assumption~(v)]. As $\Gamma$ depends quadratically on the four-velocity, it must be $1$ in the linear theory, and therefore the constraint is trivially satisfied.
%%%%%%%%%%%%%%%%%%%%%%%%%%%%%%%%%%%%%%%%%%%%%%
\section{Numerical convergence}\label{app:convergence}
\begin{figure}
    \centering
    \includegraphics[width = 0.5\textwidth]{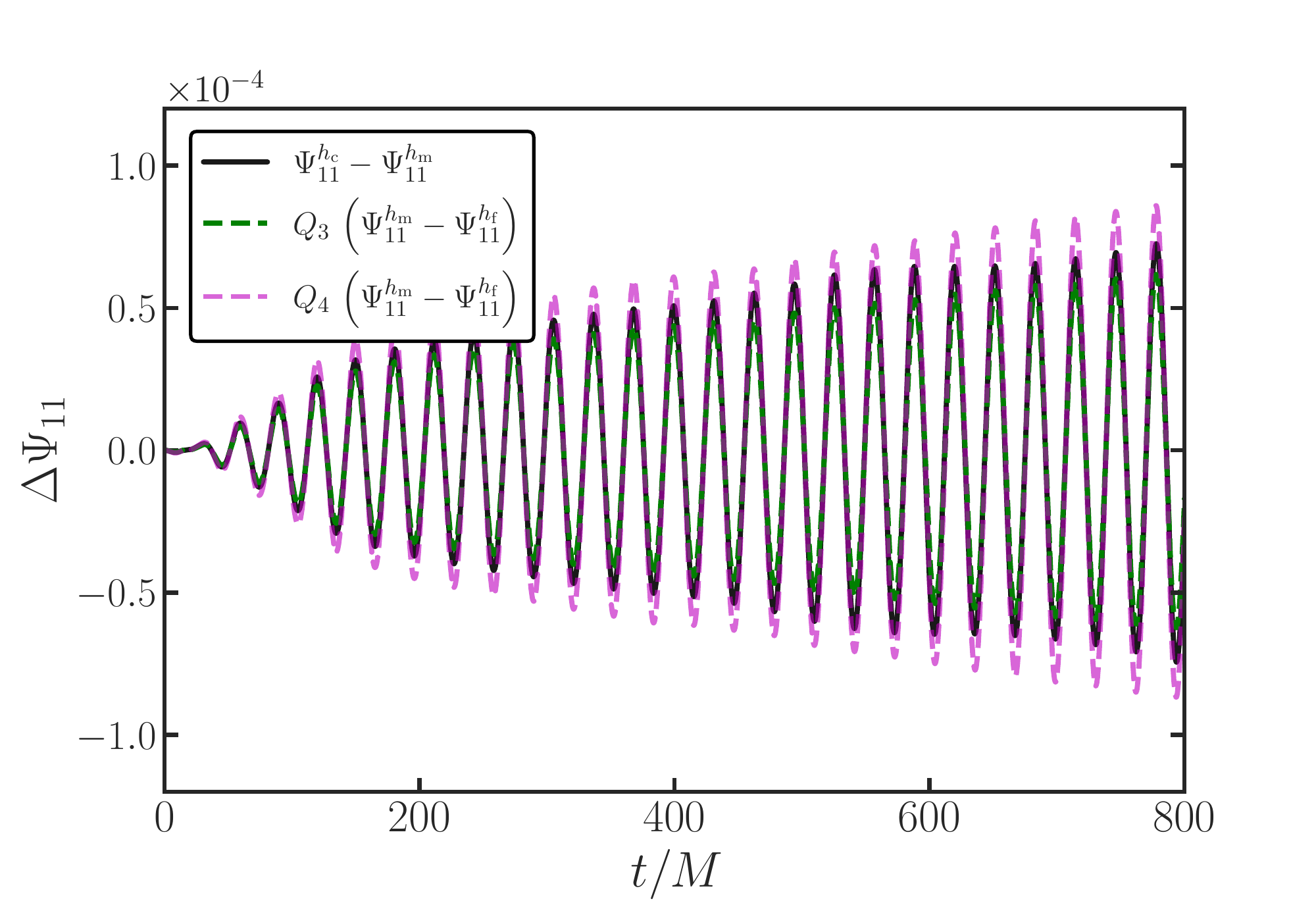}
    \caption{Convergence analysis of the $\ell = m = 1$ multipole of $\Psi$, extracted at $r_{\rm ex} = 20M$ for $\mu M = 0.2$ and $C = 10^{-3}$. The green line shows the expected result for third-order convergence ($Q_{3} = 5.64$), while the purple line is the expected result for fourth-order convergence ($Q_{4} = 7.85$).}
    \label{fig:convergencePhiC}
\end{figure}

%%%%%%%%%%%%%%%%%%%%%%%%%%%%%%%%%%%%%%%%%%%%%%
In our numerical framework, we employ the method of lines, where spatial derivatives are approximated by a fourth-order accurate finite-difference scheme and we integrate using a fourth-order Runge-Kutta method. Furthermore, Kreiss-Oliger dissipation is applied to evolved quantities in order to suppress high-frequency modes that come from the boundaries between adjacent refinement regions. The numerical simulations are performed using the open source \texttt{Einstein Toolkit}~\cite{Loffler:2011ay, Zilhao:2013hia}. For the evolution of the scalar and vector field, we extent the \texttt{ScalarEvolve}~\cite{Bernard:2019nkv, Ikeda:2020xvt, Cunha:2017wao} and \texttt{ProcaEvolve} thorns~\cite{Zilhao:2015tya, Sanchis-Gual:2022zsr}, respectively. We use \texttt{Multipatch} to interpolate between different grids in our numerical domain~\cite{Pollney:2009yz, Reisswig:2012nc}. In particular, to connect the central Cartesian grid with the spherical wave zone. Additionally, \texttt{Carpet} communicates between refinement levels with second-order and fifth-order accuracy in time and space, respectively. The Courant number in all our simulations is $0.2$, such that the Courant–Friedrichs–Lewy condition is satisfied. To check whether our numerical results respect the required convergence, we evolve the same configuration with a coarse ($h_{\rm c}$), medium ($h_{\rm m}$) and fine ($h_{\rm f}$) resolution. The convergence factor can then be calculated according to
\begin{equation}
Q_n=\frac{f_{h_{\rm c}}-f_{h_{\rm m}}}{f_{h_{\rm m}}-f_{h_{\rm f}}}=\frac{h_{\rm c}^n-h_{\rm m}^n}{h_{\rm m}^n-h_{\rm f}^n}\,,
\end{equation}
where $n$ is the expected convergence order. In our case, we take as the coarsest level $h_{\rm c}=$ $1.8 M$, then $h_{\rm m}=1.2 M$ and $h_{\rm f}=1.0 M$. As can be seen in Fig.~\ref{fig:convergencePhiC}, we obtain a convergence order between $3$ and $4$. We have performed similar tests for the other simulations in this work ({\it with} or {\it without} $C$ and {\it with} or {\it without} the plasma) and we find similar conclusions.
%%%%%%%%%%%%%%%%%%%%%%%%%%%%%%%%%%%%%%%%%%%%%%
\section{Higher multipoles}\label{appB:higherorder}
\begin{figure}
    \includegraphics[width = 0.5\textwidth]{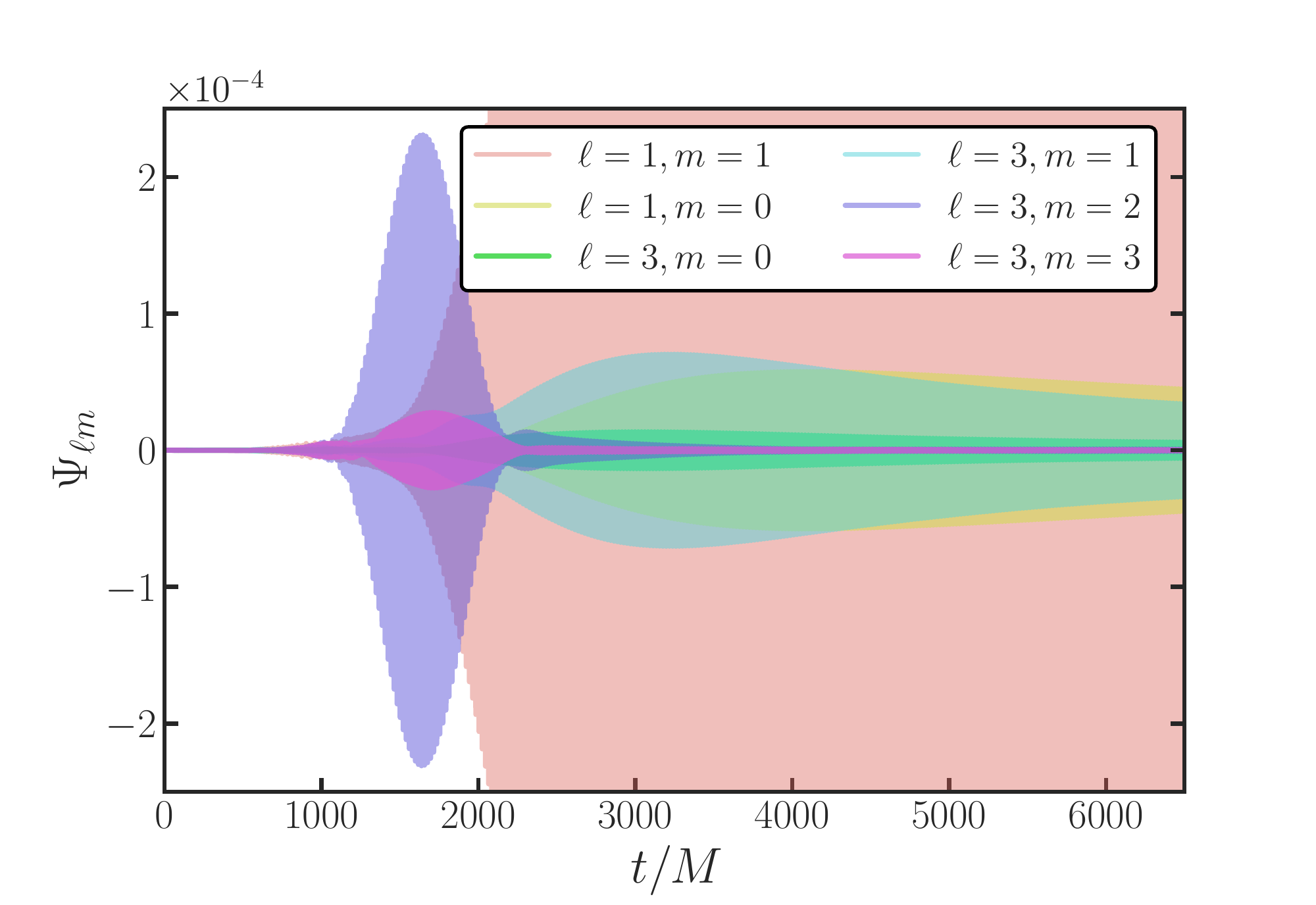}		\includegraphics[width = 0.5\textwidth]{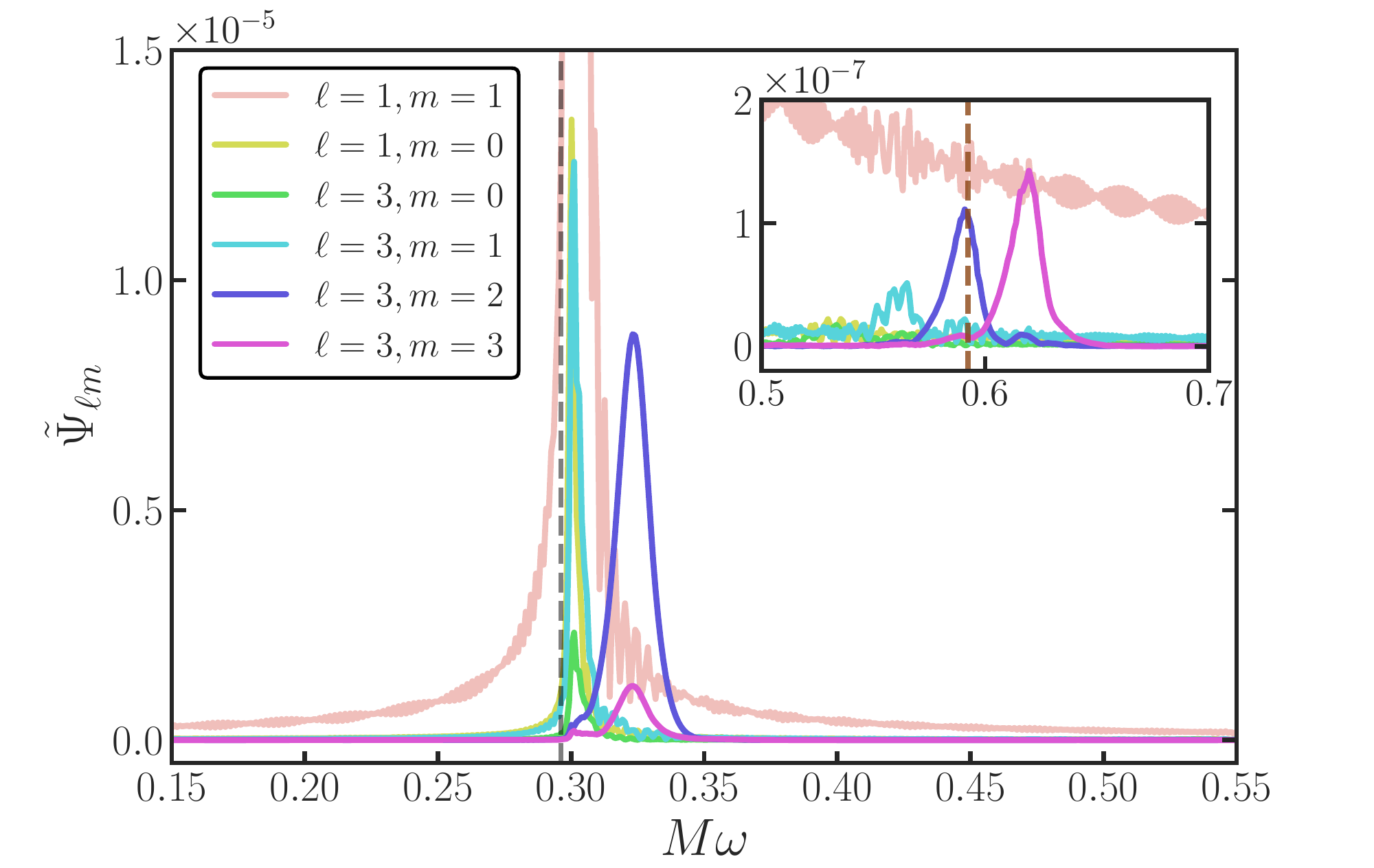}
    \caption{{\bf Top Panel:} Time evolution of various multipole
    modes of the scalar field in the supercritical case (simulation $\mathcal{I}_{3}$). The field is extracted at $r_{\rm ex} = 400M$ and $\mu M = 0.3$. Interestingly, even $\ell$ modes are not excited, while odd $\ell$ modes are, which we explain in Appendix~\ref{sec:selectionrules}.
    {\bf Bottom Panel:} The Fourier transform of the multipole modes shown in the top panel, where the gray dashed line denotes the frequency of the fundamental mode ($\omega_0$). The peaks around $\omega = 2\omega_0$ (brown dashed line) seen in the inset originate from interactions between ``up-scattered'' photons ($\omega = 3\omega_0/2$) with ``normal'' photons ($\omega = \omega_0/2$).}   
    \label{fig:scalarfieldmodes}
\end{figure}

\begin{figure}
    \includegraphics[width = 0.5\textwidth]{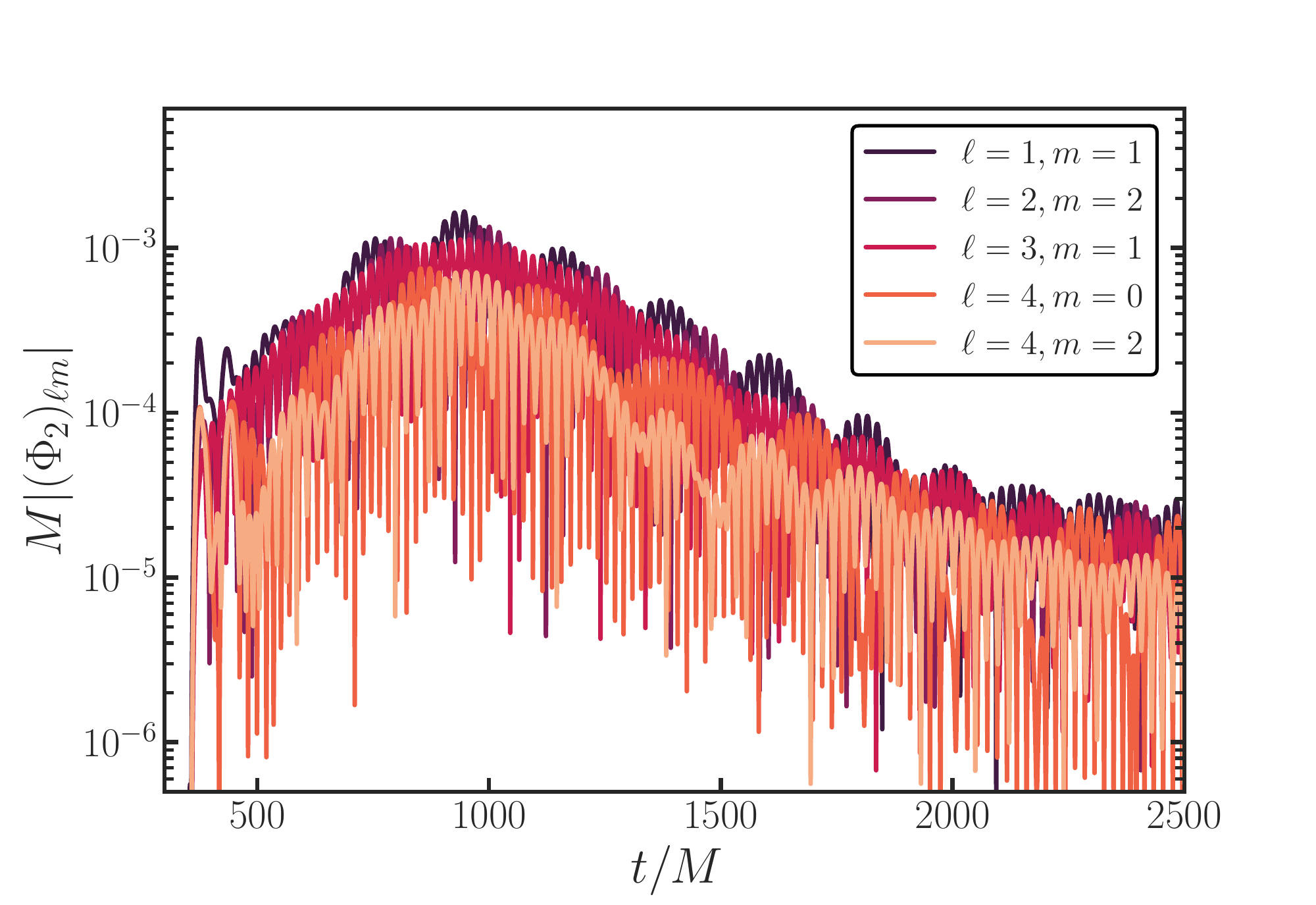}		\includegraphics[width = 0.5\textwidth]{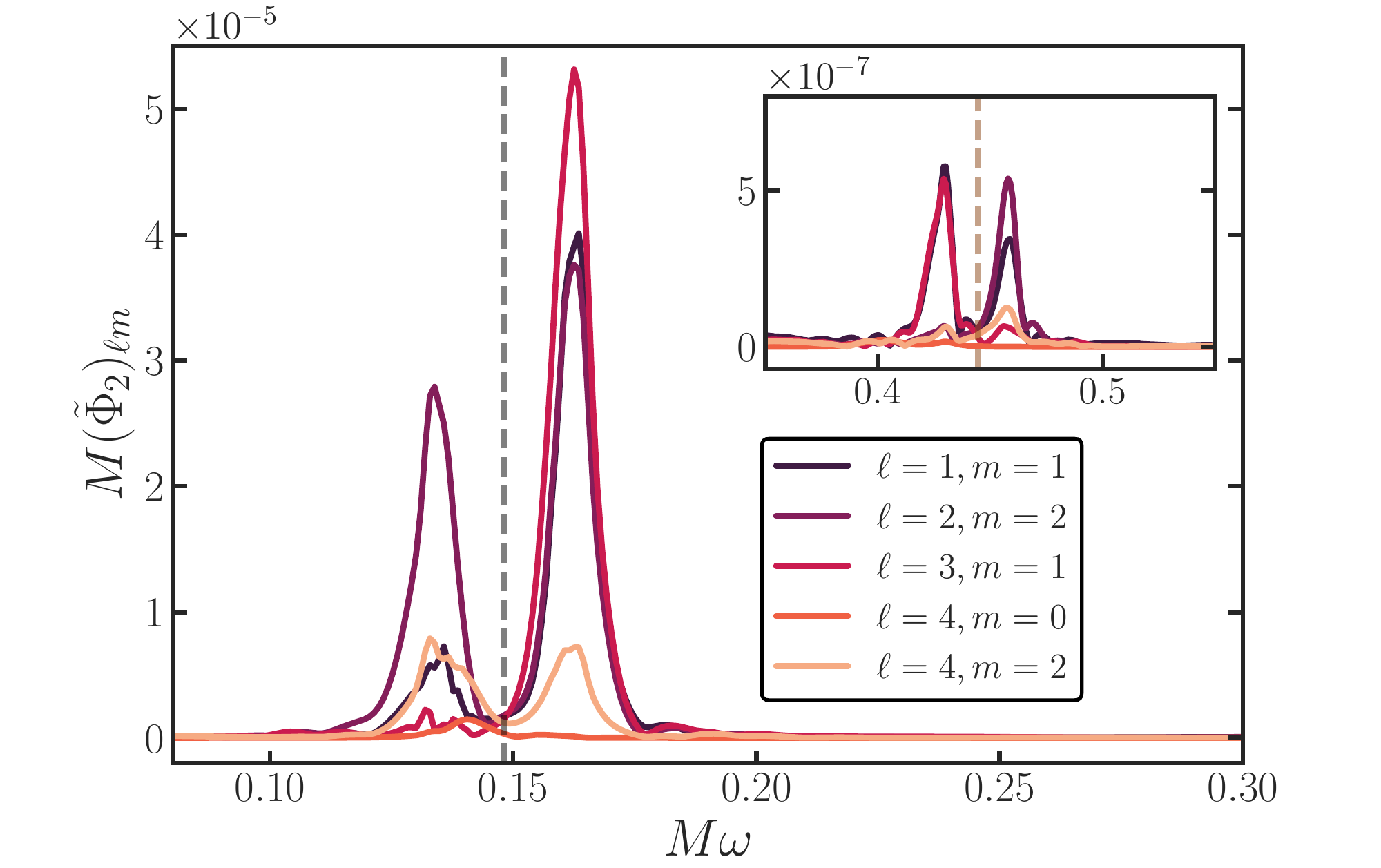}
    \caption{{\bf Top Panel:} Time evolution of various multipole
    modes of the Newman-Penrose scalar $|(\Phi_{2})_{\ell m}|$ in the supercritical regime. Considered simulation is $\mathcal{I}_{3}$. Field is extracted at $r_{\rm ex} = 400M$ and $\mu M = 0.3$.
    {\bf Bottom Panel:} Fourier transform of the multipole modes shown in the top panel. The gray and brown dashed lines indicate the frequencies at $\omega_0/2$ and $3\omega_0/2$, respectively.}    \label{fig:EMfieldmodes}
\end{figure}

%%%%%%%%%%%%%%%%%%%%%%%%%%%%%%%%%%%%%%%%%%%%%%
In the main text, we have shown the dominant contribution coming from the dipole $\ell = m = 1$ mode (cf.~Figs.~\ref{fig:BurstmuM03r20}-\ref{fig:BurstAxionmuM03r40100}). In general however, higher order multipoles are also produced. In Figs.~\ref{fig:scalarfieldmodes} and~\ref{fig:EMfieldmodes}, we show a subset of those from the scalar and vector field, respectively. In both figures, we consider simulation $\mathcal{I}_{3}$, where SR is turned off and we start in the supercritical regime. Three features are worth noting;~(i) only axion modes with odd $\ell$ can be produced from our initial data. An explanation for this selection rule is provided in Appendix~\ref{sec:selectionrules};~(ii) the Fourier transform of the vector field shows additional peaks with a frequency slightly lower than $\mu/2$ and two near $3\mu/2$. As discussed in the main text, these should be interpreted as ``photon echoes'' created by outwards traveling photons that interact with the axion cloud;~(iii) in Fig.~\ref{fig:scalarfieldmodes}, we observe that some of these up-scattered photons can recombine with ``normal'' photons ($\omega \sim \mu/2$) to form axion waves with a frequency of twice the boson mass. 
%%%%%%%%%%%%%%%%%%%%%%%%%%%%%%%%%%%%%%%%%%%%%%%%%%%%%%%%%%%%%%%
\section{Superradiant Mathieu equation}\label{sec:appendMathieu}
%%%%%%%%%%%%%%%%%%%%%%%%%%%%%%%%%%%%%%%%%%%%%%%%%%%%%%%%%%%%%%%%
By solving the superradiant Mathieu equation~\eqref{eq:modifiedMathieu} numerically for different values of $C$, we are able to find a growth rate~\eqref{eq:modifiedgrowthrate} for the EM field in flat spacetime, while assuming a homogeneous axion condensate. Remarkably, this estimate is even accurate in describing the super-exponential growth of the EM field in presence of SR, when considering the full setup on a Schwarzschild background (including the finite-size effects of the cloud with $\lambda_{\rm esc}$). In this appendix, we show a few examples of the numerical solutions to~\eqref{eq:modifiedMathieu} and we use a multiple-scale method to derive the growth rate analytically.
\begin{figure}
    \centering
    \includegraphics[width = 0.5\textwidth]{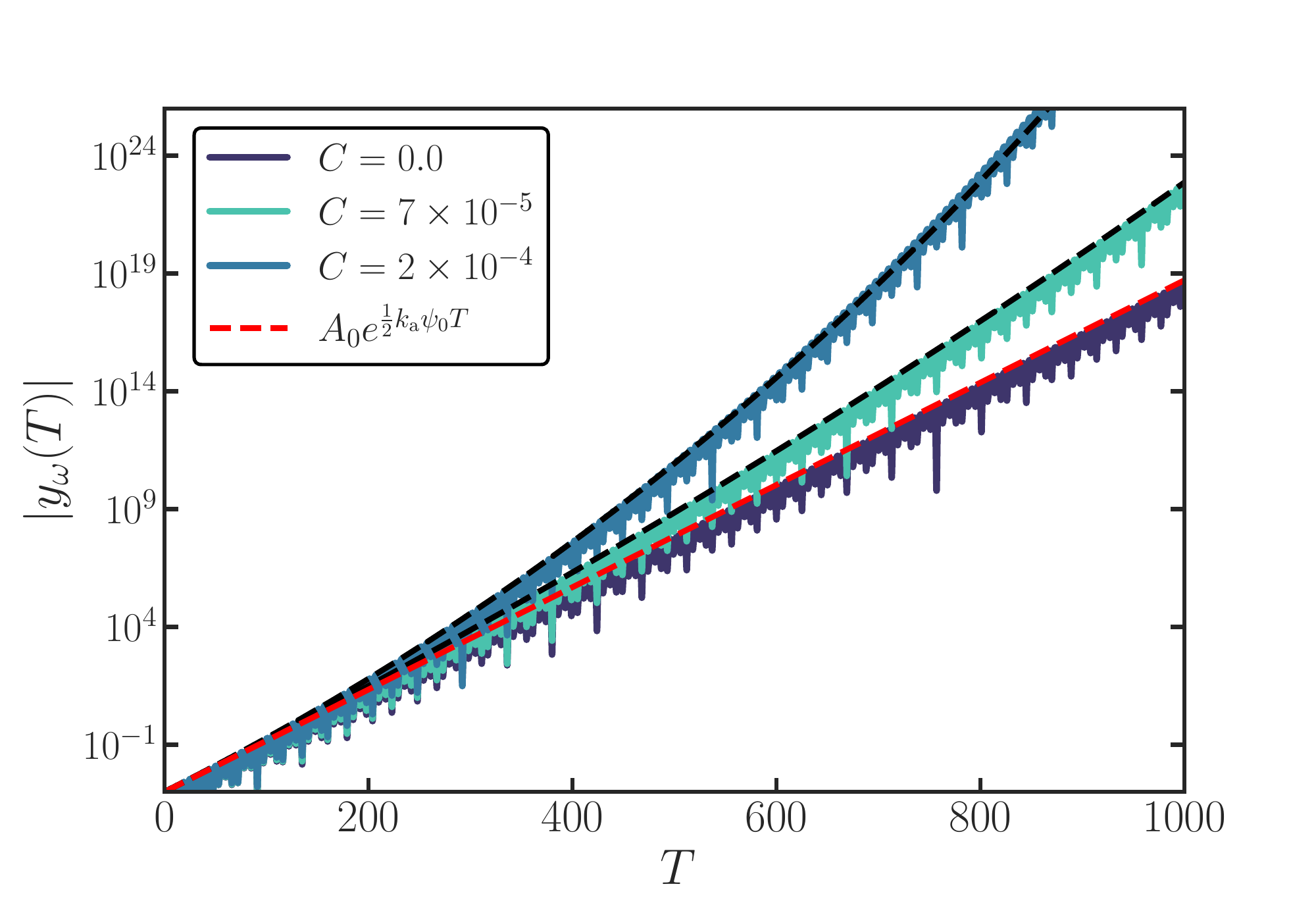}
    \caption{Numerical solutions to the superradiant Mathieu equation~\eqref{eq:modifiedMathieu} for different values of $C$. Horizontal axis shows the rescaled time $T = \mu t$. Red dashed line shows the ``standard'' Mathieu growth rate, while black dashed lines are obtained from the analytic growth rate~\eqref{eq:modifiedgrowthrate} and describe the numerical solutions well. Chosen parameters are $\mu =0.2, p_z=0.1, k_{\rm a}\psi_0 =0.1$. 
    }
    \label{fig:mathieuwithC}
\end{figure}

Figure~\ref{fig:mathieuwithC} shows various numerical solutions to the superradiant Mathieu equation~\eqref{eq:modifiedMathieu}. For $C=0$, the solution is well described by the standard Mathieu growth rate (red dashed line). For non-zero values of $C$ instead, the standard Mathieu prediction becomes inaccurate and the numerical solutions are well fitted by the super-exponential growth rate~\eqref{eq:modifiedgrowthrate} (black dashed lines).

%%%%%%%%%%%%%%%%%%%%%%%%%%%%%%
{\it Multiple-scale analysis} $-$
%%%%%%%%%%%%%%%%%%%%%%%%%%%%%%
Regular perturbation theory fails to describe some types of problems at late times. This is due to the presence of secular terms, causing non-uniformities to appear between consecutive orders in the perturbation series. An example of this happening is when solving the Mathieu equation, where a multiple-scale analysis is more suitable (see e.g.~\cite{bender78:AMM}). In the following, we will show its effectiveness for solving the superradiant Mathieu equation and with that, provide an analytical explanation of the numerically fitted growth rate~\eqref{eq:modifiedgrowthrate}. 

Let us consider the superradiant Mathieu equation~\eqref{eq:modifiedMathieu}, where we expand the exponential as
\begin{equation}
\label{eq:multiscalemathieu}
\frac{\mathrm{d}^{2}y}{\mathrm{d}T^{2}} + (b + 2 \delta (1+C T)\cos T)\,y=0\,,
\end{equation}
and we assume $\delta$ to be small. We can introduce two timescales, a {\it fast} timescale $T$, and a {\it slow} one, $\mathcal{T}=\delta T$, which we treat as independent variables. By assuming a dependence on both variables, i.e.,~$y(T)\rightarrow y(T,\mathcal{T})$ and expanding all the parameters, i.e.,~$y=y_0(T,\mathcal{T})+ \delta y_1(T,\mathcal{T})$ and $b=b_0+\delta b_1$, we can use the additional freedom from treating $\mathcal{T}$ as a new variable to eliminate secular terms. This causes the solution to hold for longer times compared to a ``normal'' perturbative approach. Assuming the SR term also to be small, i.e.,~$C=\delta \mathfrak{C}$, we can introduce yet another timescale, the {\it very slow} one, $\mathfrak{T}=\delta^2 T$. Upon expanding again $b=b_0+\delta b_1 + \delta^2 b_2$, we obtain
\begin{equation}\label{eq:multiscalefull}
     \frac{\mathrm{d}^{2}y}{\mathrm{d}T^{2}} +\left(b + \delta (b_1 + 2 \cos T)+\delta^2 (b_2 + 2 \mathfrak{C} T \cos T)\right)\,y=0\,.
\end{equation}
In the spirit of the multiple-scale method, we ``promote'' $y$ to depend on all the timescales as independent variables and then expand, i.e.,~$y(T,\mathcal{T}, \mathfrak{T})=y_0 (T,\mathcal{T}, \mathfrak{T})+\delta y_1 (T,\mathcal{T}, \mathfrak{T})+\delta^2 y_2(T,\mathcal{T}, \mathfrak{T})$. 

We can now consider~\eqref{eq:multiscalefull} order by order. At zeroth order, we have:
\begin{equation}
    \partial^2_T y_0 + b_0 y_0=0\,,
\end{equation}
where $b_0 = 1/4$ at the inset of the first unstable Mathieu band. Consequently, at this order, we obtain the solution: $y_0=A(\mathcal{T}, \mathfrak{T}) e^{i T/2}+ \mathrm{c.c.}$ Hence, the solution at the fast timescale $T$ just describes the harmonic behavior. At first-order, we have
\begin{equation}
    \partial^2_T y_1+\frac{1}{4}y_1=-2\partial_T \partial_\mathcal{T}y_0 -(b_1+2 \cos T)\,y_0\,.
\end{equation}
The right-hand side of this equation contains secular terms. However, we can use the extra dependence of $y_0$ on the slow timescale to remove it, namely by requiring $i \partial_\mathcal{T}A(\mathcal{T}, \mathfrak{T})=-b_1 A(\mathcal{T}, \mathfrak{T})+A^*(\mathcal{T}, \mathfrak{T})$. Solving this leads to the dependence of the zeroth-order solution on the slow timescale, i.e.,~$y_0=A(\mathfrak{T})e^{\sqrt{1-b_1^2}\mathcal{T}} e^{i T/2}+\mathrm{c.c.}$ Similarly, the dependence on the very slow timescale can be used to eliminate secular terms in the second-order equation.\footnote{As these computations become quite cumbersome, we do not report them here.} Following this procedure, we obtain for the zeroth-order solution: 
\begin{equation}
y_0\approx e^{\sqrt{1-b_1^2}\mathcal{T}} e^{i T/2}e^{ \mathfrak{C} T\mathfrak{T}}+\mathrm{c.c.}    
\end{equation} 
At sufficiently large times ($T \gg 1/\mathfrak{C}$), the growth rate is thus dominated by $e^{\mathfrak{C}T \mathfrak{T}}=e^{\delta C T^2}$. 

Finally, by comparing~\eqref{eq:multiscalemathieu} with~\eqref{eq:modifiedMathieu}, we identify $\delta=p_z\psi_0 k_{\rm a}/\mu^2$ and $p_z=\mu/2$, such that, after rescaling the physical time $t=T/\mu$, the growth rate is $e^{\lambda t}$ with
\begin{equation}
    \lambda t = \frac{\mu}{2} k_{\rm a} \psi_0 C t^2\,,
\end{equation}
which is the dominant growth rate at late times we found in~\eqref{eq:modifiedgrowthrateFirst} (up to a factor of 2).

The multiple-scale method thus produces a solution with three timescales;~(i) the ``fast'' timescale that corresponds to the harmonic oscillations with a frequency at half the boson mass, (ii) the ``slow'' timescale belonging to the standard Mathieu growth rate, and (iii) the ``very slow'' timescale which originates from the super-exponential growth induced by SR and becomes dominant at late times. In conclusion, the Mathieu equation provides us, once again, with a simple analytical explanation to the behavior of the full system.
%%%%%%%%%%%%%%%%%%%%%%%%%%%%%%%%%%%%%%%%
\section{Plasma Mathieu equation}\label{app:plasmaMathieu}
%%%%%%%%%%%%%%%%%%%%%%%%%%%%%%%%%%%%%%%%
In this appendix, we study the axion-photon-plasma system in flat spacetime. This analysis closely follows~\cite{SenPlasma}, yet now in the context of a Mathieu-like equation. Furthermore, we generalize their work by including a momentum equation rather than assuming Ohm's law. 

The starting point is the equations of motion~\eqref{eq:evoleqns} in Minkowski. Similar to Section~\ref{subsec:analyticalgrow}, we assume the wave vector to be along the $\hat{z}$ direction, i.e.,~$\vec{p}=(0,0,p_z)$, the EM ansatz~\eqref{eq:MathieuEM}, and a homogeneous axion condensate defined as
\begin{equation}
    \Psi=\frac{1}{2}(\psi_0 e^{-i \mu t}+ \psi_0^* e^{i \mu t})\,.
\end{equation}
By linearizing~\eqref{eq:evoleqns}, one can straightforwardly solve the momentum equation and find the velocity of the electrons with respect to the EM field. Once again, due to their large inertia, we neglect the perturbations of the ions and treat them as a neutralizing background. Since the longitudinal and transverse modes in linear photon-plasma theory in flat spacetime are decoupled, we focus on the latter, obtaining
\begin{equation}
    u^k=-\frac{q_{\rm e}}{m_{\rm e}}\alpha^k e^{i (\vec{p} \cdot \vec{x} - \omega t)}\,,
\end{equation}
where $k=\hat{x},\hat{y}$ are the transverse directions. We can now insert this expression in the current that appears on the right-hand side of Maxwell's equations, i.e.,~$j^k=q_{\rm e} n_{\rm e} u_{\rm e}^k$, to obtain two decoupled equations for the transverse polarizations, $\alpha^k$. Finally, by adopting the field redefinition $y_k =e^{i \omega t}\alpha_k$, rescaling the time as $T=\mu t$ and projecting along a circular polarization basis $e_\pm$ such that $y=y_\omega e_\pm$, we find the Mathieu equation in the presence of plasma to be~\cite{Hertzberg:2018zte}
\begin{equation}\label{eq:PlasmaMathieu}
    \partial_T^2 y_\omega +\frac{1}{\mu^{2}}\Big(p_z^2+\omega_{\rm p}^2 -2 \mu  p_z \psi_0 k_{\rm a}\text{sin}T\Big)y_\omega=0\,.
\end{equation}
One can readily see that in the absence of a plasma, i.e.,~$\omega_{\rm p}=0$, the vacuum Mathieu equation is recovered~\cite{Boskovic:2018lkj}.
\begin{figure}
    \hspace*{-0.3cm}
    \includegraphics[width = 0.48\textwidth]{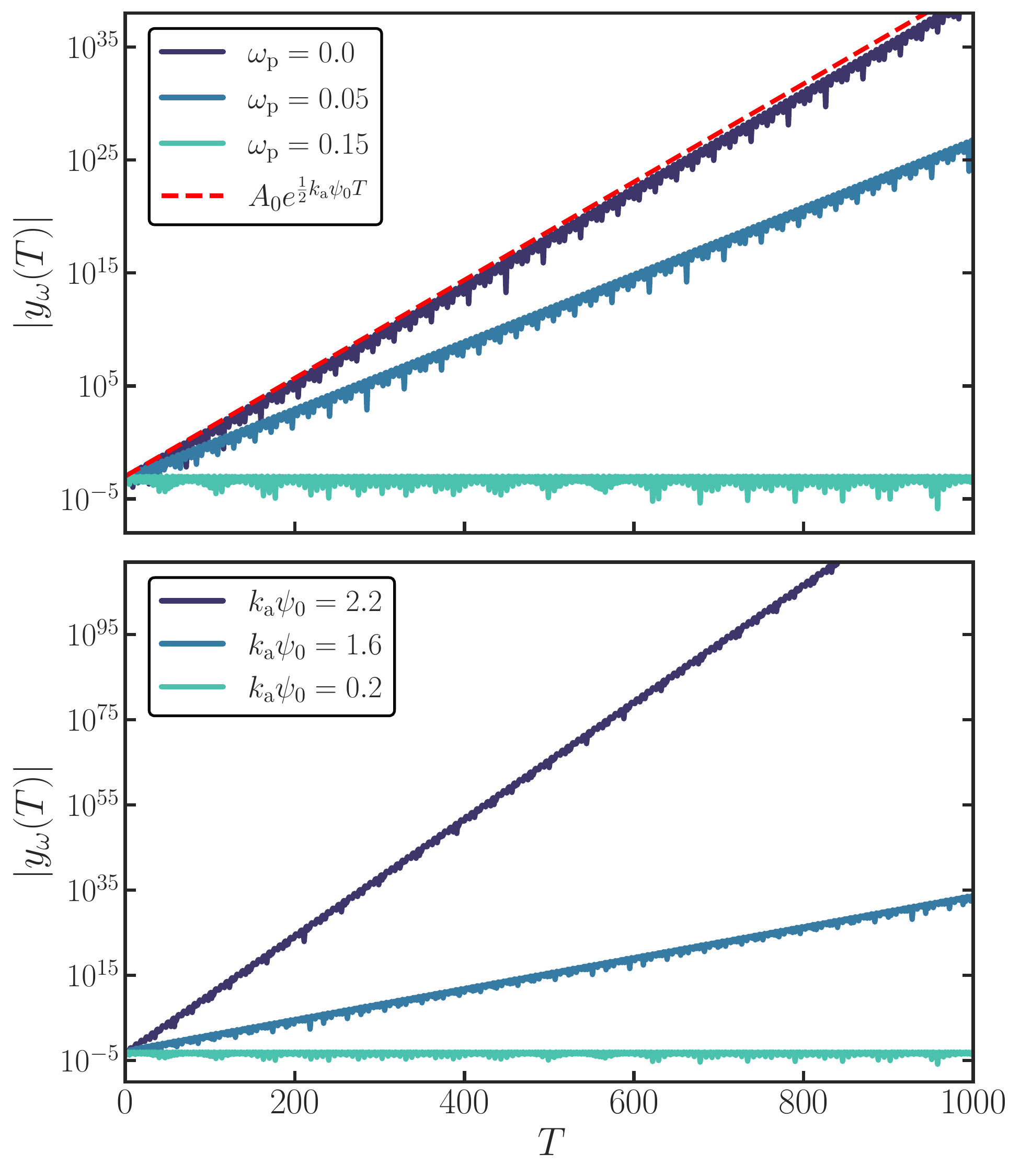}
    \caption{{\bf Top Panel:} Solutions of the plasma Mathieu equation~\eqref{eq:PlasmaMathieu} for different values of the plasma frequency. Here, we choose $\mu=0.2, p_z=0.1$ and $k_{\rm a}\psi_0 =0.2$.
    {\bf Bottom Panel:} A similar setup as above, yet now the axionic coupling is varied and $\omega_{\rm p} = 0.15 > \mu /2$. The other parameters are the same as above. As can be seen, for large values of $k_{\rm a}\psi_0$ the instability is restored.}
    \label{fig:mathieuplasma}
\end{figure}

The top panel of Fig.~\ref{fig:mathieuplasma} shows numerical solutions to the plasma Mathieu equation~\eqref{eq:PlasmaMathieu} for different values of the plasma frequency. For $\omega_{\rm p} = 0$, the solution develops an instability that is well described by the analytic solution of the vacuum Mathieu equation (red dashed line). By increasing $\omega_{\rm p}$, the growth rate of the instability becomes smaller as the interval of the momentum corresponding to the instability shrinks until the solution becomes stable when $\omega_{\rm p}\geq\mu/2$. We find good agreement with the instability interval predicted in equation~(19) from~\cite{SenPlasma}, by exploring a wide region of the parameter space. Interestingly, even when $\omega_{\rm p}\geq\mu/2$, the instability band can be widened by increasing the value of $k_{\rm a} \psi_0$, making it possible to restore the instability even for dense plasmas. This effect can be seen in the bottom panel of Fig.~\ref{fig:mathieuplasma}. By fixing $\omega_{\rm p}=0.15$ (with $\mu = 0.2$), the instability is restored for large enough values of $k_{\rm a} \psi_0$. 

%%%%%%%%%%%%%%%%%%%%%%%%
{\it Band analysis} $-$
%%%%%%%%%%%%%%%%%%%%%%%%
In~\cite{SenPlasma}, the maximum growth rate in the low axionic coupling regime is found where the condition $p_z^2 + \omega_{\rm p}^2 = \mu^2/4$ holds. While in~\cite{SenPlasma}, this result does not have an immediate interpretation, our redefinition of the system in terms of the Mathieu equation provides a simple explanation to this condition; from~\eqref{eq:PlasmaMathieu}, it follows immediately that the instability bands are located where
\begin{equation}
\label{eq:plasmabands}
 p_z^2 + \omega_{\rm p}^2= n^2 \, \frac{\mu^2}{4}\quad \text{with} \quad n \in \mathbb{N}\,.   
\end{equation}
Therefore, the maximum growth rate of~\cite{SenPlasma} can be interpreted as the first, dominant instability band, i.e.,~where $n=1$. Note that, as long as the axionic coupling is sufficiently low, the dispersion relation of the photon is not modified by the condensate, and thus $\omega^2 \approx p_z^2 + \omega_{\rm p}^2$, with $\omega$ the frequency of the photon. Therefore, similar to the vacuum case, the instability bands correspond to frequencies which are multiples of $\mu/2$. 

\begin{figure}
    \centering
    \includegraphics[width = 0.5\textwidth]{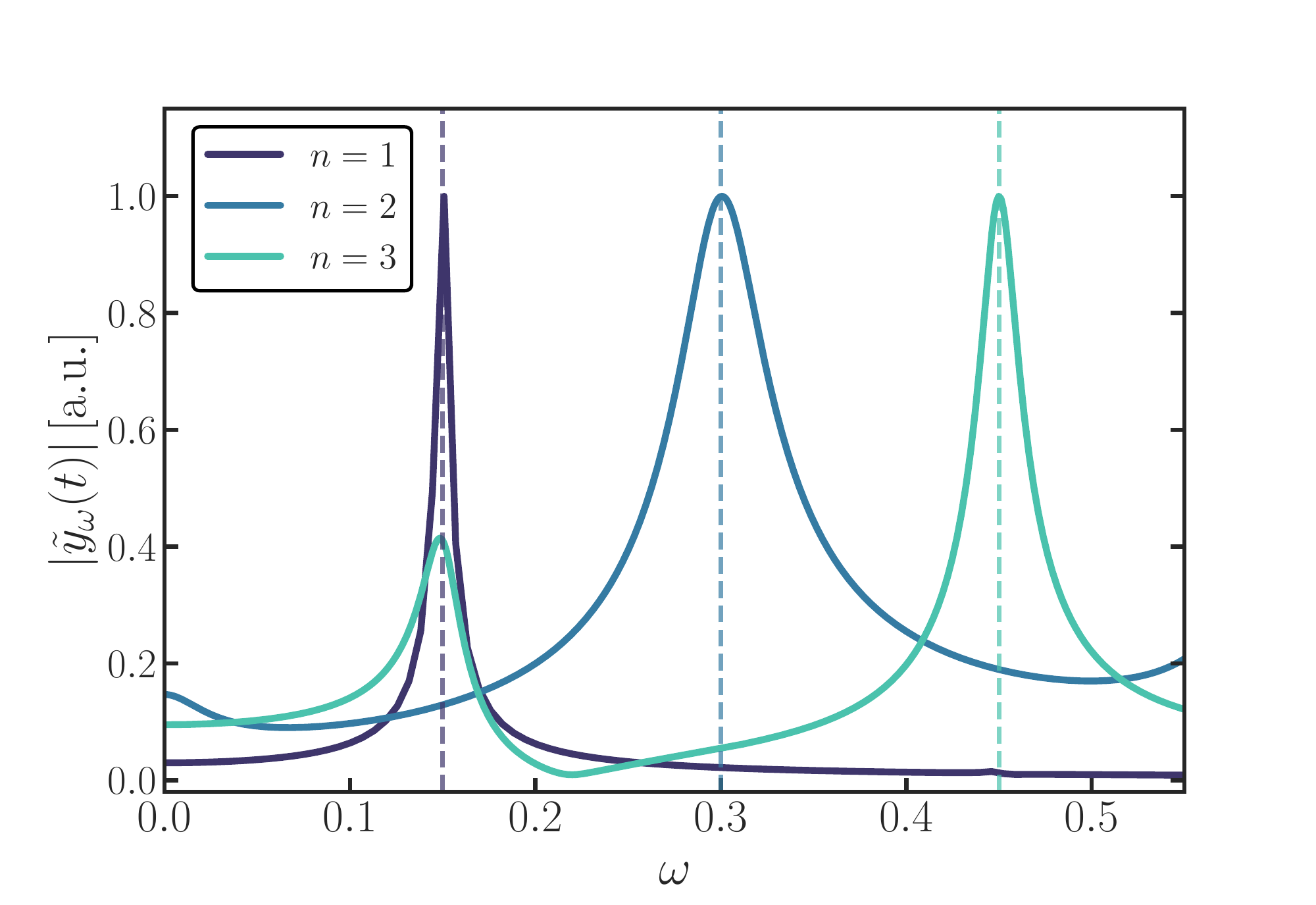}
    \caption{The Fourier transforms of three numerical solutions to the plasma Mathieu equation~\eqref{eq:PlasmaMathieu} with $\mu=0.3$ and $p_z=0.15$. The peaks have been arbitrarily normalized and we revert $t = \mu^{-1}T$ for the Fourier transform. The considered parameters for the plasma frequency and the axionic coupling for $n = 1, 2, 3$ are $\omega_{\rm p}=0, 0.2, 0.43$ and $k_{\rm a} \psi_0 = 0.02, 2, 2.8$, respectively. The dashed lines indicate $n \mu/2$ and show that each of the solutions lies in a different instability band.}
    \label{fig:bandsmathieuplasma}
\end{figure}

For large enough $\omega_{\rm p}$, i.e.,~when $\omega_{\rm p}\geq\mu/2$, the condition~\eqref{eq:plasmabands} can never be satisfied for $n=1$. However, crucially, it can still be satisfied for $n>1$, which corresponds to exciting higher bands and thus restoring the instability.\footnote{Higher bands of the Mathieu equation are narrower than the first one, making the available parameter space for an instability smaller. However, since these bands widen for large values of the axionic coupling $k_{\rm a}\psi_0$, even for $n>1$ an efficient instability can be triggered.} In Fig.~\ref{fig:bandsmathieuplasma}, we show that this is indeed the case by numerically solving~\eqref{eq:PlasmaMathieu} and taking the Fourier transform. The dark-blue line that peaks at $\mu/2$, has $\omega_{\rm p}=0$ and a moderate value of $k_{\rm a}\psi_0=0.02$, and thus behaves according to the parametric mechanism by triggering the first band. When we consider a higher plasma frequency, $\omega_{\rm p}=0.2 > \mu/2$, it is not possible to excite the first instability band anymore. However, as we increase the axionic coupling to $k_{\rm a} \psi_0=2$, the instability is restored in the second band with a frequency $\mu$ (blue line). We can continue for even higher plasma frequencies, e.g.~$\omega_{\rm p}=0.43$, which for an axionic coupling of $k_{\rm a} \psi_0=2.8$, triggers the third instability band at $\omega=3/2 \mu$ (turquoise line). Note that the chosen values of $\omega_{\rm p}$ in the three solutions approximately satisfy~\eqref{eq:plasmabands} for $n=1,2$ and $3$.
%%%%%%%%%%%%%%%%%%%%%%%%%%%%%%%%%%%%%%%%%%%%%%
\section{Selection rules}\label{sec:selectionrules}
\begin{figure}
    \centering
    \includegraphics[width = 0.5\textwidth]{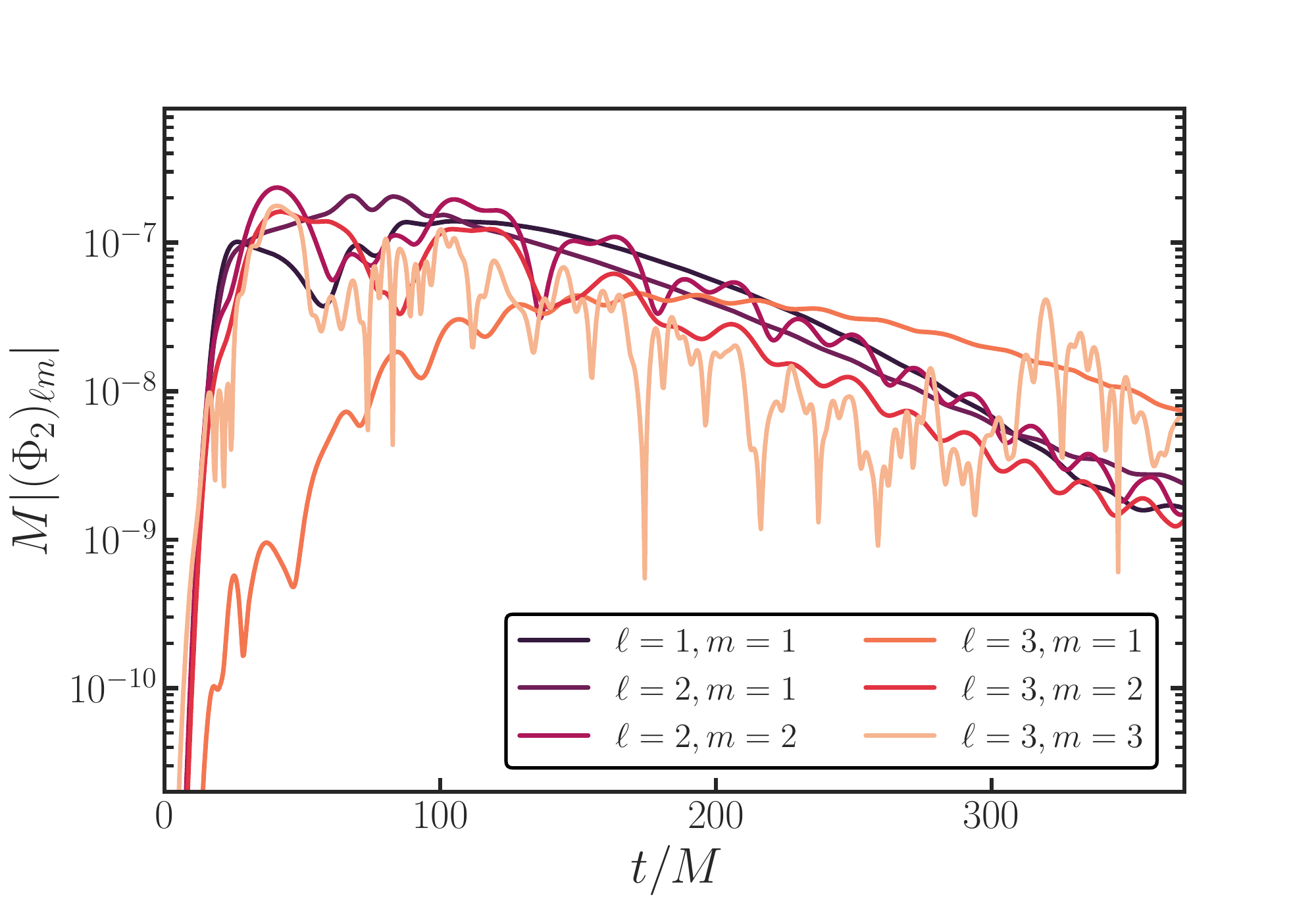}
    \caption{The time evolution of various multipole
    modes of the Newman-Penrose scalar $\Phi_{2}$ in the subcritical regime. The considered simulation is $\mathcal{I}_{2}$, where the field is extracted at $r_{\rm ex} = 20M$ and $\mu M = 0.3$. Each of the curves has been rescaled according to the order found in~\eqref{eq:selectionrules}.}
    \label{fig:EMfieldSelectionrules}
\end{figure}

%%%%%%%%%%%%%%%%%%%%%%%%%%%%%%%%%%%%%%%%%%%%%%
Using spherical harmonics, the equations of motion can be decomposed, and, as we will show, allow us to predict which modes are excited from the axionic coupling. This approach yields a consistency check of our simulations in the case where SR growth is absent. Using the electric field $E_{i}$ and the magnetic field $B_{i}$, the Maxwell equations can be written as 
\begin{equation}
\begin{aligned}
    \partial_{t}E^{i}&
    =\alpha
    KE^{i}+\beta^{j}\partial_{j}E^{i}-E^{j}\partial_{j}\beta^{i}-\epsilon^{ijk}D_{j}(\alpha B_{k})
    \\&
    +2k_{\rm a}\alpha\left(
    \epsilon^{ijk} E_{k}D_{j}\Psi +B^{i}n^{\alpha}\partial_{\alpha}\Psi
    \right)\,,\\
    \partial_{t}B^{i}&=\beta^{j}\partial_{j}B^{i}-B^{j}\partial_{j}\beta^{i}+\alpha KB^{i}+\epsilon^{ijk}D_{j}(\alpha E_{k})\,,
\end{aligned}
\end{equation}
where $\epsilon^{ijk}=-\frac{1}{\sqrt{\gamma}}E^{ijk}$ and $E^{ijk}$ is the totally anti-symmetric tensor with $E^{123}=1$. We focus on a Schwarzschild BH which has the following metric:
\begin{equation}
\mathrm{d}s^{2}=-f(r)\mathrm{d}t^{2}+\frac{1}{f(r)}(r)\mathrm{d}r^{2}+r^{2}\hat{\gamma}_{AB}\mathrm{d}x^{A}\mathrm{d}x^{B}\,,
\end{equation}
where $f(r)=1-\frac{2M}{r}$, and $\hat{\gamma}_{AB}\mathrm{d}x^{A}\mathrm{d}x^{B}=\mathrm{d}\theta^{2}+\sin^{2}\theta \mathrm{d}\varphi^{2}$.
From the spacetime symmetry, the electric and magnetic field can be decomposed using the scalar spherical harmonics $Y_{\ell m}(\theta,\varphi)$ and the vector spherical harmonics $\hat{\nabla}_{A}Y_{\ell m}$ and $V_{\ell m,A}=\hat{\epsilon}^{B}_{A}\hat{\nabla}_{B}Y_{\ell m}$. Here, $\hat{\epsilon}_{AB}$ is the anti-symmetric tensor with $\hat{\epsilon}_{\theta\varphi}=\sin\theta$. Using these harmonics functions, we expand the electric, magnetic, and scalar field as follows:
\begin{equation}
\begin{aligned}
E_{r}&=\sum_{\ell m}\mathcal{E}_{\ell m,r}Y_{\ell m}+{\rm c.c.}\,,\\
E_{A}&=\sum_{\ell m}\left\{
\mathcal{E}_{\ell m,S}\frac{\hat{\nabla}_{A}Y_{\ell m}}{\sqrt{\ell(\ell+1)}}+\mathcal{E}_{\ell m,V}V_{\ell m,A}+{\rm c.c.}
\right\}\,,\\
B_{r}&=\sum_{\ell m}\mathcal{B}_{\ell m,r}Y_{\ell m}+{\rm c.c.}\,,\\
B_{A}&=\sum_{\ell m}\left\{
\mathcal{B}_{\ell m,S}\frac{\hat{\nabla}_{A}Y_{\ell m}}{\sqrt{\ell (\ell +1)}}+\mathcal{B}_{\ell m,V}V_{\ell m,A}+{\rm c.c.}
\right\}\,,\\
\Psi&=\sum_{\ell m}\Psi_{\ell m}Y_{\ell m}+{\rm c.c.}\,,
\end{aligned}
\end{equation}
where $A = \{\theta, \varphi\}$ and $\mathcal{E}_{\ell m,r}$, $\mathcal{E}_{\ell m,S}$, $\mathcal{E}_{\ell m,V}$, $\mathcal{B}_{\ell m,r}$, $\mathcal{B}_{\ell m,S}$, $\mathcal{B}_{\ell m,V}$, and $\Psi_{\ell m}$ are all coefficients that depend on $t$ and $r$ only. Time or space derivatives are denoted with a dot or prime, respectively. Since both the scalar and vector spherical harmonics are orthogonal functions, the Maxwell equations with axionic coupling can be decomposed. We find the coefficients for the electric field to be
\begin{equation}
    \begin{aligned}
     \dot{\mathcal{E}}_{\ell m,r}&=-\frac{\ell (\ell +1)}{r^{2}}\mathcal{B}_{\ell m,V}-\frac{2k_{\rm a}}{\sqrt{f}r^{2}}\\
    &\times \sum_{\substack{\ell'm'\\ \ell''m''}}\sum_{I=1}^{4}C^{(r,I)}_{\ell m \ell' m' \ell'' m''}X_{(I),\ell m \ell' m' \ell'' m''}\,,\\
    \dot{\mathcal{E}}_{\ell m,S}&=-\frac{\sqrt{\ell(\ell+1)}}{2}(f'\mathcal{B}_{\ell m,V}+2f\mathcal{B}'_{\ell m,V})\\
    &-2k_{\rm a}\sqrt{\ell(\ell+1)}\\
    &\times \sum_{\substack{\ell'm'\\ \ell''m''}}\sum_{I=1}^{4}C^{(S,I)}_{\ell m \ell' m' \ell'' m''}X_{(I),\ell m \ell'm'\ell''m''}\,,\\
    \dot{\mathcal{E}}_{\ell m,V}&=-f\mathcal{B}_{\ell m,r}+\frac{f'\mathcal{B}_{\ell m,S}+2f\mathcal{B}'_{\ell m,S}}{2\sqrt{\ell(\ell+1)}}-2k_{\rm a}\\
    &\times \sum_{\substack{\ell'm'\\ \ell''m''}}\sum_{I=1}^{4}C^{(V,I)}_{\ell m\ell'm'\ell''m''}X_{(I),\ell m\ell'm'\ell''m''}\,.
    \label{Maxeqspherical1}
    \end{aligned}
\end{equation}
Then, we proceed with the coefficients for the magnetic field:
\begin{equation}
    \begin{aligned}
    \dot{\mathcal{B}}_{\ell m,r}&=\frac{\ell(\ell+1)}{r^{2}}\mathcal{E}_{\ell m,V}\,,\\
    \dot{\mathcal{B}}_{\ell m,S}&=\frac{\sqrt{\ell(\ell+1)}}{2}\left(f'\mathcal{E}_{\ell m,V}+2f\mathcal{E}'_{\ell m,V}\right)\,,\\
    \dot{\mathcal{B}}_{\ell m,V}&=f\mathcal{E}_{\ell m,r}-\frac{f'\mathcal{E}_{\ell m,S}+2f\mathcal{E}'_{\ell m,S}}{2\sqrt{\ell(\ell+1)}}\,,
    \label{Maxeqspherical2}
    \end{aligned}
\end{equation}
and finally the coefficients for the scalar field:
\begin{equation}
    \begin{aligned}
    \ddot{\Psi}_{\ell m}&=-f\left(\frac{2}{r}+f'\right)\Psi_{\ell m}'\\
    &-f^{2}\Psi''_{\ell m}+f\left(
    \frac{\ell(\ell+1)}{r^{2}}+\mu^{2}
    \right)\Psi_{\ell m}-2k_{\rm a}f\\
    &\times \sum_{\substack{\ell'm'\\ \ell''m''}}\sum_{I=1}^{4}C^{(\Psi,I)}_{\ell m\ell'm'\ell''m''}X_{(I),\ell m\ell'm'\ell''m''}\,,
    \label{Maxeqspherical3}
    \end{aligned}
\end{equation}
where $X_{(I),\ell m\ell'm'\ell''m''}$ is found by
\begin{equation}
\begin{aligned}
    X_{(1),\ell m\ell'm'\ell''m''}&=\int \mathrm{d}^{2}\Omega Y^{\ast}_{\ell m}Y_{\ell''m''}Y_{\ell'm'}\,,\\
    X_{(2),\ell m\ell'm'\ell''m''}&=\int \mathrm{d}^{2}\Omega Y^{\ast}_{\ell m}Y^{\ast}_{\ell''m''}Y_{\ell'm'}\,,\\
    X_{(3),\ell m\ell'm'\ell''m''}&=\int \mathrm{d}^{2}\Omega Y^{\ast}_{\ell m}\hat{\nabla}^{A}Y_{\ell''m''}V_{\ell'm',A}\,,\\
    X_{(4),\ell m\ell'm'\ell''m''}&=\int \mathrm{d}^{2}\Omega Y^{\ast}_{\ell m}\hat{\nabla}^{A}Y^{\ast}_{\ell''m''}V_{\ell'm',A}\,.
\end{aligned}
\end{equation}
Upon defining the prefactor
\begin{equation}    
Q_{\ell\ell'\ell''}=\frac{\ell(\ell+1)+\ell'(\ell'+1)-\ell''(\ell''+1)}{2}\,,
\end{equation}
we find for $C^{(r,I)}$:
\begin{equation}
    \begin{aligned}
     C^{(r,1)}_{\ell m\ell'm'\ell''m''}&=Q_{\ell''\ell'\ell}
    \mathcal{E}_{\ell''m'',V}\Psi_{\ell'm'} -r^{2}\mathcal{B}_{\ell''m'',r}\dot{\Psi}_{\ell'm'}\,,\\
    C^{(r,2)}_{\ell m\ell'm'\ell''m''}&=Q_{\ell''\ell'\ell}
    \mathcal{E}_{\ell''m'',V}^{\ast}\Psi_{\ell'm'} -r^{2}\mathcal{B}_{\ell''m'',r}^{\ast}\dot{\Psi}_{\ell'm'}\,,\\
    C^{(r,3)}_{\ell m\ell 'm'\ell''m''}&=\frac{\mathcal{E}_{\ell ''m'',S}\Psi_{\ell 'm'}}{\sqrt{\ell''(\ell''+1)}}\,,\\
    C^{(r,4)}_{\ell m\ell 'm'\ell''m''}&=\frac{\mathcal{E}^{\ast}_{\ell''m'',S}\Psi_{\ell 'm'}}{\sqrt{\ell''(\ell''+1)}}\,,
    \end{aligned}
    \label{coeff Cr}
\end{equation}
for $C^{(S,I)}$:
\begin{equation}
    \begin{aligned}
    C^{(S,1)}_{\ell m\ell'm'\ell''m''}&=-Q_{\ell''\ell\ell'}\biggl(f\mathcal{E}_{\ell''m'',V}\Psi_{\ell'm'}'\\&+\frac{\mathcal{B}_{\ell''m'',S}\dot{\Psi}_{\ell'm'}}{\sqrt{\ell''(\ell''+1)}}\biggr)\,,\\
    C^{(S,2)}_{\ell m\ell'm'\ell''m''}&=-Q_{\ell''\ell\ell'}\biggl(f\mathcal{E}^{\ast}_{\ell''m'',V}\Psi_{\ell'm'}'\\&+\frac{\mathcal{B}^{\ast}_{\ell''m'',S}\dot{\Psi}_{\ell'm'}}{\sqrt{\ell''(\ell''+1)}}\biggr)\,,\\
    C^{(S,3)}_{\ell m\ell'm'\ell''m''}&=f\mathcal{E}_{\ell''m'',r}\Psi_{\ell'm'}
    -\frac{f}{\sqrt{\ell''(\ell''+1)}}\\&\times\mathcal{E}_{\ell''m'',S}\Psi_{\ell'm'}'+\mathcal{B}_{\ell''m'',V}\dot{\Psi}_{\ell'm'}\,,\\
    C^{(S,4)}_{\ell m\ell'm'\ell''m''}&=
    f\mathcal{E}^{\ast}_{\ell''m''m,r}\Psi_{\ell'm'}
    -\frac{f}{\sqrt{\ell''(\ell''+1)}}\\&\times\mathcal{E}^{\ast}_{\ell''m'',S}\Psi_{\ell'm'}'+\mathcal{B}^{\ast}_{\ell''m'',V}\dot{\Psi}_{\ell'm'}\,,
    \end{aligned}
     \label{coeff CS}
\end{equation}
for $C^{(V,I)}$:
\begin{equation}
    \begin{aligned}
    C^{(V,1)}_{\ell m\ell'm'\ell''m''}&=-
    Q_{\ell'\ell\ell''}f\mathcal{E}_{\ell''m'',r}\Psi_{\ell'm'} \\&+
    Q_{\ell''\ell\ell'}\biggl(
    \frac{f}{\sqrt{\ell''(\ell''+1)}}\mathcal{E}_{\ell''m'',S}\Psi'_{\ell'm'}\\&-\mathcal{B}_{\ell''m'',V}\dot{\Psi}_{\ell'm'}
    \biggr)\,,\\
    C^{(V,2)}_{\ell m\ell'm'\ell''m''}&=
    -    Q_{\ell'\ell\ell''}f\mathcal{E}^{\ast}_{\ell''m'',r}\Psi_{\ell'm'}\\
    &+
    Q_{\ell''\ell\ell'}\biggl(
    \frac{f}{\sqrt{\ell''(\ell''+1)}}\mathcal{E}_{\ell''m'',S}\Psi_{\ell'm'}'\\&-\mathcal{B}_{\ell''m''}\dot{\Psi}_{\ell'm'}
    \biggr)\,,\\
    C^{(V,3)}_{\ell m\ell'm'\ell''m''}&=-\biggl(f\mathcal{E}_{\ell''m'',V}\Psi_{\ell'm'}'+\frac{\mathcal{B}_{\ell''m'',S}\dot{\Psi}_{\ell'm'}}{\sqrt{\ell''(\ell''+1)}}\biggr)\,,\\
    C^{(V,4)}_{\ell m\ell'm'\ell''m''}&=-\biggl(f\mathcal{E}^{\ast}_{\ell''m'',V}\Psi_{\ell'm'}'+\frac{\mathcal{B}^{\ast}_{\ell''m'',S}\dot{\Psi}_{\ell'm'}}{\sqrt{\ell''(\ell''+1)}}\biggr)\,,
    \end{aligned}
     \label{coeff CV}
\end{equation}
and finally for $C^{(\Psi,I)}$:
\begin{equation}
\begin{aligned}
&C^{(\Psi,1)}_{\ell m\ell'm'\ell''m''}=f\mathcal{E}_{\ell'm',r}\mathcal{B}_{\ell''m'',r}+\frac{Q_{\ell''\ell'\ell}}{r^{2}} \\ & \times 
\biggl(
\frac{\mathcal{E}_{\ell'm',S}\mathcal{B}_{\ell''m''S,}}{\sqrt{\ell'\ell''(\ell'+1)(\ell''+1)}}+\mathcal{E}_{\ell'm',V}\mathcal{B}_{\ell''m'',V}
\biggr)\,,\\
&C^{(\Psi,2)}_{\ell m\ell'm'\ell''m''}=f\mathcal{E}_{\ell'm',r}\mathcal{B}^{\ast}_{\ell''m'',r}
\frac{Q_{\ell''\ell'\ell}}{r^{2}}\\& \times \biggl(
\frac{\mathcal{E}_{\ell'm'S}\mathcal{B}^{\ast}_{\ell''m'',S}}{\sqrt{\ell'\ell''(\ell'+1)(\ell''+1)}}+\mathcal{E}_{\ell'm',V}\mathcal{B}^{\ast}_{\ell''m'',V}
\biggr)\,,\\
&C^{(\Psi,3)}_{\ell m\ell'm'\ell''m''}=\frac{1}{r^{2}}\left(\mathcal{E}_{\ell'm',V}\frac{\mathcal{B}_{\ell''m'',S}}{\sqrt{\ell''(\ell''+1)}}
\right.\\&\left.
~~~~~~+\frac{\mathcal{E}_{\ell'm'S}}{\sqrt{\ell'(\ell'+1)}}\mathcal{B}_{\ell''m''V}
\right)\,,\\
&C^{(\Psi,4)}_{\ell m\ell'm'\ell''m''}=\frac{1}{r^{2}}\left(
\mathcal{E}_{\ell'm',V}\frac{\mathcal{B}^{\ast}_{\ell''m'',S}}{r^{2}\sqrt{\ell''(\ell''+1)}}
\right.\\&\left.
~~~~~~+\frac{\mathcal{E}_{\ell'm'S}}{\sqrt{\ell'(\ell'+1)}}\mathcal{B}_{\ell'm'V}^{\ast}
\right)
\,.
\end{aligned}
\label{coeff CPsi}
\end{equation}
In our simulations, we monitor the Newman-Penrose variable $\Phi_{2}$, and the coefficient for spin-weighted spherical harmonics is
\begin{equation}
\begin{aligned}
(\Phi_{2})_{\ell m}=\frac{\sqrt{\ell(\ell+1)}}{2r}\left\{
-\left(\mathcal{B}_{\ell m,V}+\frac{\mathcal{E}_{\ell m,S}}{\sqrt{\ell(\ell+1)}}
\right)
\right.\\
\left.+i\left(
-\mathcal{E}_{\ell m,V}+\frac{\mathcal{B}_{\ell m,S}}{\sqrt{\ell(\ell+1)}}
\right)\right\}\,.
\end{aligned}
\end{equation}
The non-vanishing components of our initial data (see Appendix~\ref{app:evolwithoutplasma}) are 
\begin{equation}
\begin{aligned}
    \Psi_{1,\pm 1}&\sim \Psi_{0}\,,\\
    \mathcal{E}_{10,V}(t=0,r)&=\frac{1}{2}\sqrt{\frac{\pi}{3}}E^{\varphi}(r)\,,
\end{aligned}
\label{eq:initalSelection}
\end{equation}
where $E^{\varphi}(r)$ is defined in~\eqref{eq:InitialElectric}. Since this is a perturbative approach, we focus on the subcritical regime and assume $E^{\varphi}(r)$ is order $\mathcal{O}(\epsilon)$. Using the above equations, we can obtain the order of each mode of $(\Phi_{2})_{\ell m}$ as
\begin{equation}
    \begin{aligned}
    (\Phi_{2})_{1,0}&\sim \mathcal{O}(\epsilon)\,,\\
    (\Phi_{2})_{1,\pm 1}&\sim \mathcal{O}(k_{\rm a}\Psi_{0}\epsilon)\,,\\
    (\Phi_{2})_{2,\pm 1}&\sim \mathcal{O}(k_{\rm a}\Psi_{0}\epsilon)\,,\\
    (\Phi_{2})_{2,\pm 2}&\sim \mathcal{O}((k_{\rm a}\Psi_{0})^{2}\epsilon)\,,\\
    (\Phi_{2})_{3,\pm 3}&\sim \mathcal{O}((k_{\rm a}\Psi_{0})^{3}\epsilon)\,.
    \label{eq:selectionrules}  
    \end{aligned}
\end{equation}
In Fig.~\ref{fig:EMfieldSelectionrules}, we show $|(\Phi_{2})_{\ell m}|$ in the subcritical regime $(\mathcal{I}_{2})$, and rescale all multipoles according to~\eqref{eq:selectionrules}. As can be seen, using the rescaling, all curves are on the same order, demonstrating that our simulations show consistent behavior.

As alluded to in Section~\ref{sec:withoutSR}, from our initial data only odd $\ell$ scalar multipoles can be produced. We can now proof this. From $Y_{\ell m}(\pi-\theta,\varphi+\pi)=(-1)^{\ell}Y_{\ell m}(\theta,\varphi)$, we find that $X_{(1),\ell m\ell'm'\ell''m''}$ and $X_{(2),\ell m\ell'm'\ell''m''}$ are non-zero when $\ell+\ell'+\ell''$ is an {\it even} number, and $X_{(3),\ell m\ell'm'\ell''m''}$ and $X_{(4),\ell m\ell'm'\ell''m''}$ are non-zero when $\ell+\ell'+\ell''$ is an {\it odd} number.

Then, equations~\eqref{coeff Cr}-\eqref{coeff CV} show that the non-vanishing modes of our initial data, $\Psi_{\ell m}$ and $\mathcal{E}_{\ell m,V}$ [see~\eqref{eq:initalSelection}] with odd $\ell$ can {\it only} excite $\mathcal{E}_{\ell m,r}$ and $\mathcal{E}_{\ell m,S}$ with even $\ell$, while it excites $\mathcal{E}_{\ell m,V}$ with odd $\ell$. 

Next,~\eqref{Maxeqspherical2} implies that the non-vanishing component of $\mathcal{E}_{\ell m,r}$ and $\mathcal{E}_{\ell m,S}$ with even $\ell$ excites $\mathcal{B}_{\ell m,V}$ with even $\ell$, while the non-vanishing component of $\mathcal{E}_{\ell m,V}$ with odd $\ell$ excites $\mathcal{B}_{\ell m,r}$ and $\mathcal{B}_{\ell m,S}$ with odd $\ell$. 

Finally,~\eqref{coeff CPsi} shows that the non-vanishing component of $\mathcal{E}_{\ell m,r}$, $\mathcal{E}_{\ell m,S}$, and $\mathcal{B}_{\ell m,V}$ with even $\ell$, and $\mathcal{E}_{\ell m,V}$, $\mathcal{B}_{\ell m,r}$, and $\mathcal{B}_{\ell m,S}$ with odd $\ell$ {\it only} excites $\Psi_{\ell m}$ with odd $\ell$. Therefore, the non-vanishing components of the simulation starting from initial data~\eqref{eq:initalSelection}, only excite $\Psi_{\ell m}$ with odd $\ell$. These results are consistent with our simulations, see Fig.~\ref{fig:scalarfieldmodes}. 
\bibliography{References}
\end{document}